\documentclass[twocolumn]{aastex701}

\usepackage{amsmath}

\newcommand{\jn}{\ensuremath{j_{n}}}
\newcommand{\jt}{\ensuremath{j_{T}}}
\newcommand{\jp}{\ensuremath{j_{p}}}
\newcommand{\jk}{\ensuremath{j_{K}}}

\graphicspath{{./}{figures/}}

\shorttitle{Sloshing cold fronts in galaxy clusters}
\shortauthors{Li et al.}

\begin{document}

\title{An azimuthally resolved study of sloshing cold fronts in three nearby galaxy clusters}

\correspondingauthor{Shutaro Ueda and I-Non Chiu}

\author[0009-0003-5645-4212]{I-Hsuan Li}
\affiliation{Department of Physics, National Cheng Kung University, No.1, University Road, Tainan City 70101, Taiwan}
\email{L26134314@gs.ncku.edu.tw}

\author[0000-0001-6252-7922]{Shutaro Ueda}
\affiliation{
Faculty of Mathematics and Physics, Institute of Science and Engineering, Kanazawa University, Kakuma, Kanazawa, Ishikawa, 920-1192, Japan
}
\affiliation{
Advanced Research Center for Space Science and Technology (ARC-SAT), Kanazawa University, Kakuma, Kanazawa, Ishikawa, 920-1192, Japan
}
\email{shutaro@se.kanazawa-u.ac.jp}

\author[0000-0002-5819-6566]{I-Non Chiu}
\affiliation{Department of Physics, National Cheng Kung University, No.1, University Road, Tainan City 70101, Taiwan}
\email{inchiu@phys.ncku.edu.tw}

\author[0000-0002-7196-4822]{Keiichi Umetsu}
\affiliation{Academia Sinica Institute of Astronomy and Astrophysics (ASIAA), No. 1, Section 4, Roosevelt Road, Taipei 106216, Taiwan}
\email{keiichi@asiaa.sinica.edu.tw}

\begin{abstract}
We present a detailed analysis of sloshing cold fronts in a sample of three nearby galaxy clusters (Abell\,496, Abell\,2029, and Abell\,1644) observed with the {\em Chandra} X-ray Observatory. 
Cold fronts manifest as sharp edges in the X-ray surface brightness of the intracluster medium (ICM) in galaxy clusters. In the residual X-ray surface brightness maps, where the global ICM distribution has been subtracted, cold fronts generated by gas sloshing are observed at the boundaries of the spiral excesses. 
We perform a systematic and comprehensive study of the surface brightness edges along the spiral excesses. 
We find the deficit of the thermal pressure radially inward of the brightness edges, in contrast to stripping cold fronts that typically exhibit higher thermal pressure in brightness edges.
Assuming that the non-thermal pressure support entirely comes from the gas motions, we estimate the velocity dispersions across the edges and do not find statistically significant azimuthal variations along the sharpest edges.
With an assumption that the non-thermal pressure support is completely attributed to the magnetic pressure, we derive the upper bounds on the strength of magnetic fields, yielding $6\,\mu\mathrm{G}$, $15\,\mu\mathrm{G}$, and $5\,\mu\mathrm{G}$ for Abell\,496, Abell\,2029, and Abell\,1644, respectively.
We discuss several systematic uncertainties in characterizing the sloshing cold fronts and find that the choice of spectral extraction regions has an effect for subtle cold fronts observed with shallow X-ray data.
\end{abstract}




\section{Introduction}
\label{sec:Intro}

Galaxy clusters, the largest gravitationally collapsed systems in the Universe, accrete their masses by merging with other smaller systems, as well as continuous mass accretion from their surrounding environments. Because the intracluster medium (ICM), which is the diffuse and hot X-ray emitting gas, dominates the baryonic component of galaxy clusters, the mass accretion leaves imprints on the X-ray surface brightness of ICM, such as disturbed morphologies and clumpiness.

High angular-resolution observations performed with the {\em Chandra} X-ray Observatory have revealed sharp contact discontinuities in the X-ray surface brightness of ICM, recognized as cold fronts or shock fronts \citep[see a complete review in][]{Markevitch2007}. 
Both kinds of fronts are associated with cluster mergers. 
While shock fronts typically originate from the kinetic energy of major-merger events dissipating in the ICM, cold fronts are considered to be generated by gas motions at an interface between two phases of the ICM with different entropies. 
Specifically, a cold front is characterized by a sharp discontinuity across an edge, where the gas on the inner side (typically a remnant core or substructure) has higher density and lower temperature, but maintains the pressure equilibrium with hotter and more diffuse ambient ICM.
Cold fronts are more frequently detected than shock fronts primarily due to their significantly longer observable lifetimes, which can persist for several Gyrs \citep{Markevitch2007}.
Given their sharp discontinuities and observability, cold fronts are considered as an ideal place to study the microphysical properties of ICM \citep{Markevitch2007}.

Cold fronts are broadly classified into two types: stripping and sloshing cold fronts. 
Stripping cold fronts are typically observed at the fronts of the collisional gas component of the infalling subclusters \citep{Markevitch2000, Vikhlinin2001}. 
When an infalling subcluster hosting a cooler and denser ICM moves through the surrounding gas, a contact discontinuity between two different phases of the gas forms a stripping cold front. 
For example, stripping cold fronts have been prominently observed in several nearby merging clusters \citep{Hallman2004, Ichinohe2017, Erdim2019, Ichinohe2021}. 

On the other hand, sloshing cold fronts are generated by gas sloshing that is frequently observed in the central region of the cool-core clusters. 
A widely accepted picture is that minor mergers of substructures with a non-zero impact parameter induce the sloshing motion of the gas \citep{Ascasibar2006, Roediger2011, Ichinohe2015}. 
As the gravitational potential of a cluster is disturbed by the motion of substructures falling into the core, the central ICM gains the
angular momentum and hence starts to slosh around the cluster center, leading to the spiral motion of gas observed in an X-ray image.

\begin{figure}
\centering
\resizebox{0.45\textwidth}{!}{\includegraphics[scale=1]{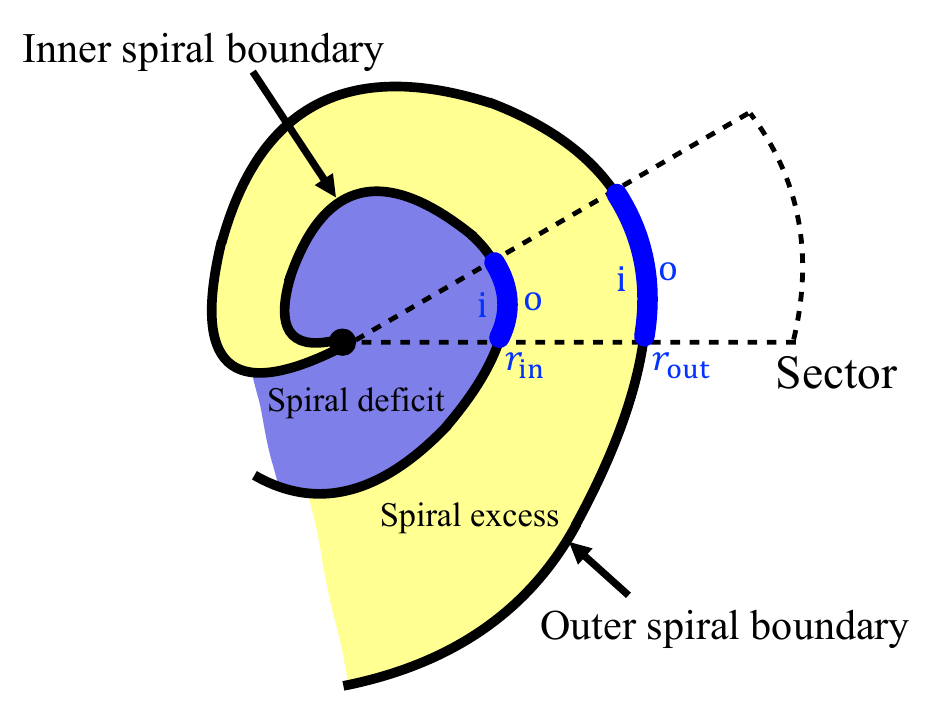}}
\caption{
This schematic illustration represents substructures seen in a residual X-ray surface brightness map.
The spiral excess (the yellow region) is the primary focus of this work.
For each detected surface brightness edge ($r_{\rm{in}}$ and $r_{\rm{out}}$), we denote the side closer to the cluster center with the subscript ${\rm{i}}$, and the farther side with the subscript ${\rm{o}}$.
Note that the subscripts ``$\rm{i}$'' and ``$\rm{o}$'' strictly refer to the radial position relative to the edge. For the outer spiral boundary, the spiral excess lies in region ``$\rm{i}$'', whereas for the inner spiral boundary, the spiral excess lies in region ``$\rm{o}$''.
}
\label{fig:Notation}
\end{figure}

Observationally, the spiral excess of sloshing cold fronts 
(see Figure~\ref{fig:Notation} for the schematic illustration) can be identified in the residual X-ray surface brightness map, from which the global ICM distribution is subtracted \citep{Churazov2003,Clarke2004,Blanton2011}.
Sloshing cold fronts typically manifest as multiple concentric arc-shaped discontinuities along the edge of the sloshing spiral, for example, Abell\,496 \citep{Ghizzardi2014} and Abell\,85 \citep{Ichinohe2015}. Additionally, the brightness edges of sloshing cold fronts are relatively subtle in contrast to the striking brightness edges observed in stripping cold fronts, leading to challenges in detecting the sloshing pattern. 
Hence, studies of sloshing cold fronts are still greatly in need to further explore the ICM physics.

Assuming that the ICM behaves as an ideal gas with a sound speed $c_{\mathrm{s}}\approx 1000~\mathrm{km}/\mathrm{s}$ and mass density $\rho$, we consider a local pressure non-equilibrium $\Delta P$ across a cold front surface of width $W\approx50~\mathrm{kpc}$. This non-equilibrium induces a gas acceleration $a = \frac{\Delta P}{\rho W}$.
Assuming adiabatic gas motion ($c_{\mathrm{s}}^2 \sim \frac{P}{\rho}$), the equilibrium time is then $t \sim
\frac{W}{c_{\rm s}}
\frac{1}{
\sqrt{\frac{\Delta P}{P}}
}$. For a significant pressure gradient $\frac{\Delta P}{P} \geq 1$, this timescale is $\sim 50~\mathrm{Myr}$, significantly shorter than the lifetime of a cold front.
That is, cold fronts are expected to be close to total pressure equilibrium. However, the main feature of stripping cold fronts is the significant gradient of the thermal pressure across the X-ray surface brightness edges \citep{Kempner2002, Rossetti2007, Ueda2024}. 
This phenomenon indicates that the non-thermal pressure substantially contributes to the total pressure support to maintain the sharpness of stripping cold fronts. 
Similar thermal pressure gradients have also been found in some sloshing cold fronts \citep{Ghizzardi2014}. 
However, there is also evidence in some galaxy clusters revealing the lack of thermal pressure gradients in sloshing cold fronts \citep{Markevitch2001,Chen2017,Paterno-Mahler2013}, indicating the absence of the non-thermal pressure support.
Several studies of simulations have shown that the turbulent gas motions, magnetic pressures, and the viscosity of the ICM may be responsible for the non-thermal pressure support \citep{ZuHone2010, ZuHone2011, Ichinohe2019, Gatuzz2022}.
Therefore, it is important to investigate the thermodynamics of sloshing cold fronts in a homogeneous manner. 
To this end, azimuthally resolved measurements provide a comprehensive assessment of the mechanism of gas sloshing.

In this work, we carry out an azimuthally resolved study of sloshing cold fronts in three nearby galaxy clusters with prominent surface brightness edges close to the cluster center. 
Our sample of three clusters includes Abell\,496, Abell\,2029, and Abell\,1644, which are referred to as A496, A2029, and A1644, respectively. 
With the goal of investigating the thermodynamic properties along the sloshing cold fronts, we azimuthally measure the density and temperature profiles across the X-ray surface brightness edges.
Based on these measurements, we further estimate the upper bounds of the velocity dispersions and magnetic field strength of the ICM required by the non-thermal pressure support at the detected surface brightness edges.

When referring to the sides of the cold front (see Figure~\ref{fig:Notation}), we use the subscript ``$\rm{i}$'' (inner) to denote the region radially inward of the edge (closer to the cluster center) and ``$\rm{o}$'' (outer) to denote the region radially outward of the edge.
When there are two edges in the same position angle from the cluster center, measurements corresponding to the edge closer to the center are denoted with the subscript ``$\rm{in}$'', while those for the farther edge are denoted with the subscript ``$\rm{out}$''.
We refer to the overdense spiral pattern in the residual X-ray surface brightness map as the spiral excess, and the underdense spiral pattern as the spiral deficit.
In this paper, the contrasts of density (\jn), temperature (\jt), thermal pressure (\jp), and entropy (\jk) are defined as the ratios of the quantities measured just radially inward of the X-ray surface brightness edge to those measured just radially outward of the edge, i.e., 
\[
j_{\mathrm{X}} \equiv \frac{\mathrm{X}_{\mathrm{i}}}{\mathrm{X}_{\mathrm{o}}} \, ,
\]
where $\mathrm{X}\in\left\lbrace n, T, p, K\right\rbrace$.
Figure~\ref{fig:Notation} provides a schematic illustration of a typical sloshing spiral feature, along with the notation adopted in this study.

This paper is organized as follows. 
A brief summary of previous studies related to the sloshing cold fronts in our sample is provided in Section~\ref{sec:SampleData}. 
In Section~\ref{sec:Analyses}, we describe the method and results of our measurements. 
We discuss the implications and systematic uncertainties in Section~\ref{sec:Discussion}. 
Finally, the conclusions are made in Section~\ref{sec:Summary}. 

Throughout this paper, we assume a flat $\Lambda$CDM cosmology with a Hubble constant of
$H_{0} = 70\,\mathrm{km}\,\mathrm{s}^{-1}\,\mathrm{Mpc}^{-1}$ and a matter density of $\Omega_{\rm{m}} = 0.3$. 
Unless otherwise stated, quoted errors stand for the $68\%$ confidence level ($1\sigma$). 
All position angles are measured in the counterclockwise direction, from west ($0^{\circ}$) to north ($90^{\circ}$).

\begin{figure*}[ht!]
    \begin{center}
        \includegraphics[width=5.4cm]{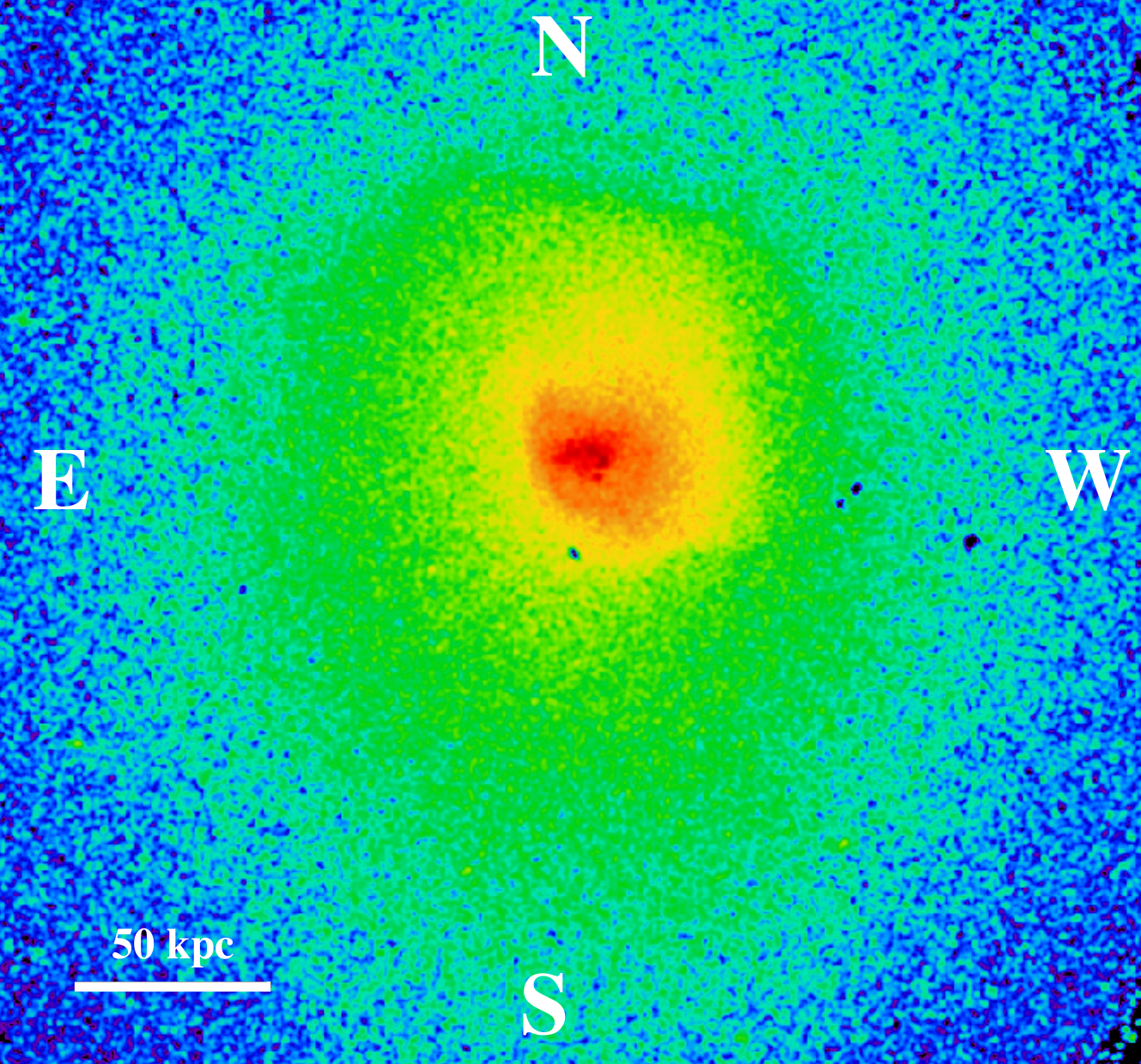}
        \includegraphics[width=5.4cm]{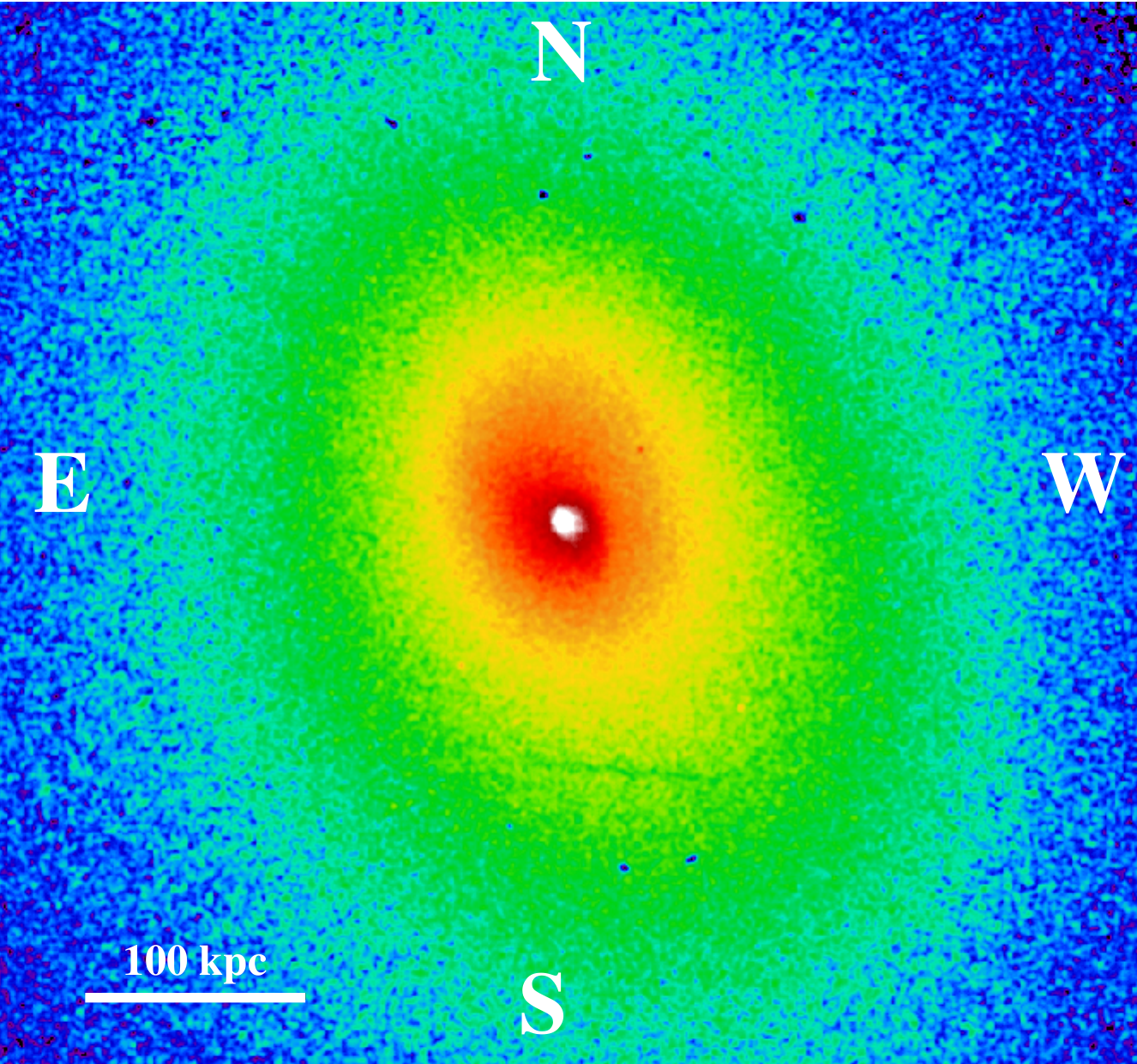}
        \includegraphics[width=5.4cm]{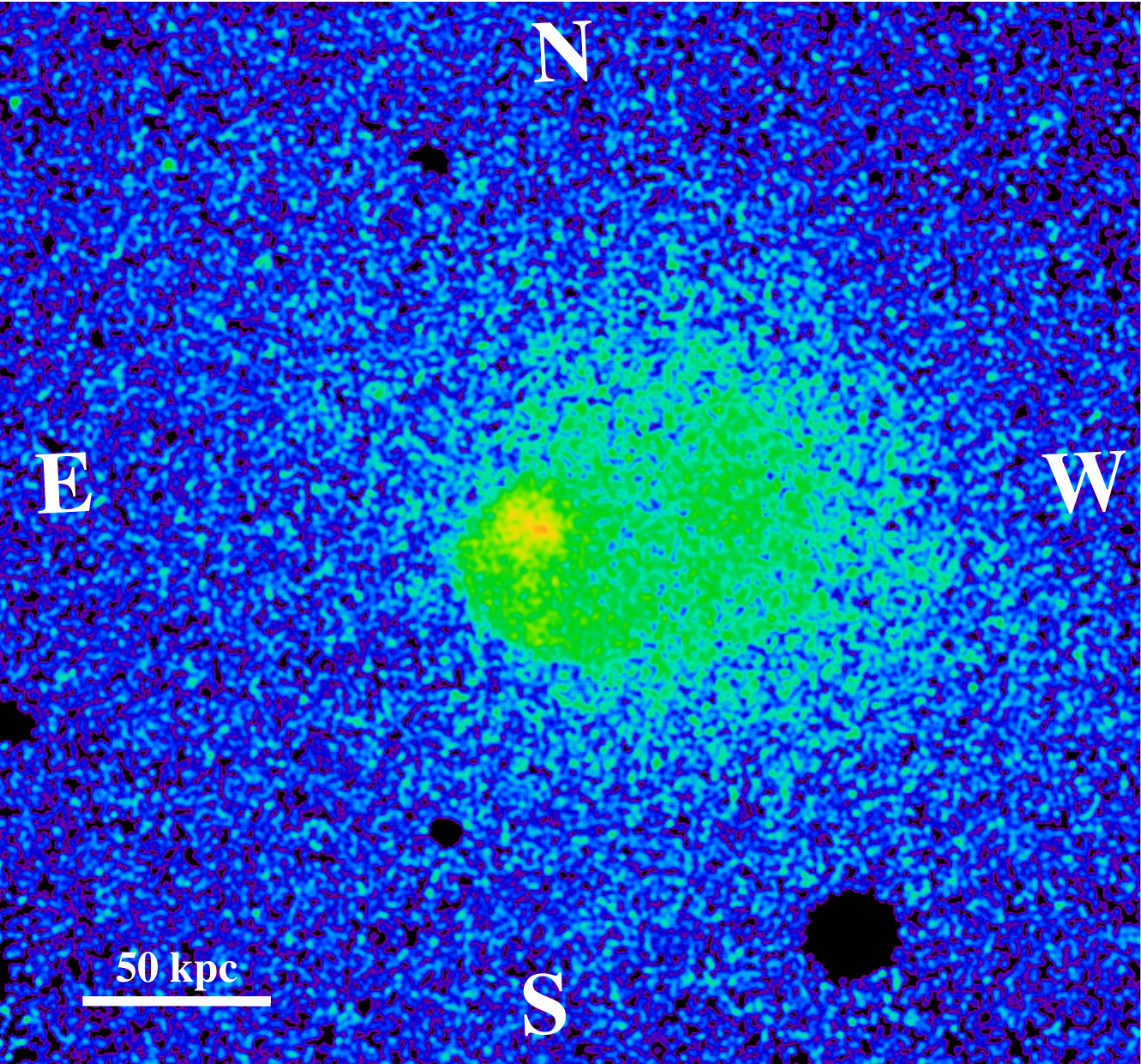}
        \includegraphics[width=5.4cm]{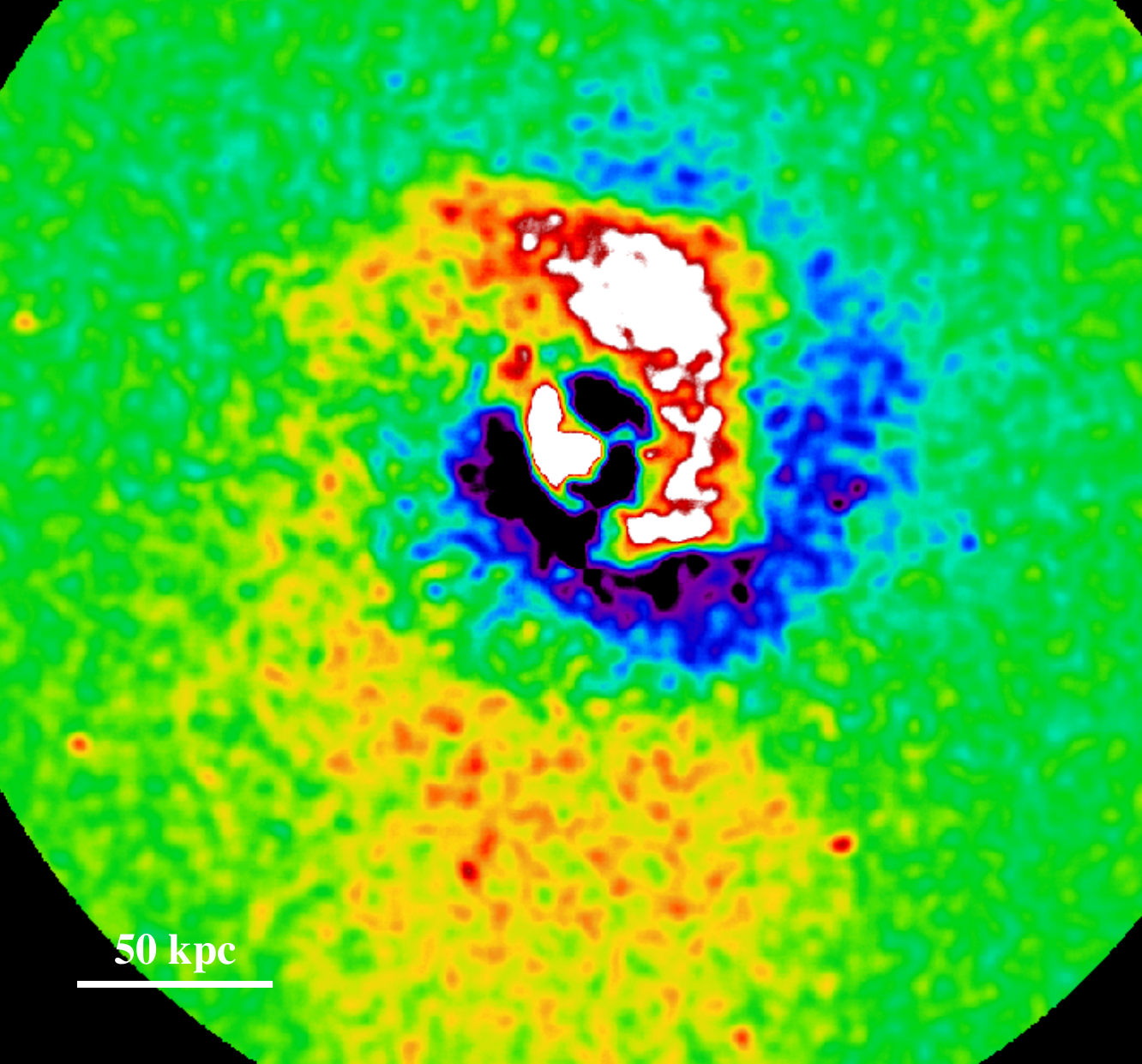}
        \includegraphics[width=5.4cm]{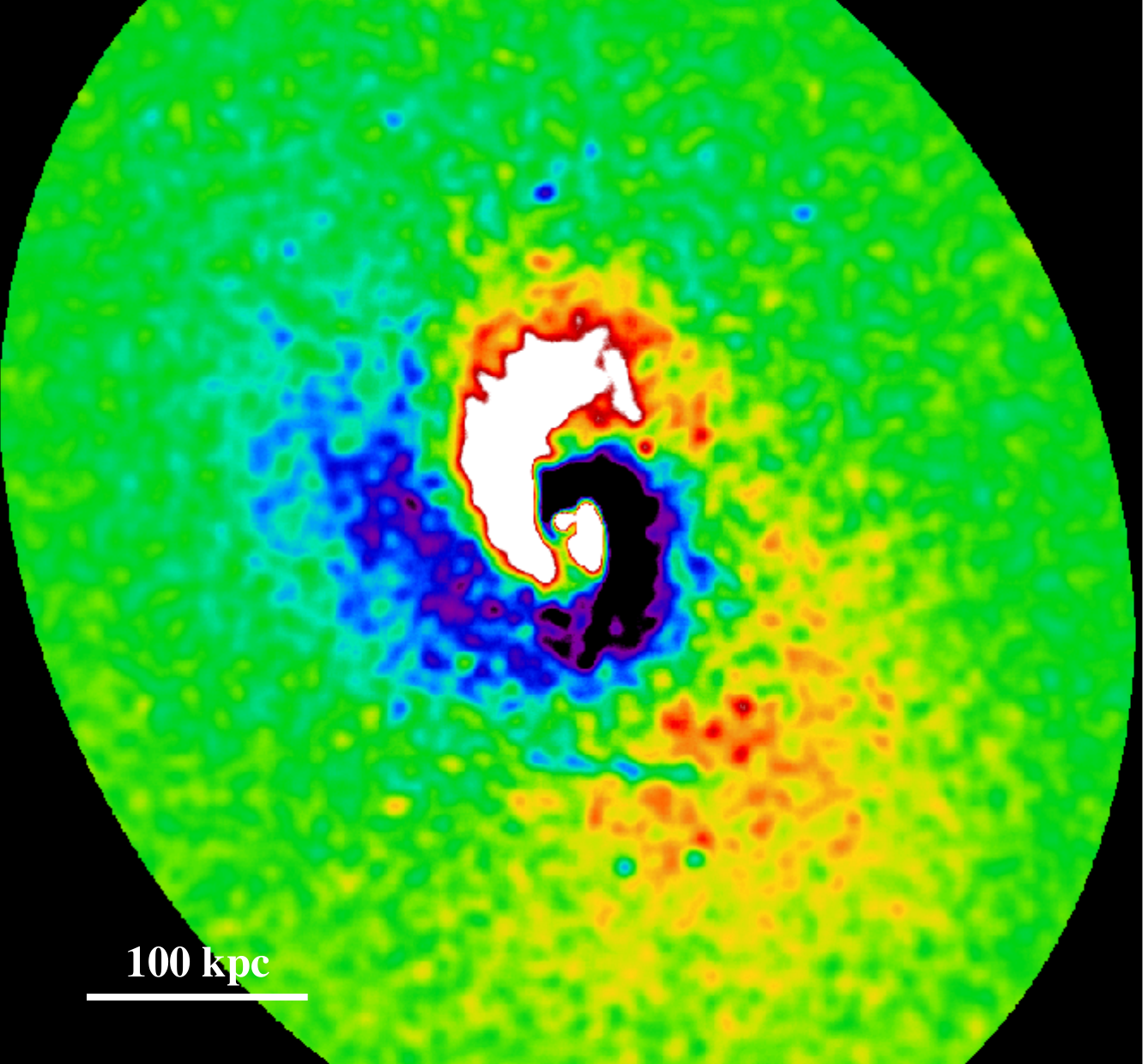}
        \includegraphics[width=5.4cm]{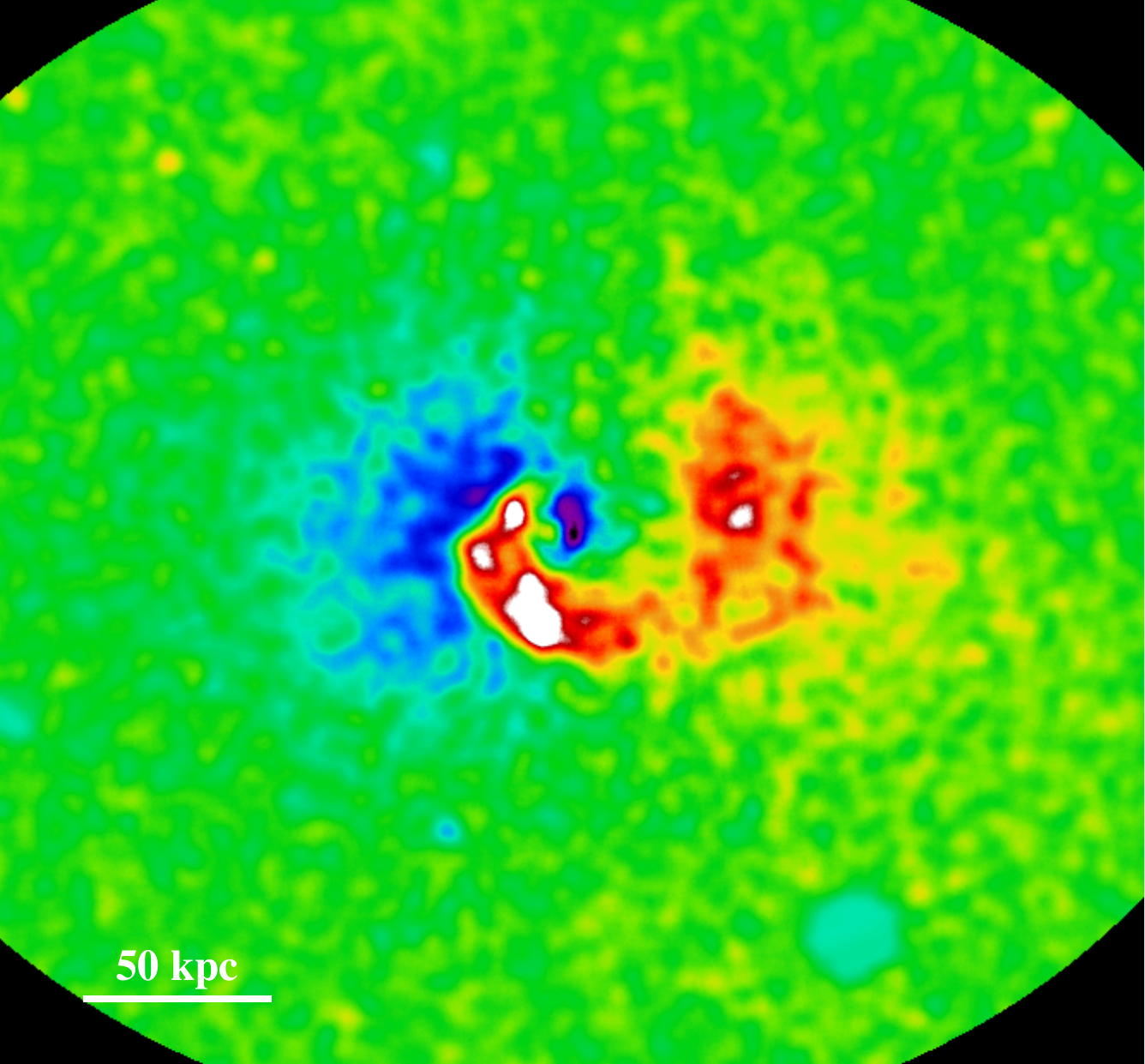}
        \includegraphics[width=5.4cm]{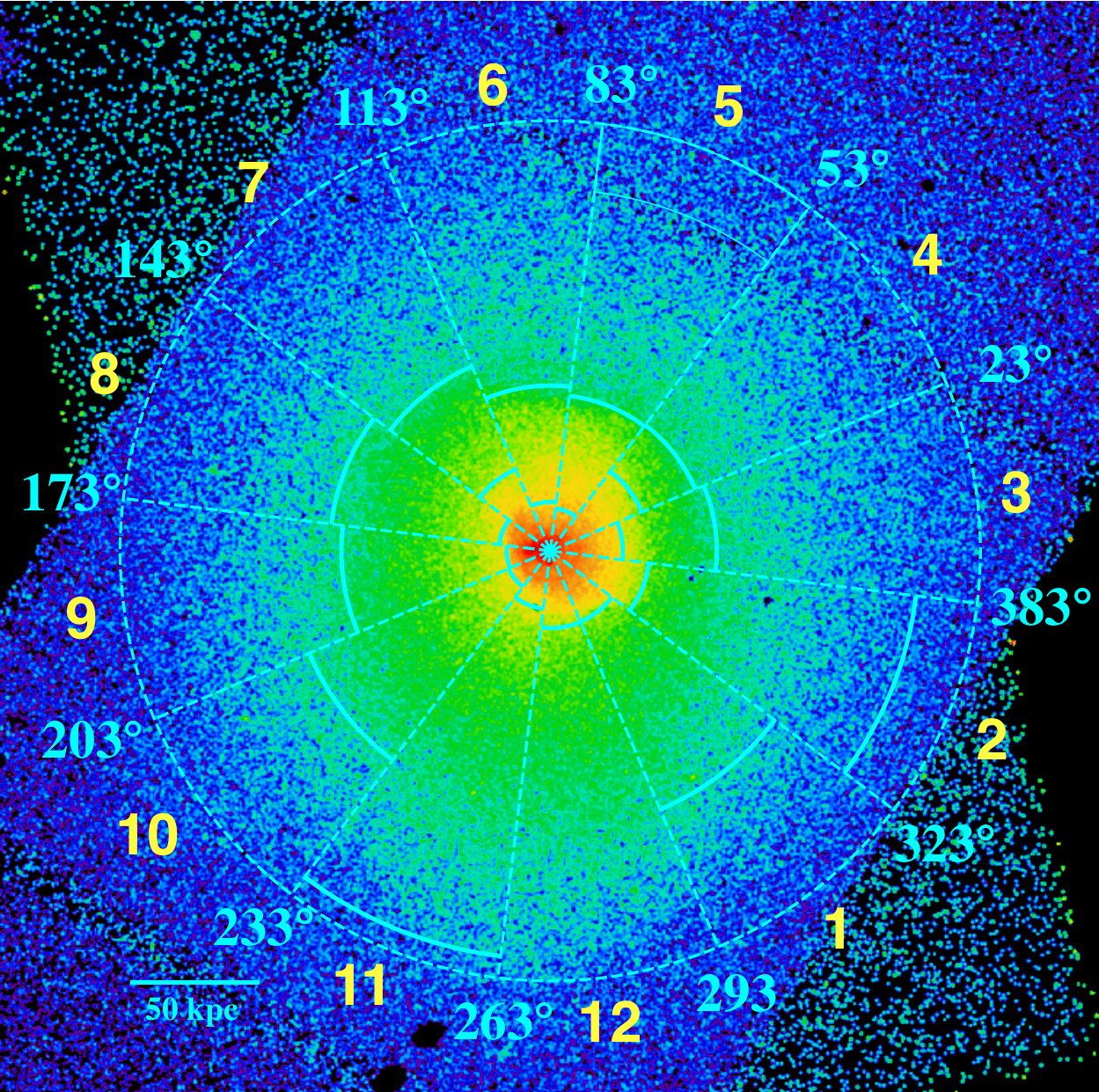}
        \includegraphics[width=5.4cm]{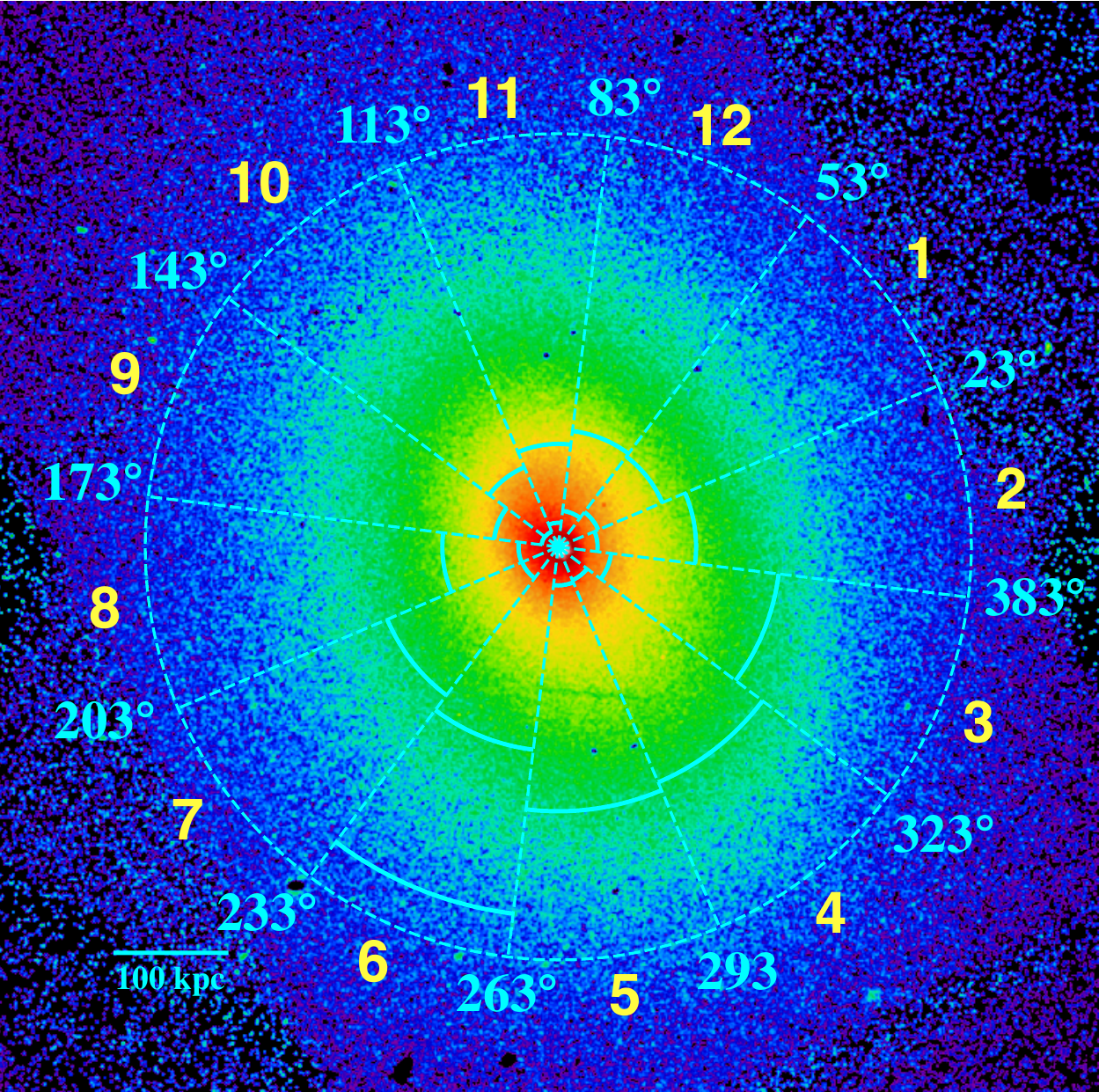}
        \includegraphics[width=5.4cm]{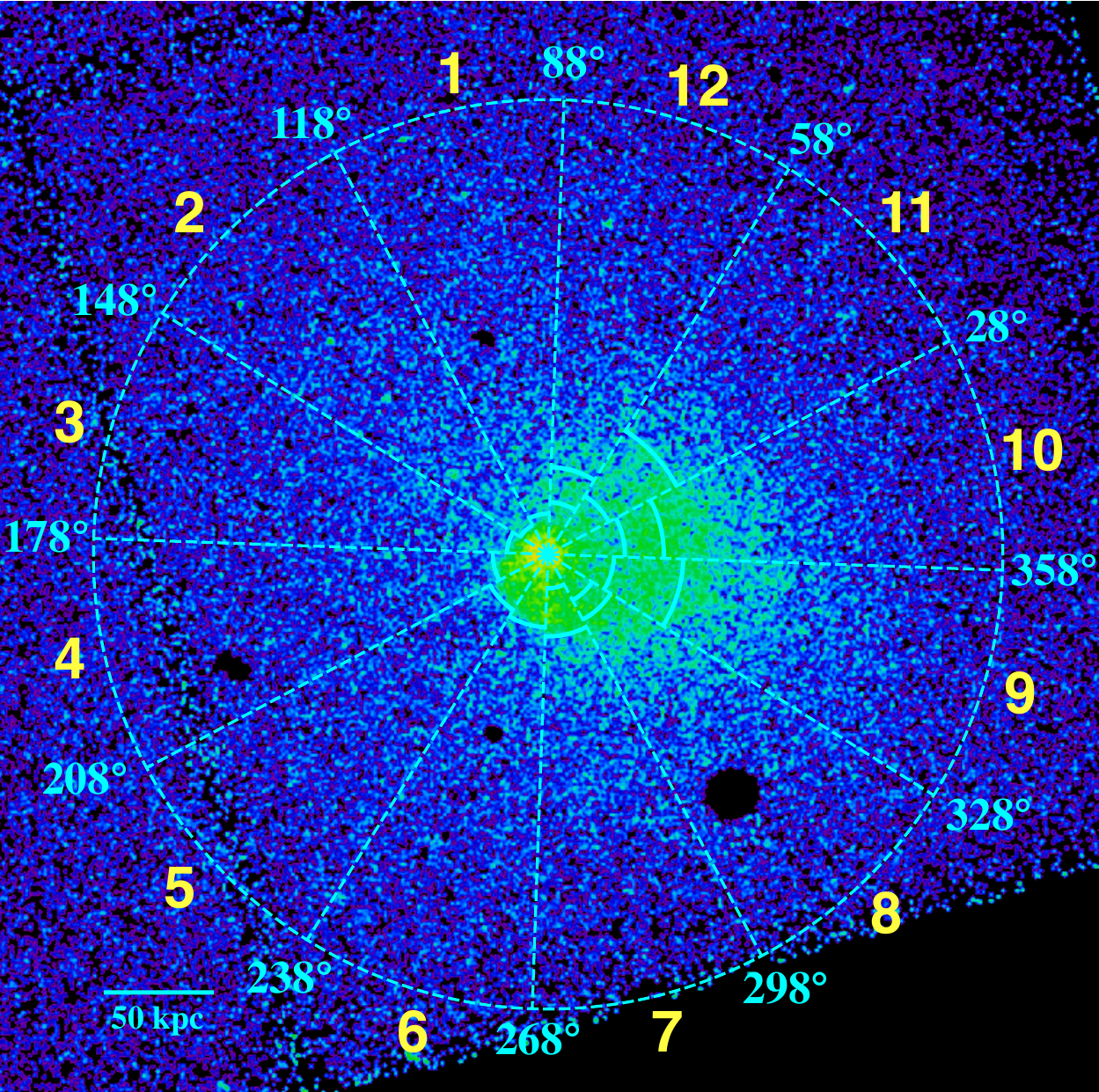}
        \includegraphics[width=5.4cm]{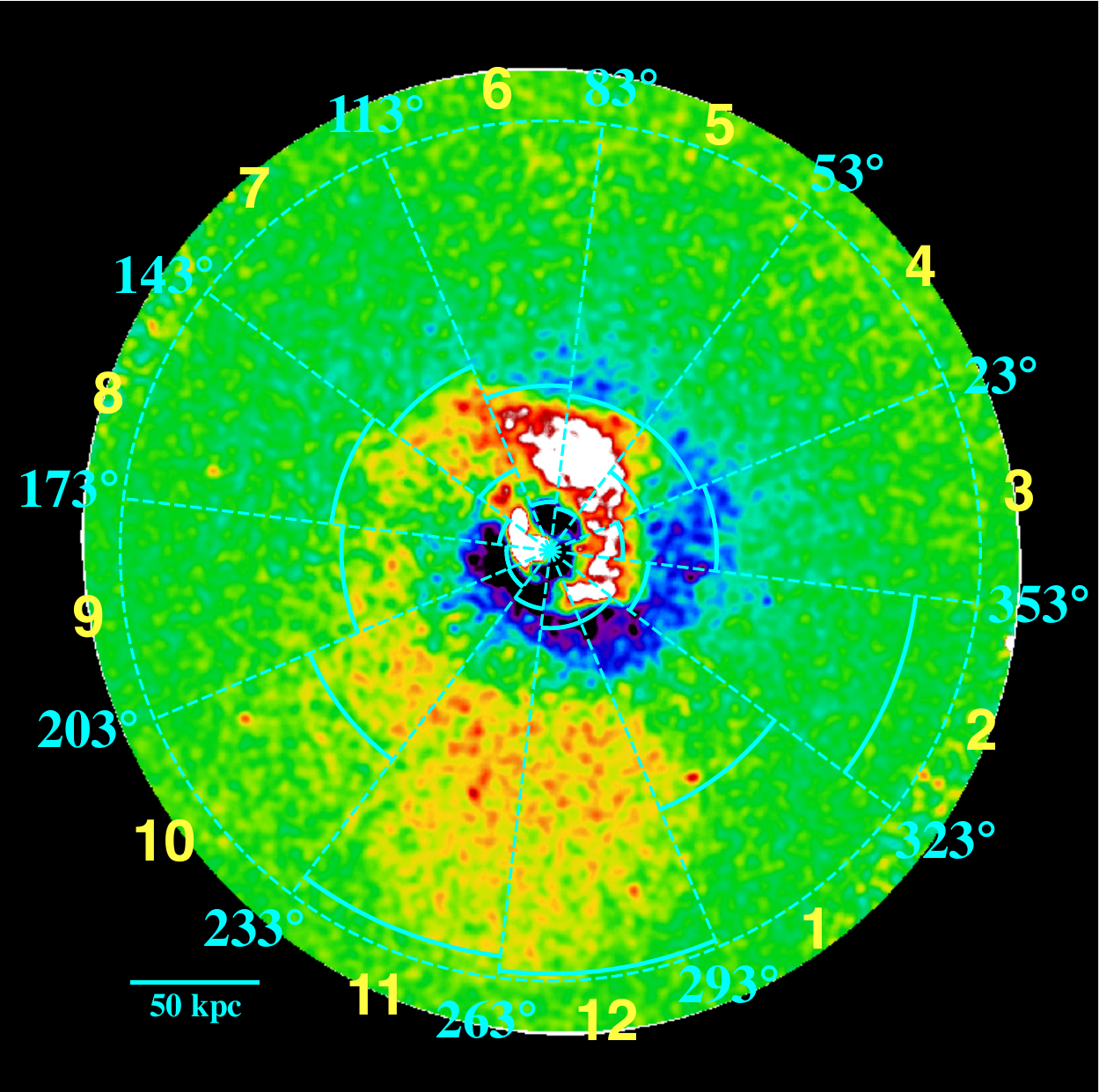}
        \includegraphics[width=5.4cm]{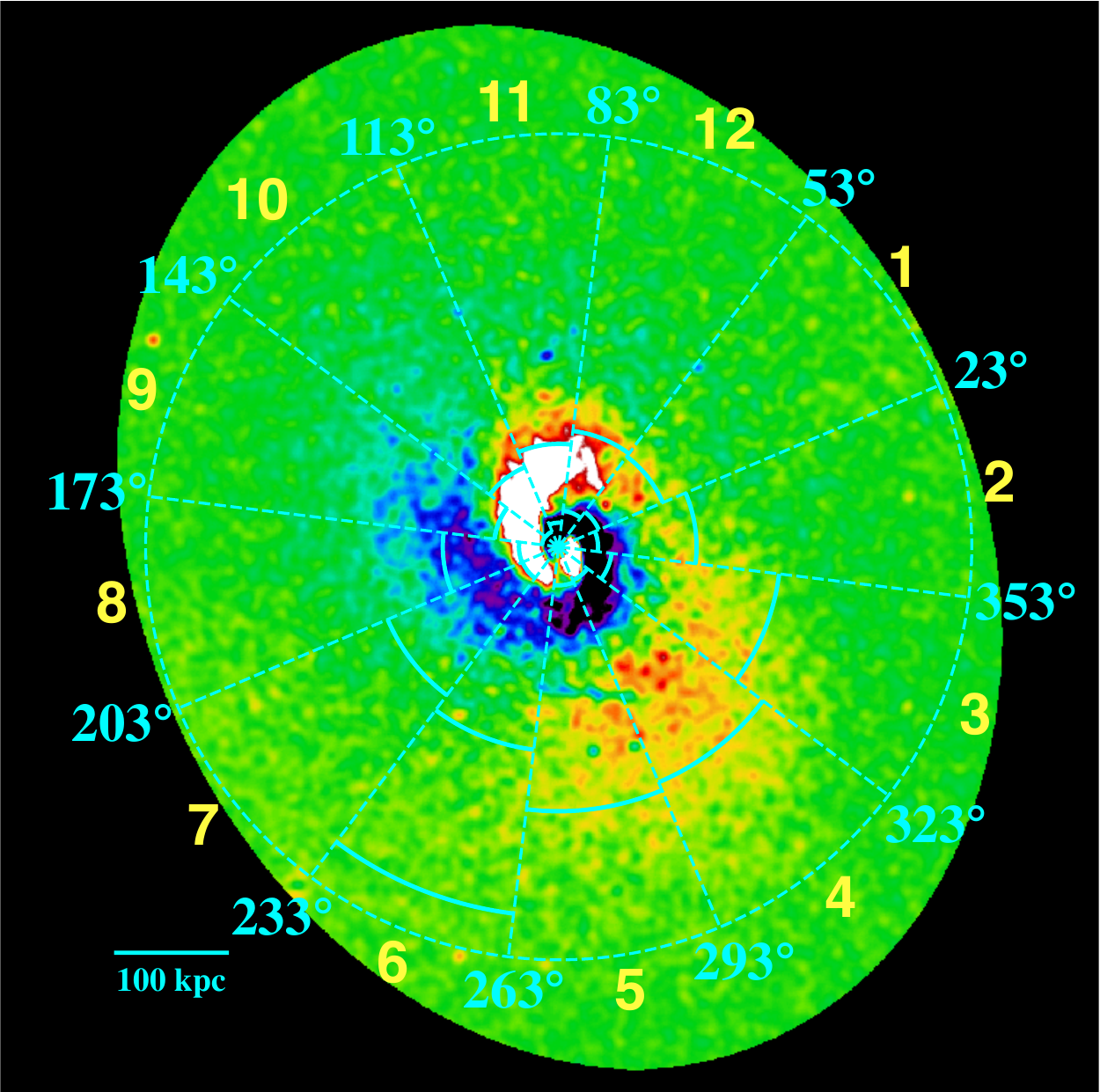}
        \includegraphics[width=5.4cm]{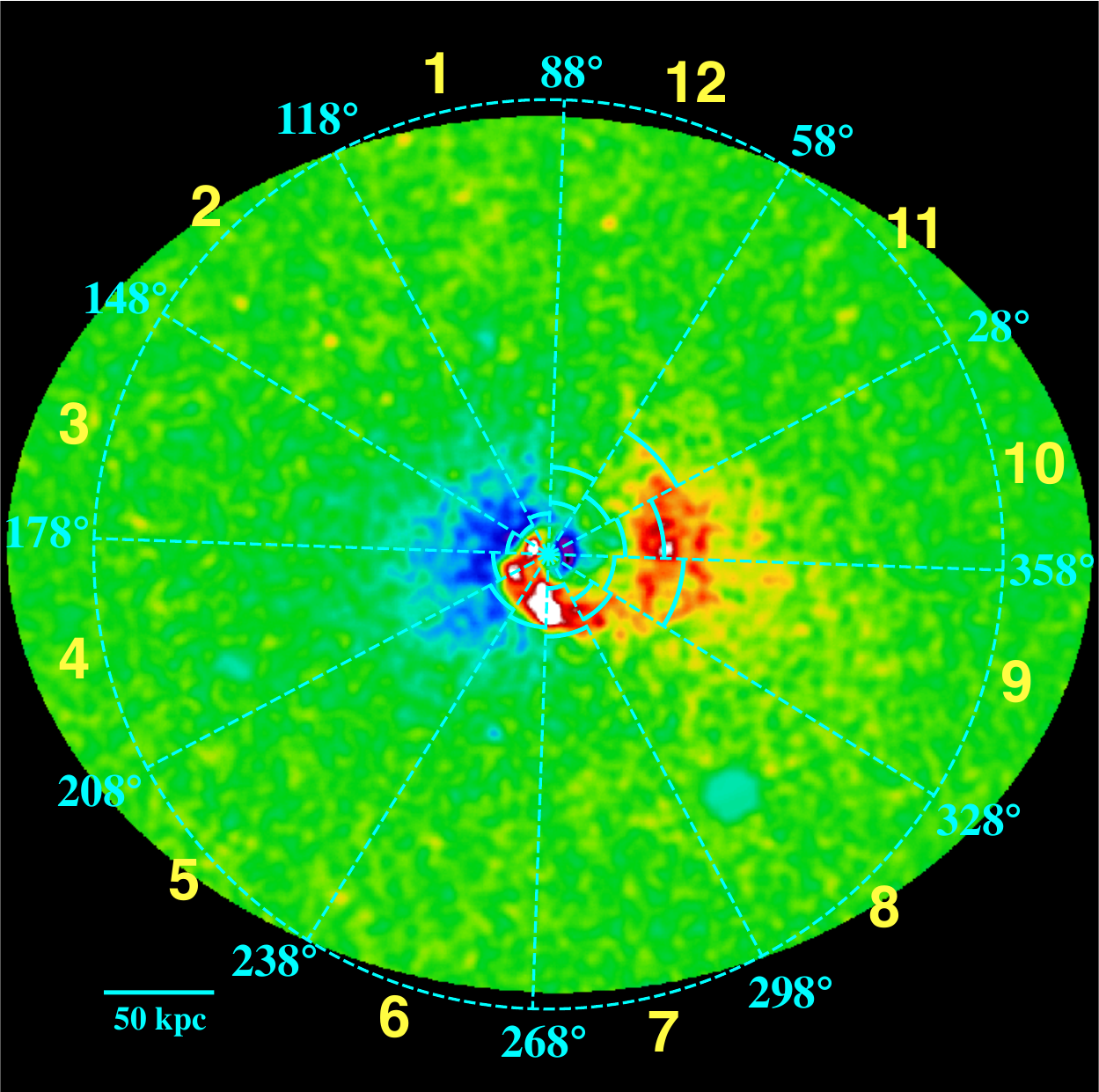}
    \end{center}
\caption{
The {\em Chandra} X-ray surface brightness images (first and third rows) and residual maps (second and fourth rows) of A496 (left), A2029 (middle), and A1644 (right).
The zoom-out version of the first and second rows are shown as the third and fourth rows, respectively. 
The surface brightness images and residual maps are smoothed by a Gaussian kernel with $3\arcsec$ and $10\arcsec$\,FWHM, respectively.
Point sources and backgrounds are masked by black ellipses in the X-ray surface brightness maps. 
The bars in the lower-left corner of each panel show the corresponding spatial scales. 
The numbers in yellow denote the indices of the sectors defined in Table~\ref{tab:AnalysesSxsectors}.
The sectors in cyan with an opening angle of $30^{\circ}$ show the regions defined for the azimuthal study.
The arcs within the sectors denote the positions of X-ray surface brightness edges, determined by the X-ray surface brightness fitting described in Section~\ref{sec:AnalysesSx}.
}
\label{fig:AnalysesImaging}
\end{figure*}


\section{Sample and observation data}
\label{sec:SampleData}

\subsection{Sample}
\label{sec:Sample}

We study the three nearby galaxy clusters: A496, A2029, and A1644, within the redshift range $0.03 < z < 0.08$.
They are included in the X-ray-selected cluster catalog, known as the HIghest X-ray FLUx Galaxy Cluster Sample \citep[HIFLUGCS;][]{Reiprich2002}. 
The basic properties of the clusters are contained in Table~\ref{tab:DataObservations}.
The three clusters are bright and exhibit clearly large excesses of sloshing spirals in their residual X-ray surface brightness maps. 
Since the spiral excess indicates the presence of the gas sloshing in the plane of the sky, the clear and large spiral excesses in our samples suggest that they are ideal targets to conduct a systematic and azimuthally resolved study.
Here, we briefly summarize previous results related to our sample.

\subsubsection*{A496}
\label{sec:SampleA496}

A496 is a nearby and bright galaxy cluster ($z = 0.03309$ and $1'' = 0.660\,\mathrm{kpc}$) with a highly symmetric X-ray morphology. With {\em Chandra} observations, \cite{Dupke2003} and \cite{Dupke2007} studied the density and temperature profiles, along with the metal distribution across the cold fronts, and concluded that the cold fronts in this cluster could be generated by gas sloshing. \cite{Tanaka2006} found the cold front approximately $240''$ south of the cluster center, and estimated the upper limit of the Mach number of the ICM to be $\approx 0.2$.
They also suggested that A496 has undergone a merger in the past.
Additionally, \cite{Ghizzardi2014} observed that the sloshing spiral has a lower temperature and entropy distribution compared to the ambient gas.
Moreover, they observed the thermal pressure gradients 
of cold fronts within A496: one between $30^{\circ}$ and $75^{\circ}$ at a radius of $92''$ away from the cluster center, another between $75^{\circ}$ and $120^{\circ}$ at a radius of $102''$,  
and the third between $240^{\circ}$ and $285^{\circ}$ at a radius of $252''$; each of them was measured with a thermal pressure contrast significantly larger than unity.

\subsubsection*{A2029}
\label{sec:SampleA2029}

A2029 is one of the most relaxed clusters and located at redshift $z = 0.07872$ ($1'' = 1.488\,\mathrm{kpc}$). Several studies with {\em Chandra} observations have been done:
Based on spectral analyses, \cite{Clarke2004} suggested that the cooling flow in this cluster is in an early stage. They also discovered extended spiral excesses in the residual X-ray surface brightness map, which are associated with the stripping gas from infalling substructures or gas sloshing motions.
\cite{Paterno-Mahler2013} analyzed the northern and southern part of the spiral excesses and obtained smooth thermal pressure profiles in both regions.
However, \cite{Naor2020} performed a deprojected analysis of the southwest cold front at $22''$, and found that the thermal pressure radially inward of the cold front is 
lower than the opposite side at a level of $\approx 1.5\sigma$.
In addition, they measured thermal pressure contrasts of seven sloshing cold fronts and confirmed the 
deficit of the thermal pressure on the side with higher density.

\subsubsection*{A1644}
\label{sec:SampleA1644}

A1644 ($z = 0.04740$ and $1'' = 0.930\,k\mathrm{pc}$) is a post-merging system consisting of two systems: 
a primary halo (A1644-S) located in the southwest of the sky and a subcluster (A1644-N) in the northeast, each connected by a band of warm gas \citep{Reiprich2004}.
\cite{Reiprich2004} compared simulations to observations in X-rays, optical, and radio wavelengths, suggesting that the subcluster has passed through the primary halo.
Additionally, \cite{Monteiro-Oliveira2020} measured the physical distance between the two components at $723\,\mathrm{kpc}$.
Regarding cold fronts, \cite{Johnson2010} studied the X-ray surface brightness edges in A1644 with {\em Chandra} observations. 
They measured the cold fronts in the main cluster (A1644-S) at $12.1''$ in the northeast and $32.6''$ in the southeast. 
We only focus on A1644-S in this work, because A1644-N does not exhibit a spiral excess in its residual X-ray surface brightness map. 
Note that \cite{Johnson2010} studied the western region of A1644-N and made the same conclusion.

\subsection{Observations and data reduction}
\label{sec:Data}

\begin{table*}
    \centering
        \caption{Summary of {\em Chandra} X-ray observations and basic properties for our sample.}
        \label{tab:DataObservations}
            \begin{tabular}{cccccccc}
            \hline
            \hline
            Cluster & Redshift & Physical scale 
                    & Expo. time\tablenotemark{a}
                    & ObsID\tablenotemark{b}
                    & Core type
                    & $M_{500}$\tablenotemark{c} & Temperature\tablenotemark{d} \\
                    &          &$\mathrm{kpc}/''$ & $k\mathrm{s}$
                    & &
                    & $10^{14}\, M_\odot$
                    &$\mathrm{keV}$ \\
            \hline
            A496  & $0.03309$ & $0.660$ & $60.2$  & $3361, 4976, 931$ & cool-core & $3.9$ & $6.2$ \\
            A2029 & $0.07872$ & $1.488$ & $100.4$ & $4977, 6101, 891$ & cool-core & $8.4$ & $8.1$ \\
            A1644 & $0.04740$ & $0.930$ & $63.9$  & $2206, 7922$      & cool-core & $4.4$ & $5.1$ \\
            \hline       
            \end{tabular}\\
    \tablenotemark{a}{Total net exposure time of {\em Chandra} observations after masking flare-time intervals.} \\
    \tablenotemark{b}{{\em Chandra} observation identification (ObsID) numbers.} \\
    \tablenotemark{c}{Masses are taken from \cite{Markevitch1999}, \cite{Li2023}, and \cite{Johnson2010} for A496, A2029, and A1644, respectively.} \\
    \tablenotemark{d}{The mean deprojected temperature of the outermost radial bins.}
\end{table*}

\begin{table*}
    \centering
        \caption{Summary of the elliptical model used for generating the residual maps.}
        \label{tab:AnalysesImagingModel}
            \begin{tabular}{cccccc}
            \hline
            \hline
            Cluster & R.A.\tablenotemark{a} & Decl.\tablenotemark{a} & PA\tablenotemark{b} ($\mathrm{deg}$) & AR\tablenotemark{c} & Semi-major axis ($''$) \\
            \hline
            A496  & $04:33:37.80$ & $-13:15:40.05$ & $113$ & $0.96$ & $252$ \\
            A2029 & $15:10:56.09$ & $05:44:41.10$  & $113$ & $0.79$ & $242$ \\
            A1644 & $12:57:11.61$ & $-17:24:33.09$ & $178$ & $0.81$ & $222$ \\
            \hline       
            \end{tabular} \\
    \tablenotemark{a}{Sky coordinates (J2000.0) of the X-ray brightness peak.} \\
    \tablenotemark{b}{Position angle measured north of west.} \\
    \tablenotemark{c}{Axis ratio.}
\end{table*}

\begin{table*}
    \centering
        \caption{Summary of the head of the positive spiral excesses in the residual X-ray surface brightness maps.}
        \label{tab:AnalysesImagingSpiral}
            \begin{tabular}{cccc}
            \hline
            \hline
            Cluster & Azimuthal angle \tablenotemark{a} ($\mathrm{deg}$) & Radius ($''$) & Direction \tablenotemark{b} \\
            \hline
            A496  & $293$ & $45.13$ & counterclockwise \\
            A2029 & $53$  & $25.52$ & clockwise \\
            A1644 & $88$  & $19.86$ & counterclockwise \\
            \hline       
            \end{tabular} \\
    \tablenotemark{a}{Azimuthal angle measured north of west.} \\
    \tablenotemark{b}{The direction for each cluster was chosen based on the winding direction of the spiral excess.}
\end{table*}

We use the archival X-ray datasets of our sample taken with the Advanced CCD Imaging Spectrometer \citep[ACIS;][]{Garmire2003, Grant2024} on board the {\em Chandra} X-ray Observatory. 
All datasets are summarized in Table~\ref{tab:DataObservations}. 
We follow the data reduction process in \cite{Ueda2024}, to which we refer readers for more details.
We use the versions of $4.15.1$ and $4.10.4$ for {\em Chandra} Interactive Analysis of Observations \citep[\texttt{CIAO};][]{Fruscione2006} and the calibration database (CALDB), respectively. 
After performing the \texttt{lc\_clean} task in \texttt{CIAO} 
to exclude the duration of the flare, we obtain the net exposure time for each cluster summarized in Table~\ref{tab:DataObservations}. 
Finally, we use \texttt{wavdetect} from \texttt{CIAO} to detect and remove point sources.


\section{Analyses and Results}
\label{sec:Analyses}

In this section, we describe the procedure to detect and analyze the sloshing cold fronts.
The detection of the sloshing cold fronts in the residual X-ray surface brightness maps
is given in Section~\ref{sec:AnalysesImaging}, followed by the fitting of the 
X-ray surface brightness profiles in Section~\ref{sec:AnalysesSx}.
The spectral analyses for the temperature profiles are presented in Section~\ref{sec:AnalysesSpectra}.
Finally, we estimate the 
thermal pressure and entropy
across the detected X-ray surface brightness edges and present the results in Section~\ref{sec:AnalysesAzimuthalVariations}.

\subsection{The X-ray imaging and the sloshing spirals}
\label{sec:AnalysesImaging}

The background-subtracted and exposure-corrected X-ray surface brightness images of A496, A2029, and A1644 are shown in the first row of Figure~\ref{fig:AnalysesImaging}. 
These images are taken in $0.5 \text{-} 7.0\,\mathrm{keV}$ band. A496 (the upper left panel) and A2029 (the upper middle panel) reveal a clear and highly axial-symmetric morphology in X-rays, suggesting that they are relatively relaxed.
In A496, the X-ray surface brightness distribution appears less continuous in the southwest yellow-to-green and the north green-to-blue regions, which is an indication of cold fronts. 
For A2029, the strip at $85''$ south from the X-ray peak 
is considered to be the X-ray absorption of a foreground edge-on spiral galaxy \citep{Clarke2004a}. 
In contrast to the former two clusters, a spiral-like pattern 
can already be visually identified
in the X-ray surface brightness of A1644 (the upper right panel), before making its residual map.

We extract the residual X-ray surface brightness maps to detect
the gas sloshing features in our samples, as shown in the second row of Figure~\ref{fig:AnalysesImaging}. 
The residual maps are generated by subtracting an elliptically mean surface brightness distribution from the observed surface brightness map (the first row in Figure~\ref{fig:AnalysesImaging}). 
The parameters characterizing the mean surface brightness elliptical distribution are taken from \cite{Ueda2021}, which we summarize in Table~\ref{tab:AnalysesImagingModel}. 
In \cite{Ueda2021}, they derived those parameters by using a concentric-ellipse-fitting algorithm \citep{Ueda2017} to fit the X-ray surface brightness, with the elliptical model centering at the X-ray brightness peak. Based on these parameters, we evaluate the elliptical model at each position by the mean of its corresponding concentric elliptical ring with a width of $2''$. 
In order to cover the largest {\em Chandra} observed area, we set the semi-major axes of the models to $252''$, $242''$, and $222''$ for A496, A2029, and A1644, respectively. We emphasize that the choice of the subtraction model does not affect our subsequent results, as the primary goal of generating these residual maps is to visually identify the starting points and direction of the sloshing patterns.

We observe large and head-on striking spiral excesses, the feature of gas sloshing, in the residual maps of all samples as seen in the second row of Figure~\ref{fig:AnalysesImaging}. 
We visually identify these spiral excesses, and define their head starting from the brightness excess with the orientation  
as summarized in Table~\ref{tab:AnalysesImagingSpiral}.

We stress that the detection of these spiral excesses is unequivocal even with a different method to create the residual maps---for example, 
the spirals were also seen in the residual map of A2029 by subtracting a two-dimensional beta model from the image \citep{Paterno-Mahler2013}. 
In addition, we observe that the spiral excesses become less distinct after a turn of approximately $180^{\circ}$ from their origin, beyond which the spiral excesses become clear again, a feature also 
seen in \cite{Ghizzardi2014} and \cite{Clarke2004}. The locations of these gaps are identified around sectors $8$ and $9$ in A496, sectors $1$ and $2$ in A2029, and sectors $8$ and $9$ in A1644.
In \cite{Ghizzardi2014}, they suggested that this phenomenon is caused by Kelvin-Helmholtz instabilities generated by shear flows along the sloshing gas, as confirmed in simulations \citep{ZuHone2010, Roediger2011, Roediger2012, Roediger2013}.
Another feature is that all three clusters exhibit a pair of spirals with a brightness excess and a corresponding deficit.
The white-to-yellow regions in the residual maps represent the excesses of the surface brightness, while the blue-to-black regions represent the deficits. This feature was also found in \cite{Ueda2021}, in which they
analyzed $28$ cool-core clusters, including those in our sample, and showed that all clusters exhibit a pair of over- and under-density patterns in their residual X-ray surface brightness maps.
We note that the pair-spiral pattern naturally emerges with the subtraction of any azimuthally averaged model.
This is because the over-density sloshing spiral increases the azimuthal mean of the corresponding elliptical annulus, which leads to a relative deficit on the other side of the same annulus after the subtraction.
For example, \cite{Zheng2025}
used a different global ICM model in the subtraction for Virgo cluster and observed a paired pattern qualitatively similar to our result.



\begin{figure*}
    \centering
    \resizebox{!}{\textheight}{
        \includegraphics[scale=1]{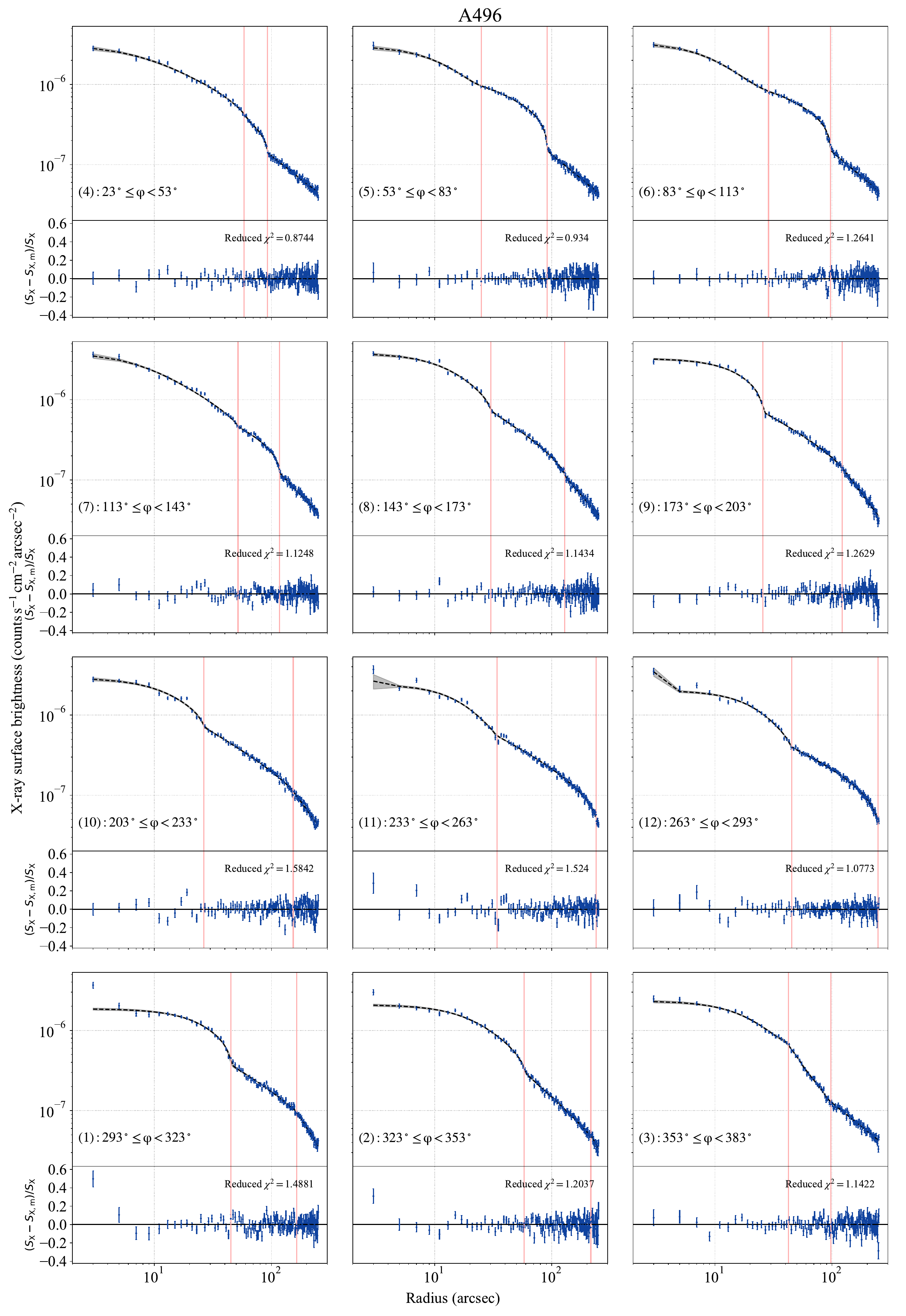}
   }
\caption{
The radial profiles of the X-ray surface brightness extracted from sectors in A496. The blue vertical bars show the $1\sigma$ confidence range of the X-ray surface brightness. The black dashed lines show the best-fit X-ray surface brightness profiles.
The pink vertical lines denote the best-fit positions of the surface brightness edges (i.e., $r_{\rm{in}}$ and $r_{\rm{out}}$). 
In addition, the panels below the profiles show the residuals with respect to the best-fit models, computed as $\left(S_{\rm{X}} - S_{\rm{X}, m}\right)/S_{\rm{X}}$, where $S_{\rm{X}}$ is the observed surface brightness and $S_{\rm{X}, m}$ is the model prediction.}
\label{fig:AnalysesSxA496}
\end{figure*}

\begin{figure*}
    \centering
    \resizebox{!}{\textheight}{
        \includegraphics[scale=1]{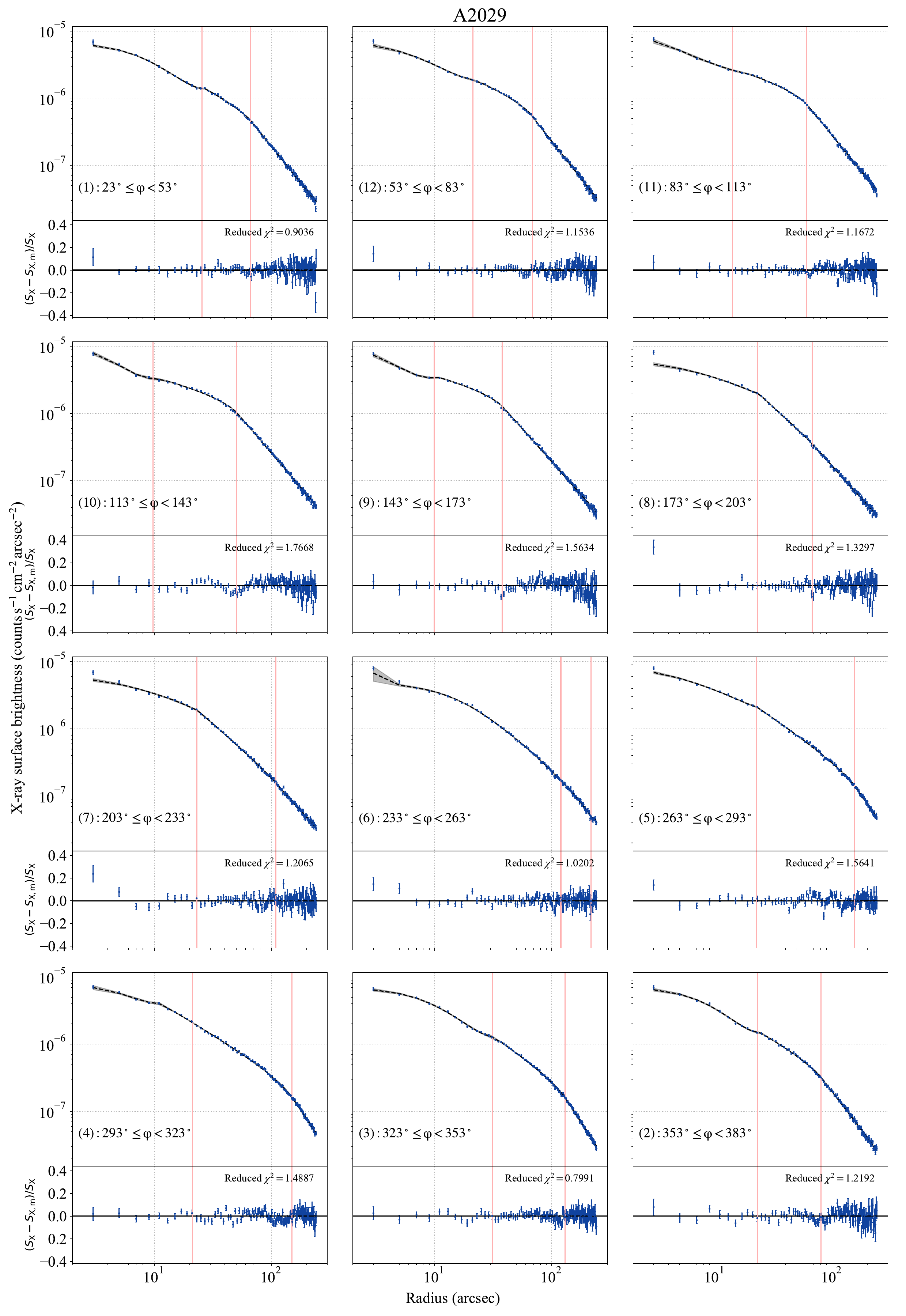}
   }
\caption{Same as Figure~\ref{fig:AnalysesSxA496}, but for A2029.}
\label{fig:AnalysesSxA2029}
\end{figure*}

\begin{figure*}
    \centering
    \resizebox{!}{\textheight}{
        \includegraphics[scale=1]{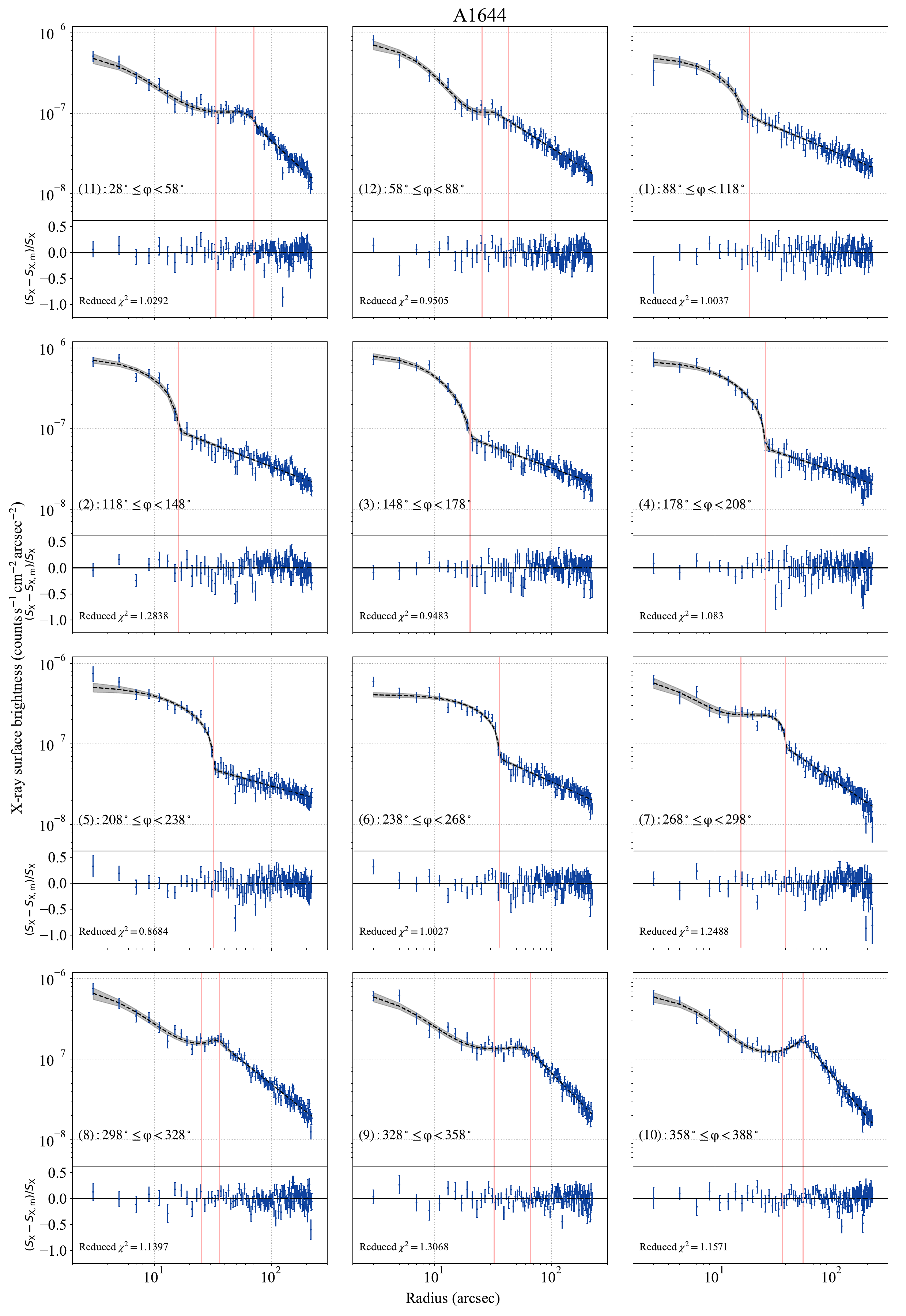}
   }
\caption{Same as Figure~\ref{fig:AnalysesSxA496}, but for A1644.}
\label{fig:AnalysesSxA1644}
\end{figure*}

\subsection{The X-ray surface brightness profiles}
\label{sec:AnalysesSx}

\begin{table}
    \begin{center}
        \caption{Summary of the azimuthal angle $\phi$ ranges of the sectors defined in the sample.}
        \label{tab:AnalysesSxsectors}
            \begin{tabular}{cccc}
            \hline
            \hline
            Label & A496 & A2029 & A1644 \\
            \hline
             1 & $          293^{\circ} -           323^{\circ}$ & $\phantom{0}23^{\circ} - \phantom{0}53^{\circ}$ & $\phantom{0}88^{\circ} -           118^{\circ}$ \\
             2 & $          323^{\circ} -           353^{\circ}$ & $          353^{\circ} -           383^{\circ}$ & $          118^{\circ} -           148^{\circ}$ \\
             3 & $          353^{\circ} -           383^{\circ}$ & $          323^{\circ} -           353^{\circ}$ & $          148^{\circ} -           178^{\circ}$ \\
             4 & $\phantom{0}23^{\circ} - \phantom{0}53^{\circ}$ & $          293^{\circ} -           323^{\circ}$ & $          178^{\circ} -           208^{\circ}$ \\
             5 & $\phantom{0}53^{\circ} - \phantom{0}83^{\circ}$ & $          263^{\circ} -           293^{\circ}$ & $          208^{\circ} -           238^{\circ}$ \\
             6 & $\phantom{0}83^{\circ} -           113^{\circ}$ & $          233^{\circ} -           263^{\circ}$ & $          238^{\circ} -           268^{\circ}$ \\
             7 & $          113^{\circ} -           143^{\circ}$ & $          203^{\circ} -           233^{\circ}$ & $          268^{\circ} -           298^{\circ}$ \\
             8 & $          143^{\circ} -           173^{\circ}$ & $          173^{\circ} -           203^{\circ}$ & $          298^{\circ} -           328^{\circ}$ \\
             9 & $          173^{\circ} -           203^{\circ}$ & $          143^{\circ} -           173^{\circ}$ & $          328^{\circ} -           358^{\circ}$ \\
            10 & $          203^{\circ} -           233^{\circ}$ & $          113^{\circ} -           143^{\circ}$ & $          358^{\circ} -           388^{\circ}$ \\
            11 & $          233^{\circ} -           263^{\circ}$ & $\phantom{0}83^{\circ} -           113^{\circ}$ & $\phantom{0}28^{\circ} - \phantom{0}58^{\circ}$ \\
            12 & $          263^{\circ} -           293^{\circ}$ & $\phantom{0}53^{\circ} - \phantom{0}83^{\circ}$ & $\phantom{0}58^{\circ} - \phantom{0}88^{\circ}$ \\
            \hline
            \end{tabular}
        \label{tab:AnalysesSxsectors}
    \end{center}
\end{table}

In this section, we aim to identify the edges in the X-ray surface brightness along the radial direction by using a generalized model.
Consider an idealized case of a spiral excess illustrated in Figure~\ref{fig:Notation}.
At a given position angle with respect to the spiral head, two edges ($r_{\mathrm{in}}$ and $r_{\mathrm{out}}$) are present along the radial direction, except at the position angle aligned with the spiral-head direction.
To describe the radial distributions while capturing the edges, we adopt a three-component model, in which the edges are defined at the boundaries between adjacent components.
However, due to noise, the complex morphology of cold fronts, or the parameter degeneracy, the best-fit $r_{\mathrm{in}}$ and $r_{\mathrm{out}}$ may not well capture the true locations of the pairs of edges.
We therefore perform visual inspections to verify the best-fit edge locations.
Alternatively, we use a simplified, two-component model when the three-component model fails to provide a good fit.

To analyze the azimuthal variations of the ICM across the sloshing cold fronts, we define $12$ sectors with an opening angle of $30^{\circ}$ to extract the radial X-ray surface brightness profiles for each cluster.
The center of these sectors is set at the X-ray brightness peak of the cluster. We define the position angle of each sector with respect to the elliptical model described in Table~\ref{tab:AnalysesImagingModel}. 
The definitions of the sectors are listed in Table~\ref{tab:AnalysesSxsectors}.
Following the same elliptical model, we set the outer boundaries of the sectors in A496, A2029, and A1644 to $252''$, $242''$, and $222''$, respectively. We examine the systematic uncertainties of our measurements by adopting an alternative definition of the sectors, as detailed in Appendix~\ref{app:SysUncertain}.
The sectors overlaid on the X-ray surface brightness and residual maps are shown in the third and fourth rows in Figure~\ref{fig:AnalysesImaging}.

In order to measure the density contrast
of the X-ray surface brightness edges, 
we extract the 
X-ray surface brightness profile of each sector with a radial binning of $2''$. 
Subsequently, all profiles are fitted with either a two- or three-component model integrated along the line of sight.
For a uniform analysis, we consistently adopt the three-component model as a starting point for each sector, as it is expected to characterize both the inner and outer spiral boundaries of a single spiral excess.
We switch to the two-component model only if the former does not provide a good fit.
Importantly, unlike other studies typically done in the literature, our homogeneous and systematic analysis across the sample is a unique strength of this work, avoiding the confirmation bias and reducing the influence of the prior knowledge from our expectation.
We note here that although our three-component model identifies both spiral boundaries, we focus on the outer spiral boundaries in the discussion, as the edges at the outer spiral boundary are less disturbed by the complex dynamic of the cluster core.
As a result, our best-fit models provide excellent descriptions of the observed X-ray surface brightness of all sectors in each cluster, as shown in Figures~\ref{fig:AnalysesSxA496}, \ref{fig:AnalysesSxA2029}, and \ref{fig:AnalysesSxA1644} for A496, A2029, and A1644, respectively.

We describe the models of the density profiles as follows.
The three-component model is composed of a beta model for the inner region ($r < r_{\rm{in}}$), a first power-law model for the intermediate region ($r_{\rm{in}} \leq r < r_{\rm{out}}$), and a second power-law model for the outer region ($r \geq r_{\rm{out}}$). This model is expressed as
\begin{equation}
\label{eq:AnalysesSx3ComModel}
    n_{\rm{e}} \left(r\right) =
        \begin{cases}
        n_{\mathrm{e, c}} \left[1 + \left(\frac{r}{r_{\rm{c}}}\right)^{2}\right]^{-3 \beta/2} & \mathrm{if}~r < r_{\rm{in}} \\
        \frac{ n_{\rm{e}, r_{\rm{in}}} }{ j_{n, \rm{in}} }  \left(\frac{r}{r_{\rm{in}}}\right)^{-2 s_{1}} & \mathrm{if}~r_{\rm{in}} \leq r < r_{\rm{out}} \\
        \frac{ n_{\rm{e}, r_{\rm{out}}}}{ j_{n, \rm{out}} } \left(\frac{r}{r_{\rm{out}}}\right)^{-2 s_{2}} & \mathrm{if}~r_{\rm{out}} \leq r \\
        \end{cases}
        \, .
\end{equation}
In Equation~(\ref{eq:AnalysesSx3ComModel}), $n_{\mathrm{e, c}}$ represents the electron number density at the cluster center, and $r_{\rm{c}}$ and $\beta$ are the core radius and slope parameter of the beta model, respectively; 
meanwhile, $n_{\rm{e}, r_{\rm{in}}}$ ($n_{\rm{e}, r_{\rm{out}}}$) denotes the electron number density at the three-dimensional radius $r_{\rm{in}}$ ($r_{\rm{out}}$) of the inner (outer) X-ray surface brightness edge, and the index $s_{1}$ ($s_{2}$) corresponds to the slope parameter of the inner (outer) power-law model.
By construction, we have 
$n_{\rm{e}, r_{\rm{in}}} = n_{\mathrm{e, c}}\left[1 + \left(r_{\rm{in}}/r_{\rm{c}}\right)^{2}\right]^{-3 \beta/2}$ and 
$n_{\rm{e}, r_{\rm{out}}} =  n_{\rm{e}, r_{\rm{in}}} / j_{n, \rm{in}} \times  \left(r_{\rm{out}}/r_{\rm{in}}\right)^{-2 s_{1}}$.
In such a parameterization, the parameter $j_{n, \rm{in}}$ ($j_{n, \rm{out}}$) is the density contrast 
at the radius $r_{\rm{in}}$ ($r_{\rm{out}}$). 
If $\jn > 1$, the electron number density radially inward of the cold front (i.e., with a smaller distance to the core) is larger than that radially outward, and vice versa.
For example, the condition of $j_{n, \rm{in}} > 1$ represents a deficit in the density towards a large distance at the edge $r_{\mathrm{in}}$ in Equation~(\ref{eq:AnalysesSx3ComModel}).

If the surface brightness profile of a sector is not well fitted by the three-component model, we adopt a two-component model instead, which is found to provide a sufficiently accurate fit in this work. 
The two-component model, which consists of a beta model for the inner profile ($r < r_{\rm{in}}$) and a power-law model ($r \geq r_{\rm{in}}$) for the outer profile, is described as
\begin{equation}
\label{eq:AnalysesSx2ComModel}
    n_{\rm{e}} \left(r\right) =
        \begin{cases}
        n_{\mathrm{e, c}} \left[1 + \left( \frac{r}{r_{\rm{c}}}\right)^{2}\right]^{-3 \beta/2} & \mathrm{if}~r < r_{\rm{in}} \\
        \frac{ n_{\rm{e}, r_{\rm{in}}} }{ j_{n, \rm{in}} } \left( \frac{ r }{ r_{\rm{in}} }\right)^{-2 s_{1}} & \mathrm{if}~r_{\rm{in}} \leq r \\
        \end{cases}
        \, ,
\end{equation}
where all parameters follow the definitions in Equation~(\ref{eq:AnalysesSx3ComModel}).

We approximate the X-ray emission using an isothermal model, so that the modeling of the X-ray surface brightness is independent of temperature \citep[e.g.,][]{Ichinohe2017, Ueda2024}.
Additionally, we assume spherical symmetry in the distribution of the ICM, such that the modeling of the X-ray surface brightness profile can be parameterized by a direct integration of $n_{\rm{e}}^{2}\left(r\right)$ along the line of sight with a varying normalization.
We find the best-fit model of the observed X-ray surface brightness profiles using the affine-invariant Markov chain Monte Carlo sampling \citep{Goodman2010} implemented by the \texttt{emcee} python package \citep{Foreman-Mackey2013}.

The results of the X-ray surface brightness fitting for A496, A2029, and A1644 are shown in Figures~\ref{fig:AnalysesSxA496},~\ref{fig:AnalysesSxA2029}, and~\ref{fig:AnalysesSxA1644}, respectively.
Except for the sectors $1$ to $6$
of A1644, all the profiles extracted from the remaining regions are well-fitted with the three-component model. 
The fitting results of A496, A2029, and A1644 yield mean reduced $\chi^{2}$ values of $1.22$, $1.27$, and $1.09$ with the degrees of freedom of $125, 120, \rm{and}\ 110$, respectively.

We stress again that we do not manually refine the inferred $r_{\mathrm{in}}$ and $r_{\mathrm{out}}$.
We adopt non-informative priors on $r_{\mathrm{in}}$ and $r_{\mathrm{out}}$ in the surface brightness fitting and consistently use the medians of their posterior distributions as their best-fit values.
Notably, some of the identified surface brightness edges, especially those in the disturbed tail regions, do not sit at the visually sharp edges in the residual maps. 
The difficulty of pinpointing those edges is reflected in the broad posterior distributions, leading to higher statistical uncertainties.
Overall, the typical $1\sigma$ uncertainties of $r_{\mathrm{in}}$ and $r_{\mathrm{out}}$ are $\approx6\arcsec$ and $\approx9\arcsec$, respectively. 
In a highly disturbed tail region (such as $r_{\mathrm{out}}$ of sector 6 in A2029), the $1\sigma$ uncertainty can reach as high as $\approx30\arcsec$.
Finally, through visual inspection, we confirm that $\approx60\%$ of the sectors have the identified outer edge $r_{\mathrm{out}}$, as the main focus of this work, that coincides with the discontinuity in the residual maps.
Meanwhile, about one third of the sectors 
(specifically, sectors 5 and 6 in A496, Sectors 4, 5, and 8 to 12 in A2029, and Sector 7 in A1644) identified both  $r_{\mathrm{in}}$ and $r_{\mathrm{out}}$ that are associated with clear discontinuities in the residual maps. 
We further discuss the systematic uncertainties of the inferred density contrasts associated with temperature in Section~\ref{sec:DiscussionCF}.

For a homogeneous study of sloshing cold fronts, we uniformly define the sectors and, hence, they do not align perfectly with the curvature of the edges of the spiral excesses.
Despite a uniform analysis, doing so may not accurately capture the density contrasts across the surface brightness edges.
Therefore, we investigate the associated systematic uncertainty by additionally analyzing several sectors with the customized directions aligning with the spiral edges (see details in Appendix~\ref{app:SysUncertain}).

\begin{figure*}[ht!]
    \begin{center}
        \includegraphics[width=17cm]{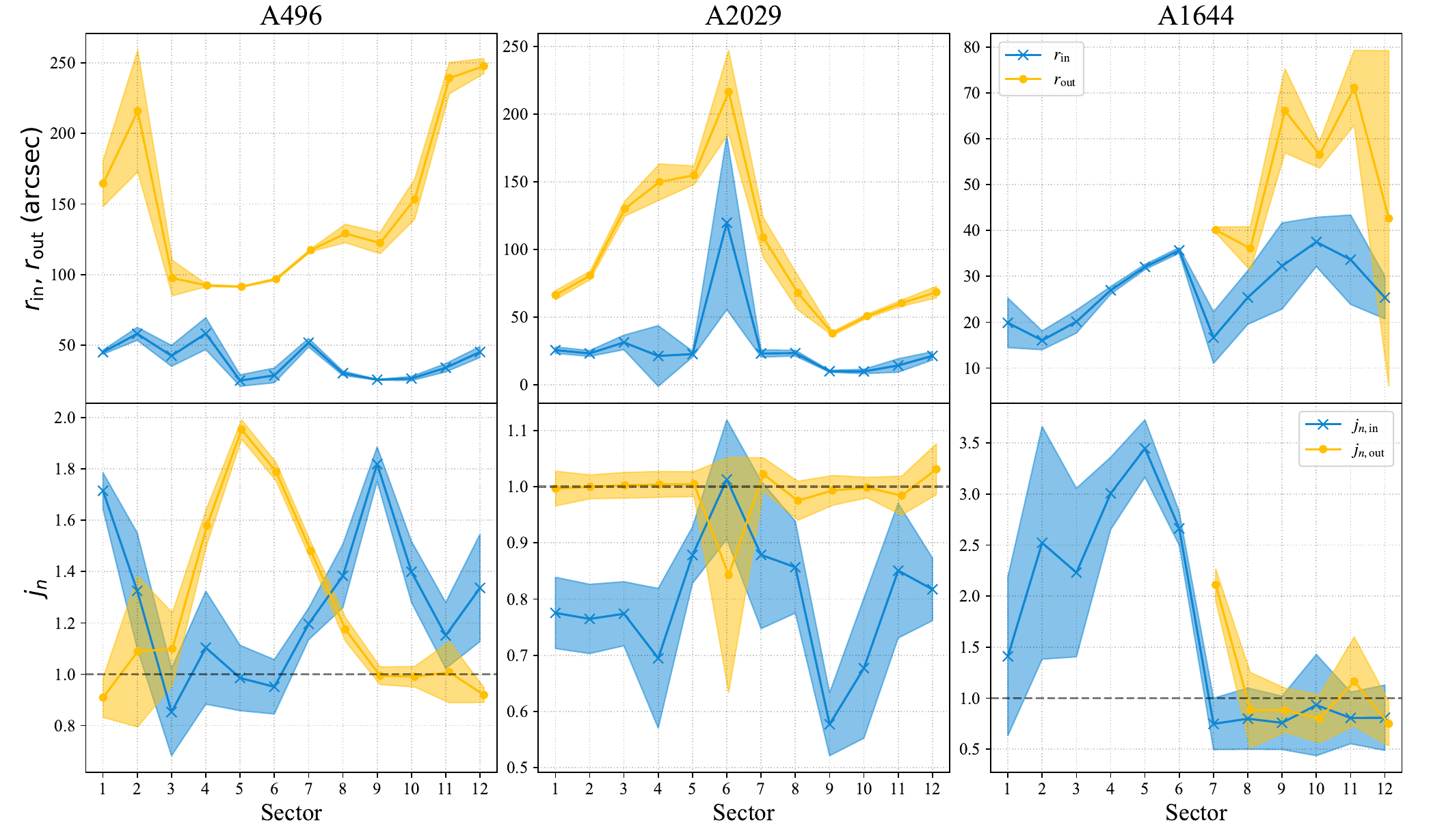}
    \end{center}
\caption{
The azimuthal variations of the best-fit three-dimensional positions $r$ (top) and density contrasts \jn\ (bottom) of the X-ray surface brightness edges for A496 (left), A2029 (middle), and A1644 (right), respectively. 
The blue and yellow data points indicate the best-fit values of the inner and outer detected edges ($r_{\rm{in}}$ and $r_{\rm{out}}$), respectively. 
The shaded areas represent the $1\sigma$ confidence ranges of the best-fit parameters. The horizontal black dashed lines in the bottom panels correspond to unity.}
\label{fig:AnalysesSx_r12_r23_n0_n1}
\end{figure*}

In Figure~\ref{fig:AnalysesSx_r12_r23_n0_n1}, we show the azimuthal variations of
the three-dimensional positions of the edges ($r_{\rm{in}}$ and $r_{\rm{out}}$) and the corresponding density contrasts (\jn). 
The measurements of the inner and outer surface brightness edges are shown in blue and yellow, respectively.
To better examine the azimuthal variations, the data points shown in Figure~\ref{fig:AnalysesSx_r12_r23_n0_n1} are plotted starting from the head of the spiral excess, as listed in Table~\ref{tab:AnalysesImagingSpiral}, and follow the direction of the spiral. 
In the following paragraphs, we present a detailed discussion of each cluster. 

\paragraph{A496} 
As seen in the fourth row of Figure~\ref{fig:AnalysesImaging}, the inner edges of the sectors $1$ to $3$, and the outer edges of the sectors $4$ to $12$, are aligned with the outer boundary of the spiral excess, suggesting that the modeling of the X-ray surface brightness correctly captures the sloshing position.
Also, as in the upper left panel of Figure~\ref{fig:AnalysesSx_r12_r23_n0_n1}, the farthest edge is located at $\approx 248''$ from the cluster center in the twelfth sector, indicating that the sloshing cold front could extend to the outskirt of the cluster. 
For the azimuthal variation of the density contrast (the lower left panel of Figure~\ref{fig:AnalysesSx_r12_r23_n0_n1}), the mean density contrast \jn\ of the sharpest edges ($r_{\rm{out}}$ of the sector $4$ to sector $8$) is $1.60 \pm 0.021$, which is similar to but marginally smaller than density contrasts typically measured in stripping cold fronts \citep[e.g., $1.68^{+0.31}_{-0.23}$ in Abell\,3667;][]{Ichinohe2017}. 
The mean density contrast of the edges along the outer boundary of the spiral excess (i.e., $r_{\rm{in}}$ in sector $1$ to sector $3$ and $r_{\rm{out}}$ in sector $4$ to sector $12$, by visual inspection) is $1.32 \pm 0.028$, similarly indicating a relatively weak contrast.
Additionally, we find that the density contrasts at the spiral tail region (the outer edges from sectors $9$ to $12$) are consistent with unity within $1\sigma$, suggesting no clear density contrast at the spiral-tail region.

\paragraph{A2029} 
As in the bottom middle panel of Figure~\ref{fig:AnalysesImaging}, the inner edges of the sectors $1$ to $3$ are aligned with the inner boundary of the spiral excess. Meanwhile, the inner edges of the sectors $4$, $5$, $7$, and $8$, as well as the outer edges of the sectors $1$ to $5$ and $9$ to $12$, are aligned with the outer boundary of the spiral excess.
Again, this demonstrates the robustness of the surface brightness modeling in identifying the sloshing position.
The sharpest edges are traced by $r_{\rm{in}}$ of the sectors $7$ and $8$ and $r_{\rm{out}}$ of the sector $9$ to sector $11$.
For sectors $1$ to $3$, the positions of $r_{\rm{in}}$ coincide with the inner boundary of the spiral excess, in contrast to the other edges, in which $r_{\rm{in}}$ is aligned with the outer boundary of the spiral excess.
Regarding the values of the density contrasts \jn\ at the edges that coincide with the outer boundary of the spiral excess, we find that \jn\ of inner edges
in sectors $4$, $5$, $7$, and $8$ are significantly below one with their mean density contrast equals to $0.83 \pm 0.05$ (see Figure~\ref{fig:AnalysesSx_r12_r23_n0_n1}).
This result shows the excess of gas density at the edge of $r_{\rm{in}}$, which is opposite to the other two clusters.

\paragraph{A1644} We show the results in the right panel of Figure~\ref{fig:AnalysesSx_r12_r23_n0_n1}.
The striking spiral excess is aligned with the inner edges $r_{\rm{in}}$ of the sectors $1$ to $6$
and the outer edges $r_{\rm{out}}$ of the sector $7$, which represent the sharpest edges.
These sectors enclose the most clearly visible part of the spiral excess in A1644. 
On the other hand, the sectors $8$ to $12$
are located at the tail of the spiral excess, such that the surface brightness edges in these sectors are associated with the disturbed morphology of the spiral tail. 
The mean density contrast ($2.48 \pm 0.24$) of the surface brightness edges along the clearly visible spiral excess is significantly higher than those typically observed in stripping cold fronts \citep[e.g., $\jn = 1.53^{+0.08}_{-0.07}$ in Abell\,2554;][]{Erdim2019}.
It is worth mentioning that the azimuthal variation of the density contrast peaks at the middle section of the sharpest edges, a feature also observed in stripping cold fronts, where the head exhibits the highest density contrast and then gradually decreases along a tail \citep{Ichinohe2017}.



\subsection{X-ray spectral analyses and the ICM temperatures}
\label{sec:AnalysesSpectra}

To measure the temperature contrasts \jt\ across the detected surface brightness edges (i.e., $r_{\rm{in}}$ and $r_{\rm{out}}$), we divide each sector into several partial-annulus-shaped regions with varying radii from the cluster center, and perform the spectral analyses.

We define the radial binning of each sector as follows. A three-component (two-component) model divides a sector into three (two) components by the positions of $r_{\rm{in}}$ and $r_{\rm{out}}$ ($r_{\rm{in}}$).
Beside the positions of $r_{\rm{in}}$ and $r_{\rm{out}}$, we further define radial binning within each component in each sector.
Specifically, we adjust the radial binning such that each radial bin contains a comparable number of photon counts.
By doing so, we effectively require that the signal-to-noise ratios of individual radial bins in all sectors are consistent within a cluster.
For A496 and A2029, the resulting mean X-ray photon counts per radial bin are $1043$ and $1537$, respectively.
On the other hand, the mean photon count for A1644 is lower due to its significantly fainter surface brightness compared to A496 and A2029. 
For A1644, the resulting radial binning has the mean photon count of $193$ for radial bins within the inner component for the two-component sectors, and the inner and intermediate components for the three-component sectors, while in both cases the outer component has a mean of $820$. 
The definitions of radial bins are summarized in Table~\ref{tab:AnalysesSpectraRadBin}.

Following the procedure described in \cite{Ueda2024}, we use the \texttt{specextract} task in \texttt{CIAO} to extract the individual spectra, responses, and ancillary response files of these radial bins. 
The spectra for analysis are calibrated using these files to account for instrumental effects.
We model the X-ray spectra using \texttt{XSPEC} \citep[version 12.13.0c;][]{Arnaud1996} with the atomic database \citep[version 3.0.9;][]{Foster2023}.

\begin{figure*}[ht!]
    \begin{center}
        \includegraphics[width=16cm]{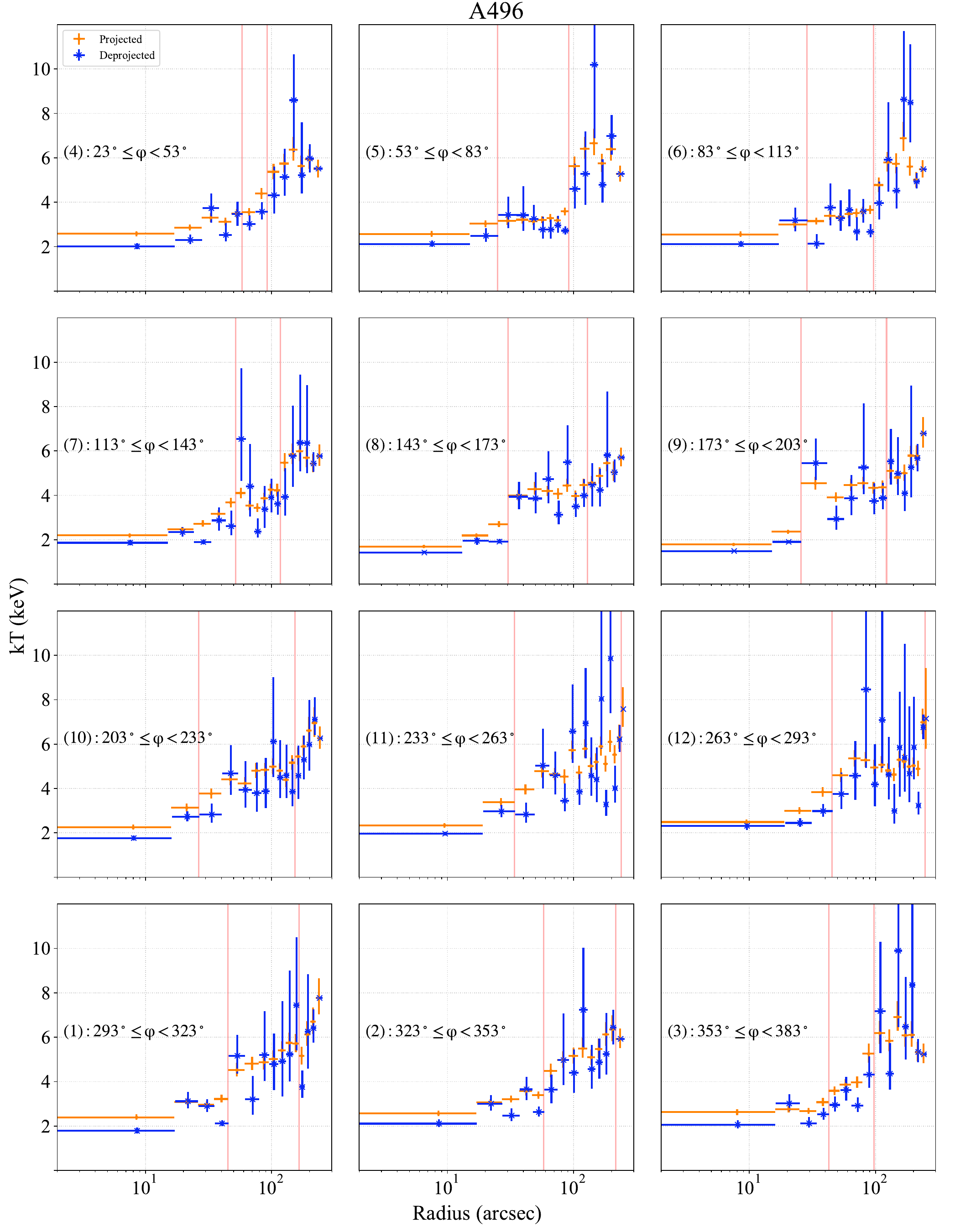}
    \end{center}
\caption{
Radial ICM temperature profiles of each sector in A496. The orange and blue data points show the projected and deprojected temperature, respectively. 
The horizontal bars represent the ranges of the radial bins and the vertical bars denote the $1\sigma$ confidence range of the temperature. 
The pink vertical lines show the positions of $r_{\rm{in}}$ and $r_{\rm{out}}$ described in Section~\ref{sec:AnalysesSx}.
}
\label{fig:AnalysesSpectraA496}
\end{figure*}

\begin{figure*}[ht!]
    \begin{center}
        \includegraphics[width=16cm]{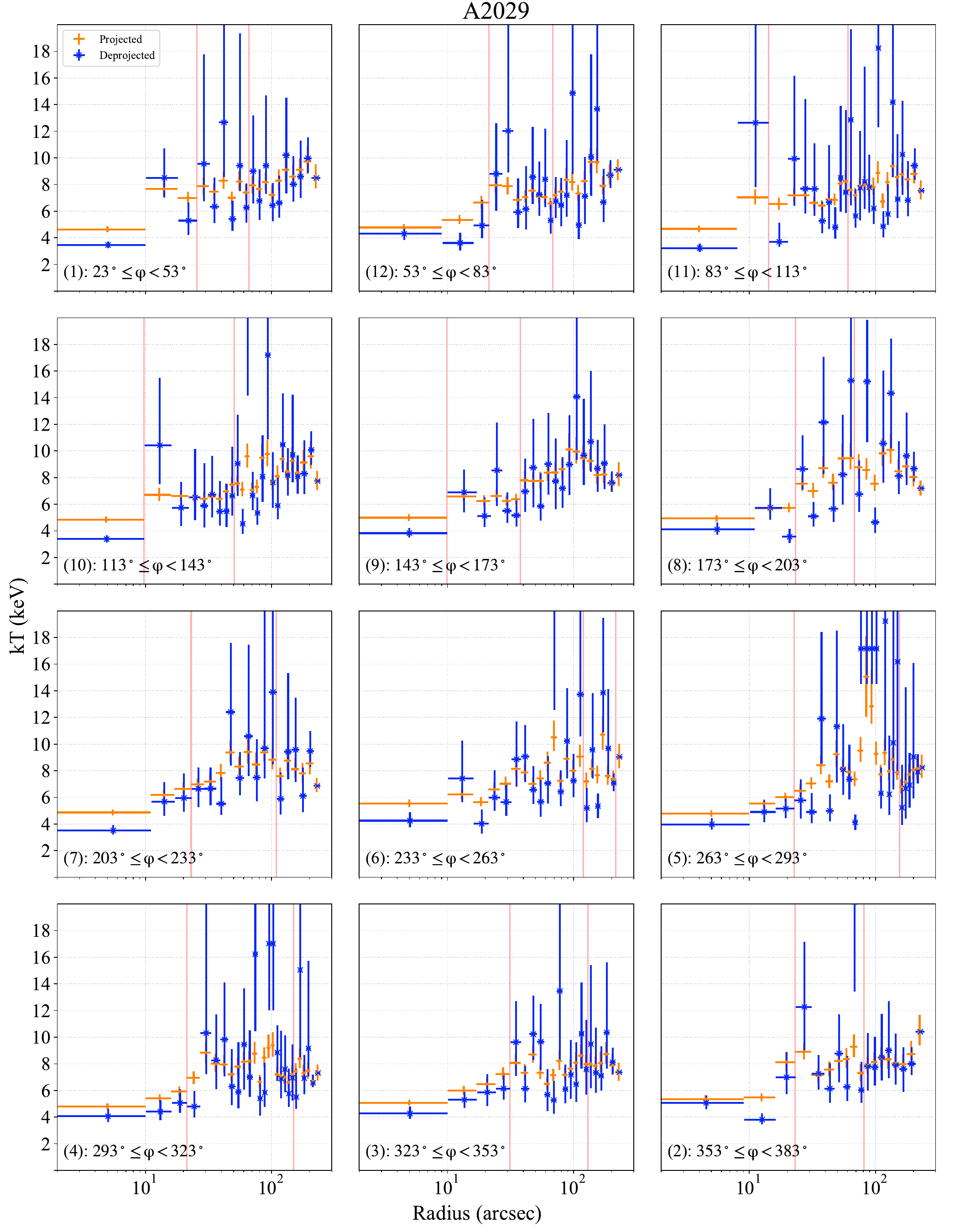}
    \end{center}
\caption{Same as Figure~\ref{fig:AnalysesSpectraA496}, but for A2029.}
\label{fig:AnalysesSpectraA2029}
\end{figure*}

\begin{figure*}[ht!]
    \begin{center}
        \includegraphics[width=16cm]{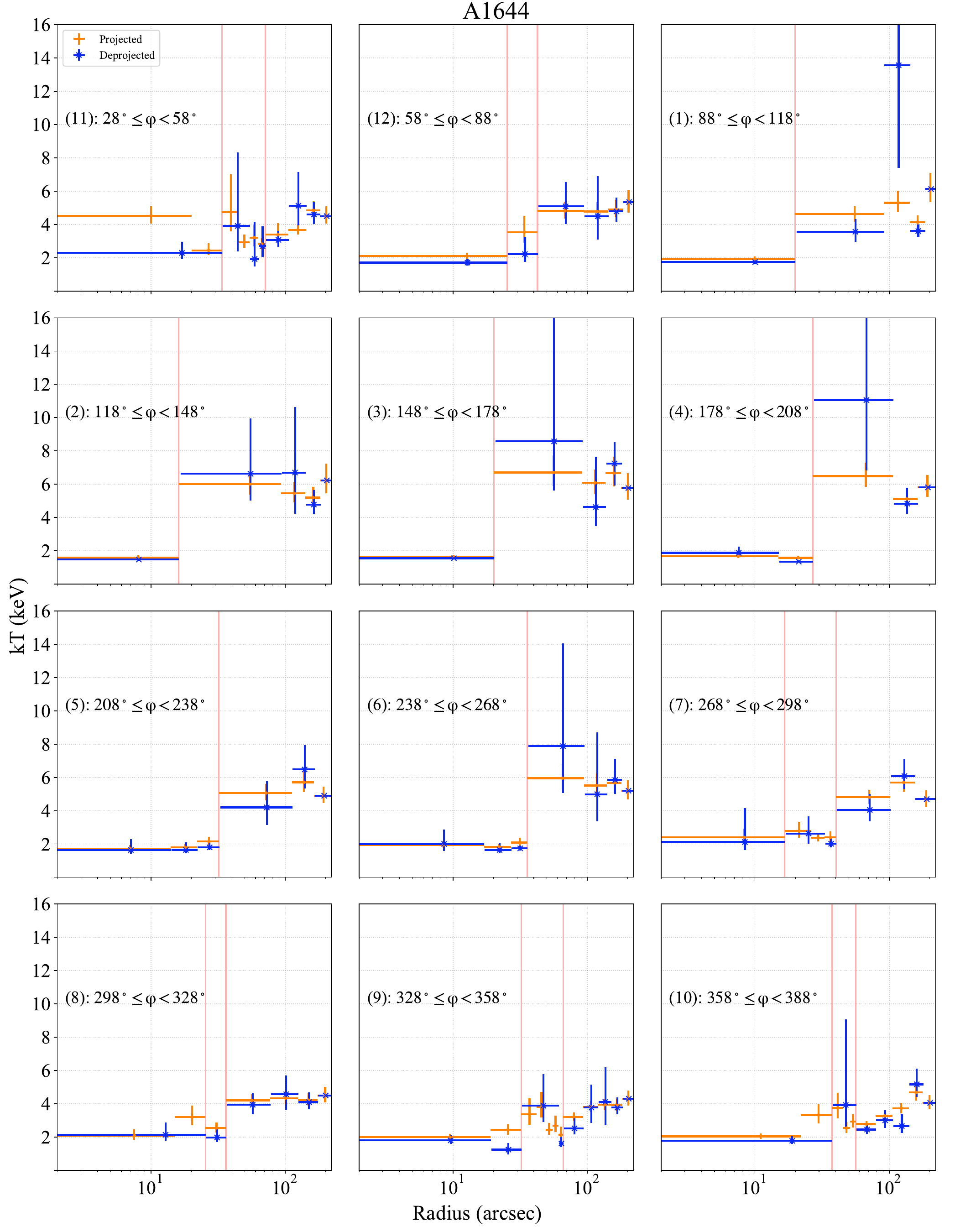}
    \end{center}
\caption{Same as Figure~\ref{fig:AnalysesSpectraA496}, but for A1644.}
\label{fig:AnalysesSpectraA1644}
\end{figure*}

As the first step of the spectral analysis, we measure the radial ICM projected temperature profiles for all sectors.
Assuming that the ICM is in collisional ionization equilibrium, 
we fit the \texttt{APEC} model \citep{Smith2001, Foster2012} to the X-ray spectra in the energy range between $0.5\,\mathrm{keV}$ and $7.0\,\mathrm{keV}$. 
We fix the column density $N_{\rm{H}}$ to $4.29 \times 10^{20}\,\rm{cm}^{-2}$, $3.01 \times 10^{20}\,\rm{cm}^{-2}$, and $4.17 \times 10^{20}\,\rm{cm}^{-2}$ for A496, A2029, and A1644, respectively, following the values taken from \cite{HI4PICollaboration2016}.
We fix the redshift of the spectra to the cluster redshift.
We vary or fix the metal abundances in the spectral models depending on whether we can obtain statistically meaningful constraints from the data.
In general, the metal abundance is treated as a free parameter for most radial bins in A496 and A2029 with a few exceptions, for which we refer readers to Appendix~\ref{app:spectral_details} for details.
On the other hand, the metal abundance is fixed to $0.3\,Z_\odot$ for A1644 due to the lack of photon counts.

As the second step, after obtaining the projected temperature profiles, we then deproject them to obtain the three-dimensional temperature profiles.
The deprojected analyses are performed using the \texttt{projct} tool in \texttt{XSPEC}, 
while we fix all parameters of spectra extracted from the outermost radial bins to the best-fit values obtained from the projected spectral analyses.
The metal abundances of individual radial bins are assigned based on the average metal abundance of the density component to which each radial bin belongs, with the component determined from the density model described in equations~(\ref{eq:AnalysesSx3ComModel}) and (\ref{eq:AnalysesSx2ComModel}).

The radial projected and deprojected temperature profiles are shown as orange and blue data points in Figures~\ref{fig:AnalysesSpectraA496}, ~\ref{fig:AnalysesSpectraA2029}, and ~\ref{fig:AnalysesSpectraA1644} for A496, A2029, and A1644, respectively. 
These profiles clearly show that these clusters are cool-core systems, with temperatures gradually decreasing towards the cluster center.

\begin{figure*}[ht!]
    \centering
        \resizebox{0.9\textwidth}{!}{\includegraphics[scale=1]{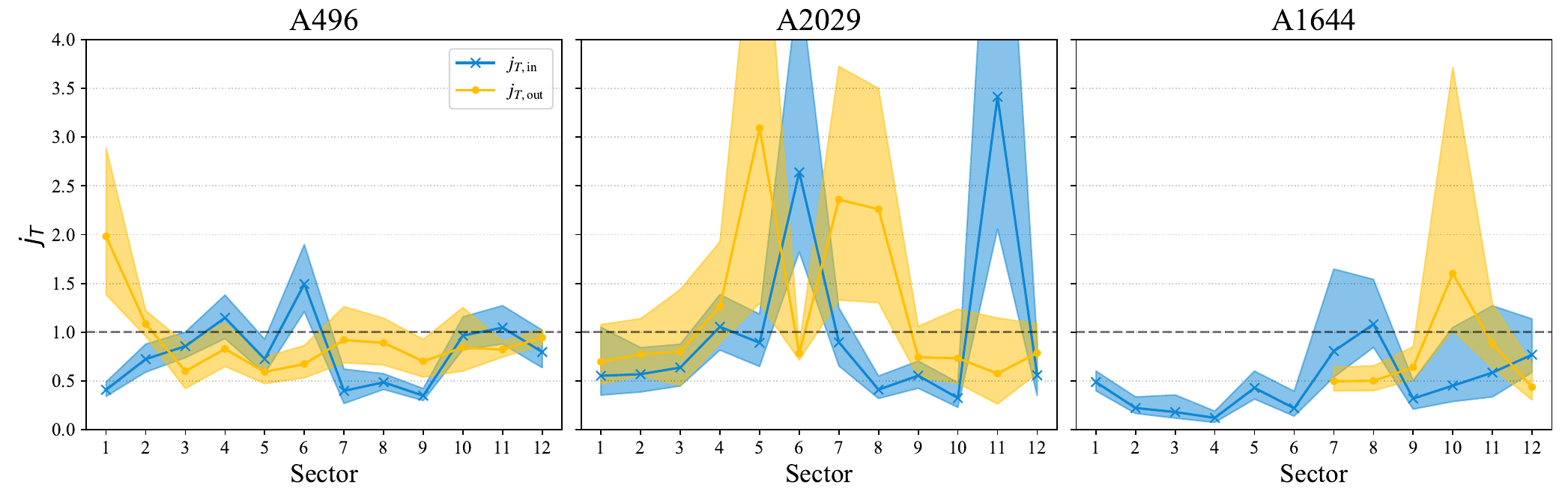}}
        \resizebox{0.9\textwidth}{!}{\includegraphics[scale=1]{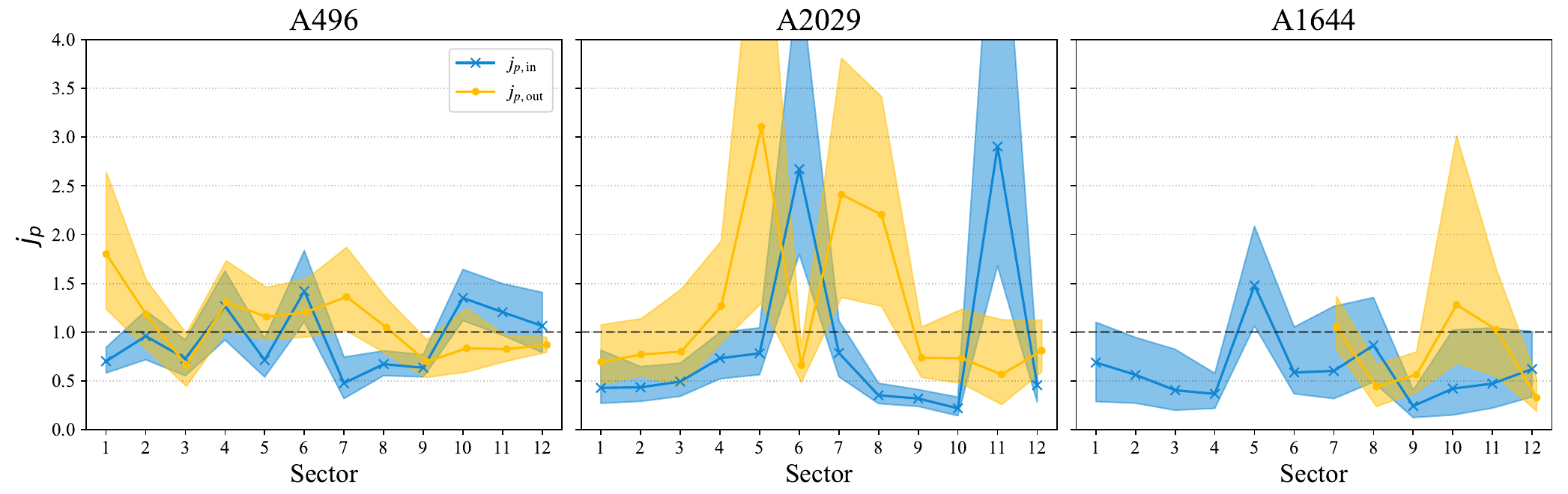}}
        \resizebox{0.9\textwidth}{!}{\includegraphics[scale=1]{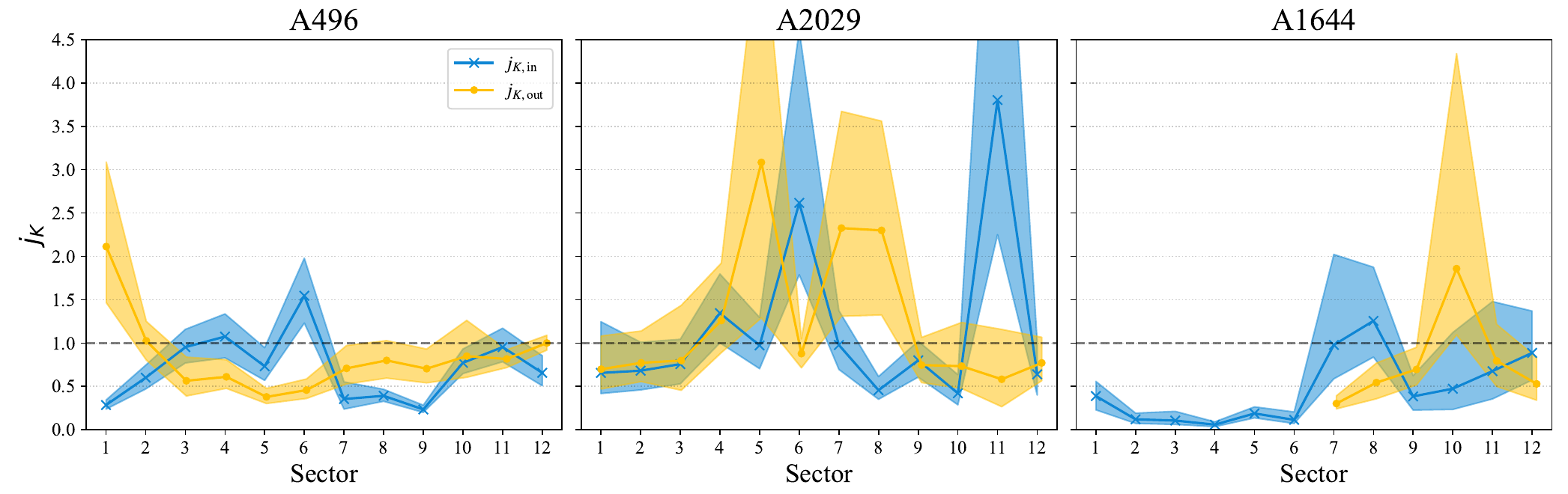}}
        \resizebox{0.9\textwidth}{!}{\includegraphics[scale=1]{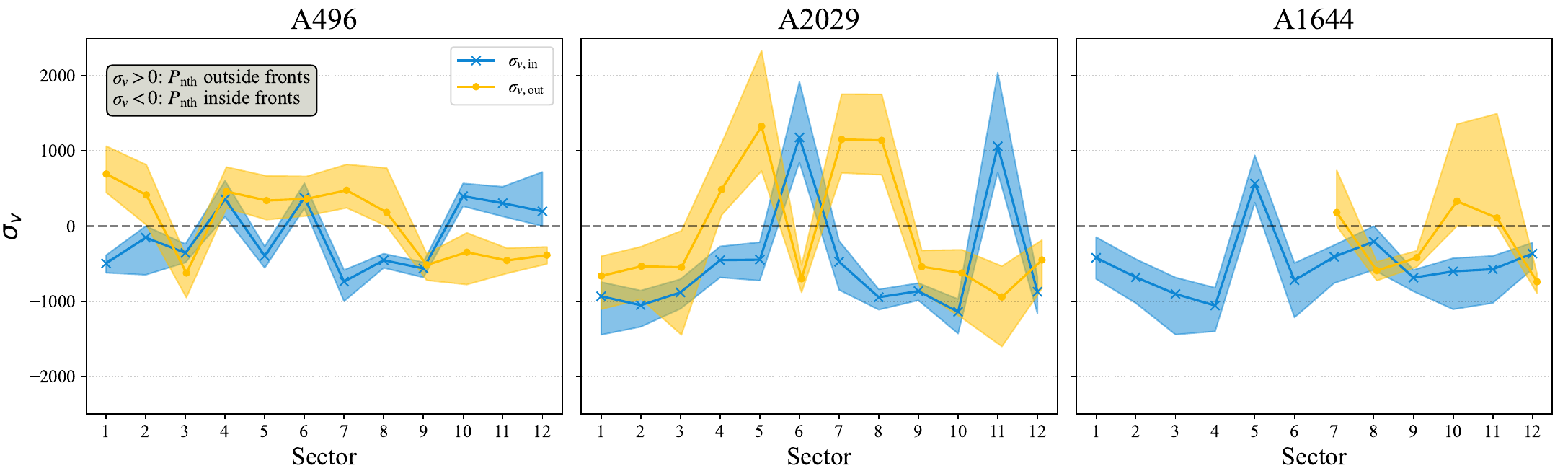}}
\caption{
The azimuthal variations of the deprojected temperature contrasts, pressure contrasts, entropy contrasts, and the velocity dispersions (from top to bottom) across X-ray surface brightness edges for A496 (left), A2029 (middle), and A1644 (right), respectively. 
The blue and yellow data points indicate the contrasts (or velocity dispersions in the bottom panels) at $r_{\rm{in}}$ and $r_{\rm{out}}$, respectively.
The shaded areas represent the $1\sigma$ confidence ranges of contrasts. The horizontal black dashed lines correspond to unity or zero (bottom panel).
}
\label{fig:AnalysesSpectraT0_T1}
\end{figure*}

We present the azimuthal variations of the deprojected temperature contrasts \jt\ across the X-ray surface brightness edges in Figure~\ref{fig:AnalysesSpectraT0_T1}. 
Considering the azimuthal variations of \jt\ along the sharpest edges described in Section~\ref{sec:AnalysesSx}, the strongest temperature contrasts occur at the middle sections of the spiral excesses, with mean temperature contrasts of $0.78^{+0.11}_{-0.083}$, $0.67^{+0.18}_{-0.10}$, and $0.31^{+0.054}_{-0.031}$ for A496, A2029, and A1644. 
We note that the temperature contrast \jt\ in A1644 is most prominent, consistent with its density contrast \jn, indicating that the gas radially inward and outward of the surface brightness edges has significantly distinct thermodynamic properties.
Additionally, a similar trend in the azimuthal variation is also found in stripping cold fronts, where the amplitude of the temperature contrast typically peaks at the leading edge of the front.

For sloshing cold fronts, the gas within the spiral excess is expected to be colder than the surrounding ambient gas.
In this picture, we expect $\jt > 1$ and $\jt < 1$ at $r_{\rm{in}}$ and $r_{\rm{out}}$, respectively.
There are some edges that yield $j_{T} > 1.0$ at a $1\sigma$ level. 
We visually inspect the positions of these edges in the residual maps (the bottom panels in Figure~\ref{fig:AnalysesImaging}) and conclude that there are three types of feature that could result in the super-unity temperature contrast:
(1) The positions of the detected edges are not aligned with the boundaries of the spiral excesses, such as $r_{\rm{out}}$ of the sector $1$ in A496, $r_{\rm{in}}$ of the sector $6$ and $r_{\rm{out}}$ of the sector $7$ in A2029, and $r_{\rm{out}}$ of the sector $10$ in A1644;
(2) The edges are located along the inner boundaries of the spiral excesses, where \jt\ is expected to be higher than unity, for example $r_{\rm{in}}$ of the sector $6$ in A496 and $r_{\rm{in}}$ in sector $11$ in A2029; (3) The edge lies at the outer boundary of the spiral deficit, where the gas is warmer and less dense than its surroundings. In this configuration, the contacting surface is also expected to have $j_{T} > 1.0$, as in $r_{\rm{out}}$ of the sector $8$ in A2029. Overall, these three cases are all attributed to the imperfect detection of the positions of the surface brightness edges.



\subsection{The thermal pressures and entropy}
\label{sec:AnalysesAzimuthalVariations}

In this section, we derive the azimuthal profiles of the thermal pressure contrasts \jp\ , defined as the ratio of the thermal pressure radially inward of the surface brightness edge ($p_{\rm{i}}$) to that radially outward of the edge ($p_{\rm{o}}$). The thermal pressure is expressed as
\begin{equation}
\label{eq:AnalysesAzimuthalVariations_p}
p = n_{\rm{e}} \times T \, ,
\end{equation}
where $n_{\rm{e}}$ represents the ICM electron number density and $T$ is the  deprojected ICM temperature.

The azimuthal variations of thermal pressure contrasts \jp\ of the sample are shown in Figure~\ref{fig:AnalysesSpectraT0_T1}.
When focusing on the sharpest edges described in Section~\ref{sec:AnalysesSx}, we find that the thermal pressure contrast is less prominent at the middle sector (sector 5) and \jp\ are generally above unity in A496.
In contrast, both A2029 and A1644 exhibit a distinct feature from that of A496, characterized by the strongest thermal pressure contrast appearing in the middle sector and remaining below one except for sector 5 in A1644.
The findings of sub-unity pressure contrasts at the outer boundaries of the spiral excesses suggest that the sloshing ICM contributes substantial non-thermal pressure support to the cold fronts, in agreement with simulations \citep{ZuHone2013}.
Furthermore, the distinct azimuthal variations of the thermal pressure contrasts in our sample relative to those in stripping cold fronts \citep{Ueda2024} suggest that these two types of cold fronts are supported by different physical mechanisms.

For reference, we derive the entropy, which is
computed as
\begin{equation}
\label{eq:AnalysesAzimuthalVariationsK}
K = n_{\rm{e}}^{-2/3} \times T \, .
\end{equation}
Similarly, we present the azimuthal variations of entropy contrasts \jk\, defined as the entropy radially inward of the surface brightness edge divided by that radially outward of it ($\frac{K_{\rm{i}}}{K_{\rm{o}}}$), 
in Figure~\ref{fig:AnalysesSpectraT0_T1}.

In the body of a sloshing spiral, where the ICM temperature is relatively low, we expect that the temperature contrast \jt\ is larger (smaller) than unity at the inner (outer) spiral boundary of the spiral excess.
Similarly, we expect the density contrast \jn\ is smaller (larger) than unity at the inner (outer) spiral boundary, because the ICM density is higher in the spiral excess of a sloshing spiral.
As a result, the entropy contrast \jk\ of a sloshing cold front is expected to have $\jk > 1$ ($\jk < 1$) at the inner (outer) spiral boundary.
For each cluster in our sample, the entropy contrasts on the sharpest edges aligned with the outer boundaries of the spiral excesses (listed in Section~\ref{sec:AnalysesSx})
are largely smaller than $1$, with A1644 showing the most significant entropy contrast ($\jk = 0.18^{+0.04}_{-0.03}$).
In the profile of A2029, however, 
the entropy contrasts across the outer edges $r_{\rm{out}}$ at the outer boundary of the spiral tail (sectors $1$ to $4$) are broadly consistent with unity; meanwhile, its thermodynamic properties of the ICM radially inward and radially outward of the edges differ only slightly in the tail region.
We note that the entropy contrast depends more strongly on the temperature contrast than the density contrast, because the latter enters with a power of $-2/3$ and the former scales linearly in the entropy.



\section{Discussion}
\label{sec:Discussion}

In Section~\ref{sec:DiscussionCF}, we present the detected X-ray surface brightness edges that satisfy the criteria of cold fronts. 
We assess the gas motion as the source of the non-thermal pressure supports for the sloshing cold fronts in Section~\ref{sec:DiscussionV}.
Alternatively, we also provide an estimate on the magnetic pressure support inferred from our measurements of the thermal pressure contrasts in Section~\ref{sec:pb}.

\subsection{Sloshing cold fronts in the sample}
\label{sec:DiscussionCF}

\begin{table*}
    \centering
        \caption{Summary of cold front detections in our sample.}
        \label{tab:key_results}
            \begin{tabular}{cccc}
            \hline
            \hline
            Cluster & Sector & Location ($''$) & Reference \\
            \hline
            A496  & $1$, $2$ & $r_{\rm in}$ & Newly discovered \\
                  & $5$, $6$ & $r_{\rm out}$ & \cite{Ghizzardi2014} \\
            \hline
            A2029 & $7$, $8$ & $r_{\rm in}$ & See Section~\ref{sec:DiscussionCF} and Appendix~\ref{app:SysUncertain} \\
                  & $9$ & $r_{\rm out}$ & See Section~\ref{sec:DiscussionCF} and Appendix~\ref{app:SysUncertain} \\
            \hline
            A1644 & $2$, $3$, $4$, $5$, $6$ & $r_{\rm in}$ & \cite{Johnson2010} \\
                  & $7$ & $r_{\rm out}$ & \cite{Johnson2010} \\
            \hline       
            \end{tabular}
\end{table*}

In this section, we describe the criteria of detecting a sloshing cold front in our cluster sample.
Because the density and temperature contrasts of a sloshing cold front are typically less prominent than those of stripping cold fronts, 
we adopt a relatively loose criterion for identifying sloshing cold fronts given the same data. 
In this work, a cold front is identified as an X-ray surface brightness edges with both 
\[
\jn > 1 \,~\mathrm{and}~\jt < 1
\]
at greater than a $1\sigma$ confidence level.
Although sloshing cold fronts may occur along either the inner or outer boundaries of the spiral excess, the inner boundary lies too close to the cluster center to be identified given the angular-resolution limits of observations.
Additionally, the central region often exhibits strong perturbations that obscure the surface brightness edges.
For these reasons, we only consider the edges aligned with the outer boundaries of the spiral excess. 
Since the gas within a spiral excess is expected to share the same physical properties, excluding the edges along the inner spiral boundary does not alter our physical interpretation, and therefore considering only the edges at the outer spiral boundary yields equivalent results. We summarize the key results, including the sectors and locations of the cold fronts, and comparisons with previous literature in Table~\ref{tab:key_results}.

In this definition, we identify the cold fronts at $r_{\rm{in}}$ in sector $1$ and $2$ and $r_{\rm{out}}$ in sector $5$ and $6$ in A496. 
The cold fronts in sector $1$ and $2$ are newly discovered, whereas those in sector $5$ and $6$ have been reported previously \citep{Ghizzardi2014}. 

For A2029, we are not able to detect sloshing cold fronts at a significance larger than $1\sigma$, even though there are clear spiral excesses in the residual map. This picture is consistent with all previous work \citep{Clarke2004, Clarke2004a, Paterno-Mahler2013}, which presented clear spiral patterns in their residual maps, temperature or entropy maps, suggesting that either sloshing cold fronts genuinely do not exist in the observed spiral excesses or that X-ray observations with higher resolution will be needed for this cluster. We note that our conclusion here is strictly based on our analysis assumptions and strategies, which include the uniform sector definition, the assumed ICM spherical distribution, the exclusion of temperature dependence in the X-ray surface brightness fitting, and our criteria for identifying cold fronts. As we demonstrate in Appendix~\ref{app:SysUncertain}, solely adjusting the sector definition can change the ultimate conclusion on an identified edge.

In A1644, we detect cold fronts at $r_{\rm{in}}$ from sector $2$ to sector $6$ and at $r_{\rm{out}}$ in sector $7$, which are consistent with the result in \cite{Johnson2010}.

In what follows, we assess the systematic uncertainties associated with the assumptions made in the detection of the cold fronts.
The first is the assumption of spherical symmetry in the spectral deprojection.
Despite the prominent anisotropy visible in the X-ray morphology, the sloshing patterns represent only a perturbation to the global ICM distribution.
To first order, the assumption of spherical symmetry affects only the geometric overlap between the iso-emissivity shells and line-of-sight cylinders.
The observed X-ray flux is the integral of the emissivity along the line of sight.
Because the X-ray emissivity is proportional to the square of the gas density, the inferred gas density $n$ depends on the assumed line-of-sight extent of the ICM distribution.
Approximately, the inferred gas density scales as $n\propto q^{-\frac{1}{2}}$, where $q$ is the ratio of the line-of-sight extent of the ICM distribution to that of a spherically symmetric distribution.
For a prolate (oblate) ICM distribution, we have $q > 1$ ($q < 1$).
As we are quantifying the ratios of the thermal properties across the edges, the geometric factors cancel in both the numerator and denominator.
Therefore, we do not expect significant impact of the spherical symmetry on the results.
For comparison, \cite{Naor2020} studied the southwest cold front in A2029 with a careful modeling accounting for the non-spherical ICM distribution and obtained the pressure contrast of $0.67^{+0.18}_{-0.13}$, in good agreement with our measurement of $0.73^{+0.27}_{-0.21}$ at $r_{\mathrm in}$ of sector 4.

The second systematic uncertainty is associated with the assumption of an isothermal gas in the surface brightness modeling.
We fit the surface brightness profile of each sector ignoring the dependence on the gas temperature in equations~(\ref{eq:AnalysesSx3ComModel})~and~(\ref{eq:AnalysesSx2ComModel}).
The X-ray emissivity $\epsilon$ is characterized by both the gas density $n$ and temperature $T$ as $\epsilon \propto n^{2} \sqrt{T}$.
Consequently, the inferred gas density scales approximately as $n \propto T^{-\frac{1}{4}}$.
Consider a global temperature $T_{\mathrm{cl}}$.
The inferred gas density can be further reduced as $n \propto T_{\mathrm{cl}}^{-\frac{1}{4}} \left(1 + \delta_T \right)^{-\frac{1}{4}}$, where the temperature fluctuation $\delta T$ is given by $\delta_T = \frac{\Delta T}{ T_{\mathrm{cl}} }$.
Because we derive the density contrasts as the ratios across the edges in the parameterized density profiles, a density contrast \jn\ between the inner and outer regions of a sloshing cold front becomes
\[
\frac{n_{\rm i}}{n_{\rm o}} = \left(\frac{n_{\mathrm{i}}}{ n_{\mathrm{o}} } \right)_{\mathrm{isothermal}}
\times \left(\frac{
1 + \delta_{T,\mathrm{i}}
}{
1 + \delta_{T,\mathrm{o}}
}\right)^{-1/4} \, ,
\]
where the subscript $\mathrm{isothermal}$ denotes the density contrast obtained with an isothermal model.
Taking face values of $T_{\mathrm{cl}} = 5~\mathrm{keV}$, $\Delta T_{\mathrm{i}} = -2~\mathrm{keV}$, and $\Delta T_{\mathrm{o}} = 2~\mathrm{keV}$, corresponding to a temperature contrast $\jt = T_{\mathrm{i}} / T_{\mathrm{o}} \approx 0.4$, we have a bias factor for the inferred density contrast as 
$\left(\frac{
1 + \frac{\Delta T_{\rm i}}{T_{\mathrm{cl}}}
}{
1 + \frac{\Delta T_{\rm o}}{T_{\mathrm{cl}}}
}\right)^{-1/4}
\approx 1.24$.
Similarly, this bias factor becomes $\approx1.1$ for $\left(T_{\mathrm{cl}}, \Delta T_{\mathrm{i}}, \Delta T_{\mathrm{o}}\right) = \left(5~\mathrm{keV}, -1~\mathrm{keV}, 1~\mathrm{keV}\right)$, corresponding to $\jt = 0.7$.
Therefore, we expect a systematic uncertainty of the isothermal model on the density contrast at a level of approximately $10\%$ to $20\%$ for an edge with a large temperature contrast.


\subsection{The non-thermal pressure from the gas motion}
\label{sec:DiscussionV}

In Section~\ref{sec:AnalysesAzimuthalVariations}, we observe the contrasts of the thermal pressure at the X-ray surface brightness edges (see Figure~\ref{fig:AnalysesSpectraT0_T1}), strongly indicating the presence of non-thermal pressure supports to maintain the sharpness of the cold fronts.
In fact, stripping cold fronts in nearby galaxy clusters have been studied in this context \citep[e.g., ][]{Vikhlinin2001, Ichinohe2017, Ueda2024}.

In the following, we discuss the non-thermal pressure support originating from gas motions. We begin by expressing the required non-thermal pressure support $p_{\mathrm{nth}}$ as:
\begin{equation}
\label{eq:p_nth}
    p_{\mathrm nth} =
p_{\mathrm{th, o}} \times |\jp - 1|
\, ,
\end{equation}
where $p_{\mathrm{th, o}}$ is the thermal pressure of the gas farther from the cluster center.
Assuming that the gas is adiabatic, we have $\gamma \frac{p_{\mathrm th}}{\rho} = c_{\rm s}^2$, where $\gamma$ is the adiabatic index.
We compute the sound speed as $\sqrt{\frac{\gamma T}{m_{\mathrm p} \mu}}$, where $m_{\mathrm{p}}$ is the proton mass and $\mu$ is the mean molecular weight of the ICM evaluated as $0.6$.
Assuming that the entire non-thermal pressure support is attributed to the gas motion, we quantify the required three-dimensional velocity dispersion to maintain the local pressure equilibrium as:
\begin{equation}
\label{eq:sigma_v}
\sigma_v =
\pm \sqrt{\frac{p_{\mathrm{nth}}}{\rho}} =
\pm \sqrt{\frac{c_{\mathrm{s, o}}^2}{\gamma}
      \Big|\jp - 1\Big|}
\, .
\end{equation}
Note that the sign in Equation~(\ref{eq:sigma_v}) does not represent the direction of the local gas flow.
Instead, it is introduced to indicate the side of the cold front which requires the non-thermal pressure support.
As a result, we are able to infer the azimuthal variations of the three-dimensional velocity dispersion in our sample, as shown in the bottom panel of Figure~\ref{fig:AnalysesSpectraT0_T1}. For the sharpest edges in A496 and A2029, the velocity dispersions are generally consistent, showing no clear azimuthal trend. For A1644, the amplitudes of the velocity dispersions show a similar pattern to those of the sharpest edges in the other two clusters. However, in sector $5$, the required non-thermal pressure arises from the opposite side of the cold front relative to the other sectors in A1644. Nevertheless, we currently do not visually find any significant difference in this sector compared to others under our assumption and analysis strategy.

If the gas bulk motion is a primary source for this non-thermal pressure \citep[see also][]{Vikhlinin2001}, we can describe it by the Euler equation.
Because cold fronts are highly localized perturbations, and assuming locally steady states, the local Euler equation governing the gas motions across them is given by
\begin{equation}
    (\mathbf{v} \cdot \nabla) \mathbf{v} =
-\frac{1}{\rho} \nabla p_{\mathrm{th}}
\, ,
\end{equation}
in which $p_{\mathrm{th}}$ is the thermal pressure, $\rho$ is the gas mass density, and $\mathbf{v} = \left(v_r, v_\theta, v_\phi\right)$ is the bulk velocity of the gas in the spherical coordinate system $\left( r, \theta, \phi\right)$.
We express the gas bulk motion in the tangential direction as $v_{\mathrm{t}} = \sqrt{v_{\theta}^2 + v_{\phi}^2}$.
As seen, the bulk motion of the gas forms a ram pressure $p_{\mathrm{ram}}$ as a source of the non-thermal pressure.
Specifically, the ram pressure in the $r$ direction yields an order-of-magnitude relation
$p_{\mathrm{ram}} \propto \rho \left( v_r^2 + v_{\mathrm{t}}^2\right)$, containing the contributions from both the radial and tangential flows of the gas.

In the context of sloshing cold fronts, the bulk motion of the gas follows a spiral pattern with the flow direction being predominantly tangential to the cold front edge.
Numerical simulations consistently indicate that the tangential flow of the gas dominates over the radial component in sloshing cold fronts \citep{Ascasibar2006}.
Consequently, the ram pressure associated with the tangential bulk motion is expected to be significantly larger than that associated with the radial bulk motion by a factor of approximately $v_{\mathrm{t}}^2 / v_r^2$, which may exceed an order of magnitude if $v_{\mathrm{t}} \gg v_{r}$.
That is, we expect that the ram pressure in sloshing cold fronts is primarily dominated by the tangential gas flow.
If the ram pressure of the ICM bulk motion is the dominant component for the non-thermal pressure support, we are able to infer 
the amplitude of the ICM bulk motion across the surface brightness edges from the thermal pressure contrasts (\jp).

In Appendix~\ref{app:StagPoinModel}, on the other hand, we provide an alternative estimate of the effective velocity gradient using a toy model that is commonly used to describe the gas flow near the stagnation point of a stripping cold front.
Following \cite{Ghizzardi2014}, we infer the Mach number using the stagnation-point model of \cite{Vikhlinin2001} and further estimate the amplitude of the corresponding velocity gradient.
Despite providing a useful order-of-magnitude estimate, we stress that the stagnation-point model is not physically ideal for describing the sloshing cold fronts, where the gas continues to move along the front rather than decelerating to zero velocity at a stagnation point.

\subsection{The magnetic pressure}
\label{sec:pb}

One of the possible mechanisms for the non-thermal pressure support, apart from the ICM bulk motions, is the ICM microphysical properties, such as turbulence, magnetic fields, and viscosity. 
The tangential flow of the ICM is expected to be the fastest and highly ordered at the starting position of the spiral excess. 
This highly ordered flow stretches the magnetic field along the flow direction, which is capable of supporting cold fronts.

Indeed, it was discovered in MeerKAT observations that radio jets bent by magnetic fields align well with the structure of a cold front in the merging cluster Abell\,3376 \citep{Chibueze2021}. 
Furthermore, \cite{Walker2017} found that radio mini-halos in the cool cores of galaxy clusters exhibit morphologies similar to the corresponding cold fronts observed in X-rays. 
The radio emissions are produced by the synchrotron emission of electrons which are re-accelerated by turbulence in the sloshing motions.
These features suggest that magnetic fields might play a role in shaping sloshing cold fronts, resulting in non-negligible magnetic pressures. 

In the following, we provide an order-of-magnitude estimate of the upper bound on the magnetic pressure inferred from the measured thermal pressure contrasts.
We consider the scenario that the non-thermal pressure support is entirely attributed to the magnetic field strength $B_{\mathrm{upper}}$ across the surface brightness edge.
The magnetic pressure is expressed as 
$p_{\mathrm{B}} = \frac{ B^2 }{ 8 \pi }$.
For the case of $\jp > 1$, the non-thermal pressure support $p_{\mathrm{nth}}$ to maintain pressure equilibrium is
\[
p_{\mathrm{nth}} = \frac{1}{8 \pi} B_{\mathrm{upper}}^2 \, ,
\]

and it is related to the thermal pressure $p_{\mathrm{th}}$ as $p_{\mathrm{nth}}= p_{\rm{th}} \times \left(j_{p} - 1\right) $.
Thus, we have

\begin{equation}
\label{eq:pb2}
B_{\mathrm{upper}} \approx \sqrt{ 8\pi p_{\mathrm{nth}} } = \sqrt{ 8\pi p_{\mathrm{th}} \left(j_p - 1\right)}  \, .
\end{equation}

We express the thermal pressure in terms of the gas mass fraction $f_{\mathrm{gas}}$ and temperature $T$ within the radius $R_{500}$ as
\[
p_\mathrm{th} = \frac{\rho_{\mathrm{gas}}}{\mu m_{\mathrm{p}}} k_{\mathrm{B}} T = \frac{f_{\mathrm{gas}} \times (500 \rho_{\mathrm{crit}})}{\mu m_{\mathrm{p}}} k_{\mathrm{B}} T \, ,
\]where $\rho_{\rm gas}$ is the ICM density, $k_{\mathrm{B}}$ is the Boltzmann constant and $\rho_{\rm crit}$ is the critical density at the cluster redshift.
Accounting for the dependence of the thermal pressure on $f_{\mathrm{gas}}$ and  $T$, we estimate the upper bound of the magnetic field $B_{\mathrm{upper}}$ as
\begin{equation}
\label{eq:pb3}
B < B_{\mathrm{upper}}
\approx 
10\,\left(\left(j_{p} - 1\right)\left(\frac{f_{\mathrm{gas}}}{0.1}\right)\left(\frac{T}{5\mathrm{keV}}\right) \right)^{\frac{1}{2}}\,\mu\rm{G} \, .
\end{equation}

To provide an estimation of 
$B_{\mathrm{upper}}$ for each of our samples, we take the mean of thermal pressure contrasts that are greater than unity, approximate $f_{\rm gas} = 0.1$, and set $T$ at the mean deprojected temperature of the outermost radial bins.
The resulting parameter sets $\left( j_p , T \right)$ are $\left( 1.3 , 6.2\,\rm{keV} \right)$, 
$\left( 2.4 , 8.1\,\rm{keV} \right)$, and 
$\left( 1.2 , 5.1\,\rm{keV} \right)$ for A496, A2029, and A1644, respectively. 
Consequently, the quantities $B_{\mathrm{upper}}$ of A496, A2029, and A1644 are approximately $6\,\mu\rm{G}$, $15\,\mu\rm{G}$, and $5\,\mu\rm{G}$, respectively.

We note that the resulting magnetic field $B_{\mathrm{upper}}$ represents an upper limit, under the assumption that all non-thermal pressure support comes from the magnetic pressure.
Nevertheless, our estimation serves as a good approximation, as it is comparable to the local magnetic field strength of approximately $3$ to $5\,\mu\rm{G}$ at the center of Coma cluster from Faraday rotation measurements \citep{Bonafede2010}.
Additionally, \cite{Chen2017} estimated the lower limits of the magnetic fields in Abell\,2204 at levels of approximately $20$--$30\,\mu\mathrm{G}$ to suppress the Kelvin-Helmholtz instability.
Also, \cite{Chibueze2021} found that a magnetic field of $\sim 10\,\mu\mathrm{G}$ is required to reproduce the morphology of observed radio jets in Abell\,3376.


\section{Conclusions}
\label{sec:Summary}

In this paper, we perform a study of the sloshing cold fronts in three nearby galaxy clusters A496, A2029, and A1644 using archival X-ray datasets from {\em Chandra} observatory.
Our goal is to investigate the azimuthal variations of the thermodynamic properties---the gas density $n_{\mathrm{e}}$, temperature $T$, thermal pressure $p$, and the entropy $K$---and quantify their contrast across the edges of the cold fronts.
The three clusters at low redshifts $z\lesssim 0.08$ are massive and X-ray bright, and display a clear gas sloshing pattern in the plane of sky, resulting in an ideal sample for this task.

We employ a systematic approach to study the cold fronts to minimize the dependence on visual inspections, as one of the most distinct features in this work.
To begin with, we reduce the X-ray data sets closely following the reprocessing steps detailed in previous work \cite{Ueda2024}.
For each cluster, we then subtract the global distribution from the observed surface brightness map, where the global distribution is inferred from an elliptical model.
We observe clear spiral patterns in the residual map, as the signatures of sloshing cold fronts.
Next, we determine the head and direction of the sloshing cold fronts, followed by an azimuthal binning with an open angle of $30^{\circ}$ starting from the head along the sloshing direction.
This results in $12$ sectors per cluster.
Depending on the fitting quality, we fit a two- or three-component gas density model to the observed surface brightness profile of each sector.
The parameters of the best-fit surface brightness profile of the sector define the edges and the density contrasts \jn\ of the surface brightness edges at the corresponding azumuthal angle.
Following that, we construct the per-sector radial binning, including the cold front edges, in a way that each radial bin shares roughly the same number of X-ray photons.
This promises that the signal-to-noise ratios of the measurements in individual radial bins of one sector are approximately the same.
With the defined radial bins, we extract the spectra and perform spectroscopic analyses to obtain, the (projected and deprojected) temperature $T$, the thermal pressure $p$, and the entropy $K$.
Finally, we quantify the contrasts \jn, \jt, \jp, and \jk\ in the density, temperature, thermal pressure, and the entropy, respectively, in the azimuthal angles defined by the sectors.

We classify a surface brightness edge as a sloshing cold front if it exhibits a super-unity density contrast ($\jn > 1$) and a sub-unity temperature contrast ($\jt < 1$) at levels greater than $1\sigma$.
Focusing on the edges that align with the outer boundaries of the spiral excesses, we successfully detect the northern cold front in A496 and the eastern cold front in A1644, both of which have been detected in previous literature.
Furthermore, we detect the southwest cold front in A496 for the first time.
As for A2029, while its edges do not meet our criteria for a cold front detection, the conclusion of no cold front presence remains sensitive to specific choices of the spectral extraction regions.
Notably, solely adjusting the sector definition can yield an opposite conclusion.

We observe the thermal pressure contrast for the sharpest edges in our sample, indicating the presence of the non-thermal pressure support.
We discuss and quantify the possible sources of this non-thermal pressure support.
Assuming the non-thermal pressure support entirely originates from gas motions, we provide an order-of-magnitude estimate of the velocity dispersion at each edge. 
In this scenario, we do not find significant azimuthal variations of the velocity dispersion along the sharpest brightness edges.
On the other hand, we estimate the upper limit of the magnetic field along the edges of the cold fronts, assuming that the non-thermal pressure support is entirely attributed to the magnetic pressure.
With the assumption, our thermal pressure contrasts imply upper limits of the magnetic fields as $B < 6\,\mu\mathrm{G}$, $B < 15\,\mu\mathrm{G}$, and $B < 5\,\mu\mathrm{G}$ for A496, A2029, and A1644, respectively.
These estimates are consistent with that observed in Coma cluster.

Finally, we discuss the caveats associated with the modeling assumptions, including the spherical symmetry and isothermal model.
Although the assumptions may result in systematic uncertainties on the derived contrasts of the edges, the physical interpretation remains intact in this work.
Nevertheless, our homogeneous approach minimizes the visual inspection and confirmation bias, at the cost of introducing the scatter compared to case studies where customized analyses are adopted.
Hence, one of the important future directions is to increase the sample size, which will provide a promising way for statistical studies by lowering the uncertainty of the mean.
Moreover, this work highlights the importance of studies using the ongoing X-ray mission \textit{XRISM}, which provides high-resolution X-ray spectra to uncover the velocity structure and dynamical state of ICM cold fronts.



\begin{acknowledgments}
We sincerely thank the anonymous referee for the detailed and constructive comments that led to significant improvements of this manuscript.
We thank Sheng-Chieh Lin for providing constructive comments on this manuscript.
I-Hsuan Li thanks the support from the ASIAA Summer Student Program.
I-Non Chiu thanks Yi Yang for the hospitality in the Institute of Physics (IoP), Academia Sinica.
This work is supported by the National Science and Technology Council in Taiwan (Grant NSTC 111-2112-M-006-037-MY3 and 114-2112-M-006-017-MY3).
This work makes use of the resources in National Center for High-performance Computing (NCHC) in Taiwan.
This work is supported by JSPS KAKENHI grant number JP25K23398 and JP26K00741 (S.U.). 
Shutaro Ueda acknowledges support by Program for Forming Japan's Peak Research Universities (J-PEAKS) Grant Number JPJS00420230006.
K.U. acknowledges support from the National Science and Technology Council of Taiwan (grant NSTC 112-2112-M001-027-MY3) and the Academia Sinica Investigator Award (grant AS-IA112-M04).
The scientific results in this article are based on data obtained from the Chandra Data Archive (CDA), contained in \dataset[DOI:10.25574/cdc.600]{https://doi.org/10.25574/cdc.600}.
\end{acknowledgments}

\facilities{CXO}
\software{astropy \citep{AstropyCollaboration2013}, CIAO \citep{Fruscione2006}, XSPEC \citep{Arnaud1996}}

\clearpage

\appendix
\twocolumngrid



\section{Radial binning definition}
\label{app:RadBinDef}
Table~\ref{tab:AnalysesSpectraRadBin} summarizes the radial bins adopted in the spectral analyses.

\begin{table*}
    \begin{center}
        \caption{A summary of the outer boundaries of the radial bins. (The boundaries are in units of $''$ and measured from the cluster center.)}
        \label{tab:AnalysesSpectraRadBin}
        \resizebox{\textwidth}{!}{%
            \begin{tabular}{cccccccccccccc}
            \hline
            \hline
            Cluster & Radial bin & Sector $4$ & Sector $5$ & Sector $6$ & Sector $7$ & Sector $8$ & Sector $9$ & Sector $10$ & Sector $11$ & Sector $12$ & Sector $1$ & Sector $2$ & Sector $3$ \\
                 &      & $23^{\circ} \leq \phi < 53^{\circ}$ & $53^{\circ} \leq \phi < 83^{\circ}$ & $83^{\circ} \leq \phi < 113^{\circ}$ & $113^{\circ} \leq \phi < 143^{\circ}$ & $143^{\circ} \leq \phi < 173^{\circ}$ & $173^{\circ} \leq \phi < 203^{\circ}$ & $203^{\circ} \leq \phi < 233^{\circ}$ & $233^{\circ} \leq \phi < 263^{\circ}$ & $263^{\circ} \leq \phi < 293^{\circ}$ & $293^{\circ} \leq \phi < 323^{\circ}$ & $323^{\circ} \leq \phi < 353^{\circ}$ & $353^{\circ} \leq \phi < 383^{\circ}$ \\
            \hline
            A496 & $1$  & $17.000$ & $15.000$ & $17.000$ & $15.000$ & $13.000$ & $15.000$ & $16.000$ & $19.000$ & $19.000$ & $17.000$ & $17.000$ & $16.000$ \\
                 & $2$  & $28.000$ & $25.054$ & $28.691$ & $24.000$ & $21.000$ & $25.628$ & $26.478$ & $34.125$ & $31.000$ & $26.000$ & $27.000$ & $25.000$ \\
                 & $3$  & $38.000$ & $35.000$ & $39.000$ & $33.000$ & $30.224$ & $41.000$ & $40.000$ & $49.000$ & $45.235$ & $35.000$ & $37.000$ & $34.000$ \\
                 & $4$  & $48.000$ & $44.000$ & $48.000$ & $43.000$ & $43.000$ & $56.000$ & $54.000$ & $64.000$ & $61.000$ & $45.131$ & $47.000$ & $42.558$ \\
                 & $5$  & $58.280$ & $52.000$ & $57.000$ & $51.848$ & $56.000$ & $72.000$ & $69.000$ & $78.000$ & $76.000$ & $61.000$ & $58.137$ & $52.000$ \\
                 & $6$  & $74.000$ & $61.000$ & $66.000$ & $62.000$ & $69.000$ & $88.000$ & $82.000$ & $91.000$ & $91.000$ & $79.000$ & $74.000$ & $64.000$ \\
                 & $7$  & $92.340$ & $70.000$ & $75.000$ & $72.000$ & $82.000$ & $105.000$ & $96.000$ & $104.000$ & $105.000$ & $95.000$ & $91.000$ & $79.000$ \\
                 & $8$  & $115.000$ & $79.000$ & $84.000$ & $82.000$ & $96.000$ & $122.396$ & $109.000$ & $117.000$ & $119.000$ & $112.000$ & $109.000$ & $97.602$ \\
                 & $9$  & $137.000$ & $91.512$ & $96.851$ & $93.000$ & $111.000$ & $140.000$ & $123.000$ & $130.000$ & $133.000$ & $129.000$ & $128.000$ & $119.000$ \\
                 & $10$ & $160.000$ & $112.000$ & $115.000$ & $104.000$ & $129.189$ & $159.000$ & $137.000$ & $144.000$ & $147.000$ & $147.000$ & $148.000$ & $139.000$ \\
                 & $11$ & $184.000$ & $134.000$ & $135.000$ & $117.360$ & $150.000$ & $179.000$ & $153.471$ & $158.000$ & $161.000$ & $164.590$ & $169.000$ & $161.000$ \\ 
                 & $12$ & $214.000$ & $156.000$ & $156.000$ & $136.000$ & $172.000$ & $201.000$ & $170.000$ & $172.000$ & $176.000$ & $182.000$ & $192.000$ & $183.000$ \\
                 & $13$ & $252.000$ & $180.000$ & $177.000$ & $156.000$ & $196.000$ & $225.000$ & $188.000$ & $187.000$ & $192.000$ & $202.000$ & $215.872$ & $205.000$ \\
                 & $14$ &   & $217.000$ & $198.000$ & $178.000$ & $222.000$ & $252.000$ & $207.000$ & $203.000$ & $208.000$ & $224.000$ & $252.000$ & $227.000$ \\
                 & $15$ &   & $252.000$ & $222.000$ & $201.000$ & $252.000$ &   & $229.000$ & $220.000$ & $226.000$ & $252.000$ &   & $252.000$ \\
                 & $16$ &   &   & $252.000$ & $225.000$ &   &   & $252.000$ & $239.185$ & $247.726$ &   &   &   \\
                 & $17$ &   &   &   & $252.000$ &   &   &   & $252.000$ & $252.000$ &   &   &   \\
            \hline
            Cluster & Radial bin & Sector $1$ & Sector $12$ & Sector $11$ & Sector $10$ & Sector $9$ & Sector $8$ & Sector $7$ & Sector $6$ & Sector $5$ & Sector $4$ & Sector $3$ & Sector $2$ \\
                 &      & $23^{\circ} \leq \phi < 53^{\circ}$ & $53^{\circ} \leq \phi < 83^{\circ}$ & $83^{\circ} \leq \phi < 113^{\circ}$ & $113^{\circ} \leq \phi < 143^{\circ}$ & $143^{\circ} \leq \phi < 173^{\circ}$ & $173^{\circ} \leq \phi < 203^{\circ}$ & $203^{\circ} \leq \phi < 233^{\circ}$ & $233^{\circ} \leq \phi < 263^{\circ}$ & $263^{\circ} \leq \phi < 293^{\circ}$ & $293^{\circ} \leq \phi < 323^{\circ}$ & $323^{\circ} \leq \phi < 353^{\circ}$ & $353^{\circ} \leq \phi < 383^{\circ}$ \\
            \hline
            A2029 & $1$  & $10.000$ & $9.000$ & $8.000$ & $9.753$ & $9.893$ & $11.000$ & $11.000$ & $10.000$ & $10.000$ & $10.000$ & $10.000$ & $9.000$ \\
                  & $2$  & $18.000$ & $16.000$ & $14.170$ & $16.000$ & $17.000$ & $18.000$ & $17.000$ & $16.000$ & $16.000$ & $16.000$ & $17.000$ & $16.000$ \\
                  & $3$  & $25.519$ & $21.291$ & $20.000$ & $22.000$ & $22.000$ & $23.284$ & $23.031$ & $21.000$ & $22.582$ & $21.221$ & $24.000$ & $23.080$ \\
                  & $4$  & $32.000$ & $27.000$ & $25.000$ & $27.000$ & $27.000$ & $29.000$ & $29.000$ & $26.000$ & $28.000$ & $27.000$ & $31.230$ & $31.000$ \\
                  & $5$  & $38.000$ & $33.000$ & $30.000$ & $31.000$ & $32.000$ & $35.000$ & $36.000$ & $32.000$ & $34.000$ & $33.000$ & $38.000$ & $39.000$ \\
                  & $6$  & $45.000$ & $39.000$ & $35.000$ & $36.000$ & $37.841$ & $42.000$ & $43.000$ & $38.000$ & $40.000$ & $39.000$ & $44.000$ & $47.000$ \\
                  & $7$  & $52.000$ & $44.000$ & $40.000$ & $41.000$ & $44.000$ & $50.000$ & $51.000$ & $44.000$ & $46.000$ & $45.000$ & $51.000$ & $55.000$ \\
                  & $8$  & $59.000$ & $50.000$ & $45.000$ & $46.000$ & $51.000$ & $59.000$ & $60.000$ & $51.000$ & $52.000$ & $51.000$ & $58.000$ & $63.000$ \\
                  & $9$  & $66.268$ & $56.000$ & $50.000$ & $50.678$ & $58.000$ & $67.889$ & $70.000$ & $58.000$ & $58.000$ & $57.000$ & $65.000$ & $72.000$ \\
                  & $10$ & $75.000$ & $62.000$ & $55.000$ & $56.000$ & $67.000$ & $79.000$ & $81.000$ & $66.000$ & $65.000$ & $63.000$ & $73.000$ & $80.770$ \\
                  & $11$ & $84.000$ & $68.390$ & $60.326$ & $61.000$ & $76.000$ & $91.000$ & $94.000$ & $74.000$ & $72.000$ & $70.000$ & $81.000$ & $92.000$ \\
                  & $12$ & $95.000$ & $75.000$ & $66.000$ & $67.000$ & $86.000$ & $106.000$ & $108.834$ & $83.000$ & $80.000$ & $77.000$ & $90.000$ & $104.000$ \\
                  & $13$ & $107.000$ & $83.000$ & $72.000$ & $73.000$ & $98.000$ & $122.000$ & $125.000$ & $93.000$ & $89.000$ & $84.000$ & $99.000$ & $118.000$ \\
                  & $14$ & $121.000$ & $92.000$ & $78.000$ & $80.000$ & $112.000$ & $141.000$ & $143.000$ & $105.000$ & $97.000$ & $91.000$ & $109.000$ & $134.000$ \\
                  & $15$ & $138.000$ & $103.000$ & $85.000$ & $88.000$ & $127.000$ & $163.000$ & $164.000$ & $119.722$ & $105.000$ & $98.000$ & $120.000$ & $153.000$ \\
                  & $16$ & $157.000$ & $115.000$ & $92.000$ & $97.000$ & $144.000$ & $186.000$ & $187.000$ & $133.000$ & $114.000$ & $106.000$ & $130.121$ & $177.000$ \\
                  & $17$ & $179.000$ & $129.000$ & $100.000$ & $106.000$ & $163.000$ & $213.000$ & $214.000$ & $147.000$ & $123.000$ & $114.000$ & $142.000$ & $206.000$ \\
                  & $18$ & $206.000$ & $144.000$ & $109.000$ & $116.000$ & $185.000$ & $242.000$ & $242.000$ & $162.000$ & $132.000$ & $122.000$ & $156.000$ & $242.000$ \\
                  & $19$ & $242.000$ & $161.000$ & $119.000$ & $127.000$ & $213.000$ &   &   & $178.000$ & $142.000$ & $131.000$ & $172.000$ &   \\
                  & $20$ &   & $181.000$ & $130.000$ & $139.000$ & $242.000$ &   &   & $196.000$ & $154.905$ & $140.000$ & $191.000$ &   \\
                  & $21$ &   & $207.000$ & $142.000$ & $152.000$ &   &   &   & $216.470$ & $166.000$ & $149.865$ & $214.000$ &   \\
                  & $22$ &   & $242.000$ & $155.000$ & $167.000$ &   &   &   & $242.000$ & $178.000$ & $161.000$ & $242.000$ &   \\
                  & $23$ &   &   & $169.000$ & $192.000$ &   &   &   &   & $191.000$ & $173.000$ &   &   \\
                  & $24$ &   &   & $189.000$ & $217.000$ &   &   &   &   & $206.000$ & $186.000$ &   &   \\
                  & $25$ &   &   & $216.000$ & $242.000$ &   &   &   &   & $223.000$ & $202.000$ &   &   \\
                  & $26$ &   &   & $242.000$ &   &   &   &   &   & $242.000$ & $220.000$ &   &   \\
                  & $27$ &   &   &   &   &   &   &   &   &   & $242.000$ &   &   \\
            \hline
            Cluster & Radial bin & Sector $11$ & Sector $12$ & Sector $1$ & Sector $2$ & Sector $3$ & Sector $4$ & Sector $5$ & Sector $6$ & Sector $7$ & Sector $8$ & Sector $9$ & Sector $10$ \\
                 &      & $28^{\circ} \leq \phi < 58^{\circ}$ & $58^{\circ} \leq \phi < 88^{\circ}$ & $88^{\circ} \leq \phi < 118^{\circ}$ & $118^{\circ} \leq \phi < 148^{\circ}$ & $148^{\circ} \leq \phi < 178^{\circ}$ & $178^{\circ} \leq \phi < 208^{\circ}$ & $208^{\circ} \leq \phi < 238^{\circ}$ & $238^{\circ} \leq \phi < 268^{\circ}$ & $268^{\circ} \leq \phi < 298^{\circ}$ & $298^{\circ} \leq \phi < 328^{\circ}$ & $328^{\circ} \leq \phi < 358^{\circ}$ & $358^{\circ} \leq \phi < 388^{\circ}$ \\
            \hline
            A1644 & $1$  & $20.000$ & $25.364$ & $19.862$ & $16.052$ & $20.121$ & $15.000$ & $14.000$ & $17.000$ & $16.648$ & $15.000$ & $19.000$ & $22.000$ \\
                  & $2$  & $33.630$ & $42.610$ & $91.000$ & $93.000$ & $92.000$ & $26.956$ & $22.000$ & $27.000$ & $26.000$ & $25.392$ & $32.255$ & $37.498$ \\
                  & $3$  & $45.000$ & $94.000$ & $142.000$ & $141.000$ & $137.000$ & $107.000$ & $32.047$ & $35.656$ & $33.000$ & $36.077$ & $42.000$ & $45.000$ \\
                  & $4$  & $54.000$ & $143.000$ & $184.000$ & $182.000$ & $179.000$ & $163.000$ & $112.000$ & $95.000$ & $40.105$ & $77.000$ & $49.000$ & $51.000$ \\
                  & $5$  & $63.000$ & $184.000$ & $222.000$ & $222.000$ & $222.000$ & $222.000$ & $164.000$ & $140.000$ & $102.000$ & $124.000$ & $55.000$ & $56.501$ \\
                  & $6$  & $71.095$ & $222.000$ &   &   &   &   & $222.000$ & $180.000$ & $155.000$ & $174.000$ & $61.000$ & $79.000$ \\
                  & $7$  & $105.000$ &   &   &   &   &   &   & $222.000$ & $222.000$ & $222.000$ & $66.082$ & $106.000$ \\
                  & $8$  & $143.000$ &   &   &   &   &   &   &   &   &   & $93.000$ & $140.000$ \\
                  & $9$  & $181.000$ &   &   &   &   &   &   &   &   &   & $120.000$ & $178.000$ \\
                  & $10$ & $222.000$ &   &   &   &   &   &   &   &   &   & $150.000$ & $222.000$ \\
                  & $11$ &   &   &   &   &   &   &   &   &   &   & $182.000$ &   \\
                  & $12$ &   &   &   &   &   &   &   &   &   &   & $222.000$ &   \\
            \hline
            \end{tabular}
            }
    \end{center}
\end{table*}

\section{Metal abundance in projected spectral analyses and tied radial bins in the deprojected spectral analysis}
\label{app:spectral_details}

\begin{figure*}[ht!]
    \begin{center}
        \includegraphics[width=16cm]{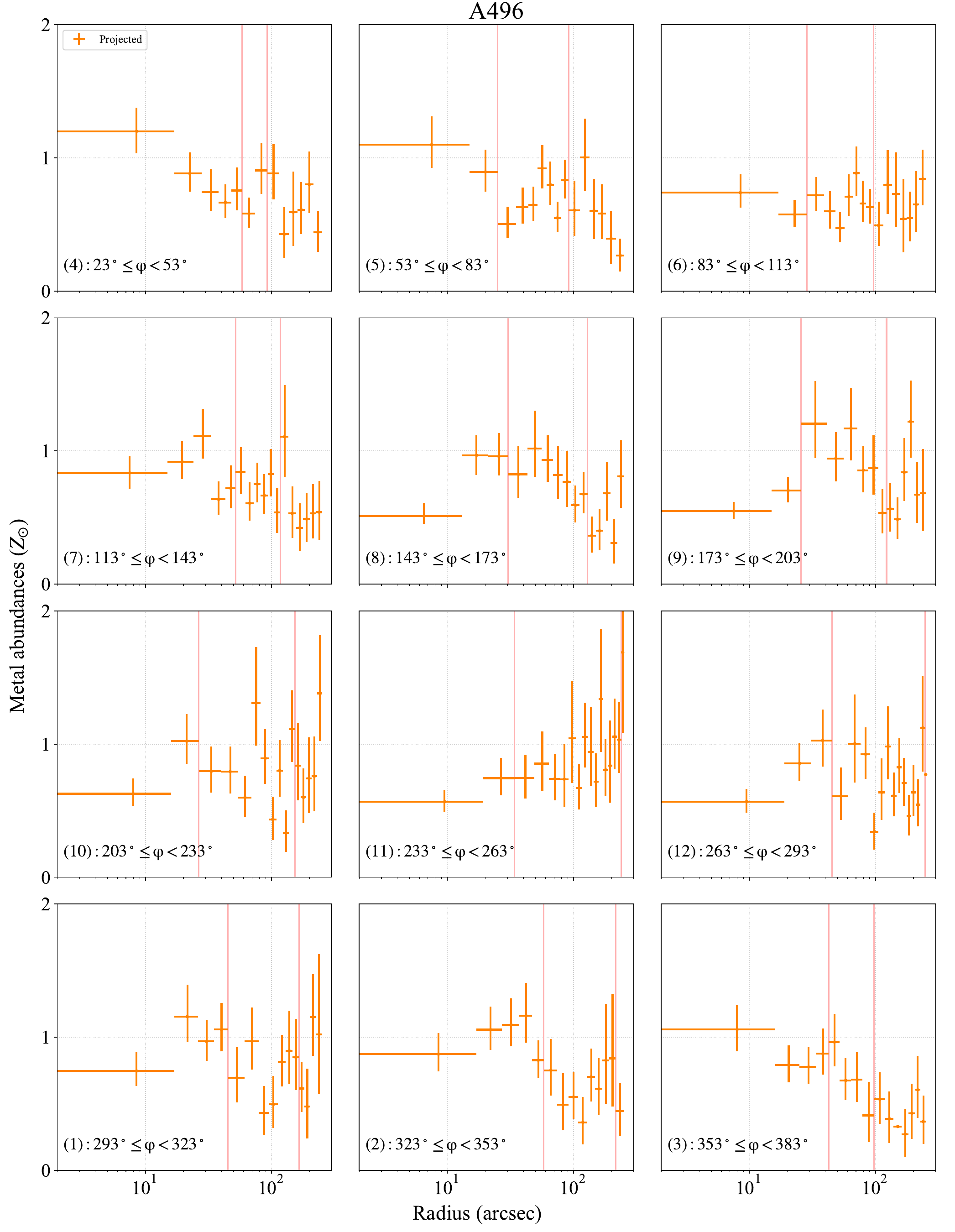}
    \end{center}
\caption{Radial ICM metal abundance profiles of each sector in A496. The orange data points show the projected metal abundances.
The horizontal bars represent the ranges of the radial bins and the vertical bars denote the $1\,\sigma$ confidence range of the metal abundance. 
The pink vertical lines show the positions of $r_{\rm{in}}$ and $r_{\rm{out}}$ described in Section~\ref{sec:AnalysesSx}.}
\label{fig:A496_M}
\end{figure*}

\begin{figure*}[ht!]
    \begin{center}
        \includegraphics[width=16cm]{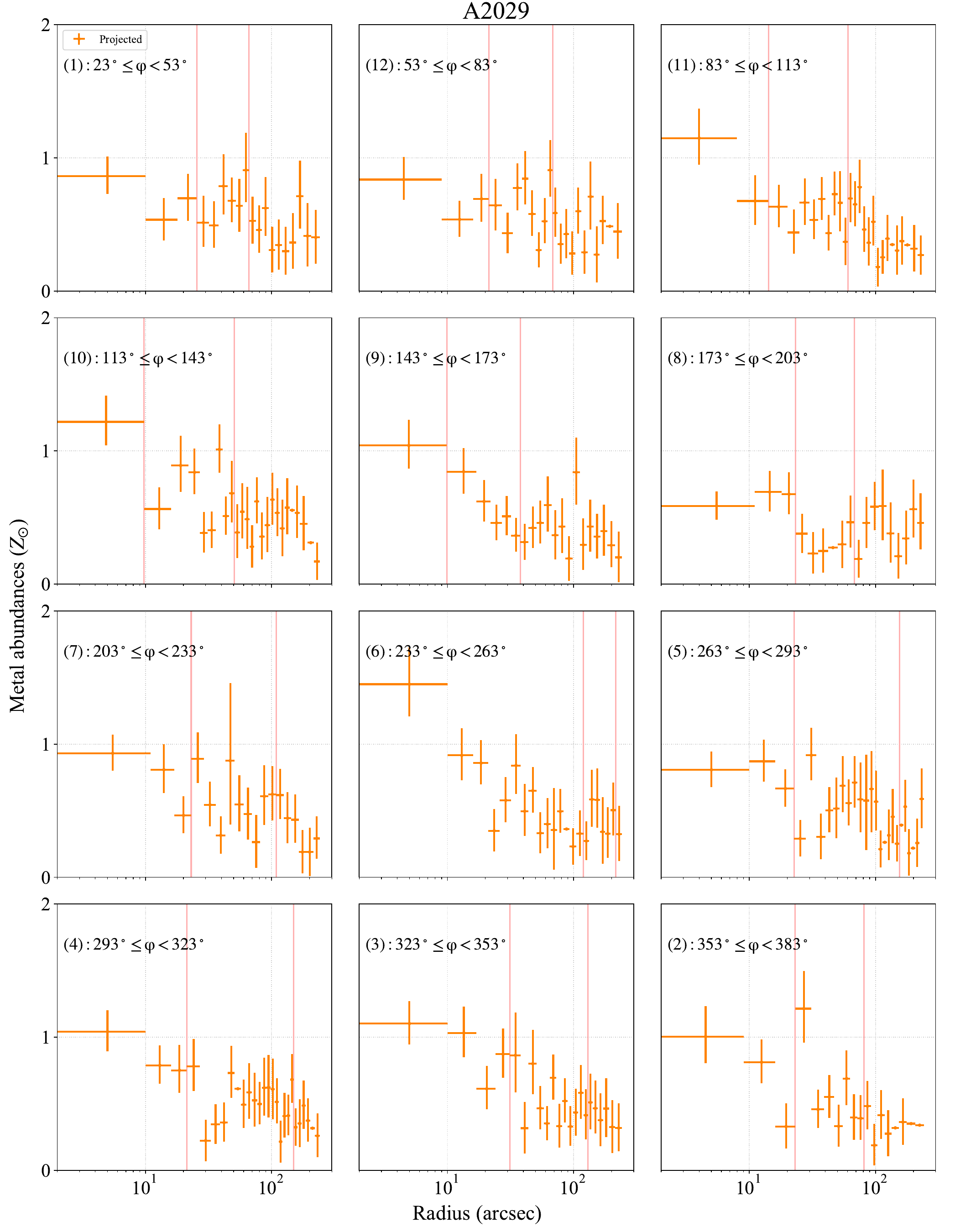}
    \end{center}
\caption{Same as Figure~\ref{fig:A496_M}, but for A2029.}
\label{fig:A2029_M}
\end{figure*}

The metal abundances are treated as free parameters for most radial bins in A496 and A2029.
A few exceptions are made, where the abundances of spectra from the radial bins listed in Table~\ref{tab:spectral_details} are set to the mean of the adjacent radial bins. 
In addition, for the outermost radial bin of the sector $7$ in A496 and sector $2$ in A2029, the metal abundances are fixed to the mean values of the outermost radial bins from other sectors within the same cluster.
On the other hand, the metal abundances of spectra extracted from A1644 are set to $0.3$, because we cannot put any statistically meaningful constraints for this cluster with a free parameter of the abundance. The radial ICM metal abundance profiles of A496 and A2029 are shown in Figures~\ref{fig:A496_M} and ~\ref{fig:A2029_M}, respectively.
Because the photon counts per bin in A1644 are insufficient to statistically constrain the deprojected temperature, we tied several adjacent radial bins together to analyze this cluster. 
The combined bin groups are listed in Table~\ref{tab:tied_bins}.

\begin{table}[h]
    \begin{center}
        \caption{Summary of the radial bins with fixed metal abundance.}
        \label{tab:spectral_details}
            \begin{tabular}{ccc}
            \hline
            \hline
            Cluster & Sector & Radial bins \tablenotemark{a} \\
            \hline
            A496  & $3$ & $11^{\rm{th}}$ \\
            \hline
            A2029 & $2$  & $15^{\rm{th}}$, $17^{\rm{th}}$ \\
                  & $4$  & $9^{\rm{th}}$, $26^{\rm{th}}$ \\
                  & $5$  & $17^{\rm{th}}$, $21^{\rm{st}}$, $24^{\rm{th}}$ \\
                  & $6$  & $13^{\rm{th}}$ \\
                  & $8$  & $7^{\rm{th}}$ \\
                  & $10$ & $21^{\rm{st}}$, $24^{\rm{th}}$ \\
                  & $11$ & $21^{\rm{st}}$, $24^{\rm{th}}$ \\
                  & $12$ & $21^{\rm{st}}$ \\
            \hline
            \end{tabular}
    \end{center}
    \tablenotemark{a}{The radial bins are counted from the cluster center.} \\
\end{table}

\begin{table}[h]
    \begin{center}
        \caption{Summary of the tied radial bins in A1644.}
        \label{tab:tied_bins}
            \begin{tabular}{ccc}
            \hline
            \hline
            Cluster & Sector & Tied radial bins \tablenotemark{a} \\
            \hline
            A1644 & $7$  & $\left(2^{\rm{nd}}, 3^{\rm{rd}}\right)$ \\
                  & $8$  & $\left(1^{\rm{st}}, 2^{\rm{nd}}\right)$ \\
                  & $9$  & $\left(3^{\rm{rd}}, 4^{\rm{th}}, 5^{\rm{th}}, 6^{\rm{th}}\right)$ \\
                  & $10$ & $\left(1^{\rm{st}}, 2^{\rm{nd}}\right)$, $\left(3^{\rm{rd}}, 4^{\rm{th}}, 5^{\rm{th}}\right)$ \\
                  & $11$ & $\left(1^{\rm{st}}, 2^{\rm{nd}}\right)$, $\left(3^{\rm{rd}}, 4^{\rm{th}}\right)$ \\
            \hline
            \end{tabular}
    \end{center}
    \tablenotemark{a}{The radial bins listed in parentheses are tied together.} \\
\end{table}

\section{Systematic uncertainty of the sector definition}
\label{app:SysUncertain}
In our analyses, we set the X-ray brightness peak as the geometric center of the sector. However, defining the sectors along the curvature of the spiral excess might improve the measurements of density and temperature contrasts. To investigate systematic uncertainties resulting from our sector definition, we additionally analyze several X-ray surface brightness profiles extracted from the sectors defined along the curvature of the spiral excess. Finally, we estimate the differences in thermal pressure or density contrasts between our fiducial results 
and those defined along the curvature of the spiral excess, and compare them with previously published studies.

\subsection*{A496}
\label{sec:SysUncertainA496}
\begin{figure*}
    \begin{center}
        \includegraphics[height=6cm]{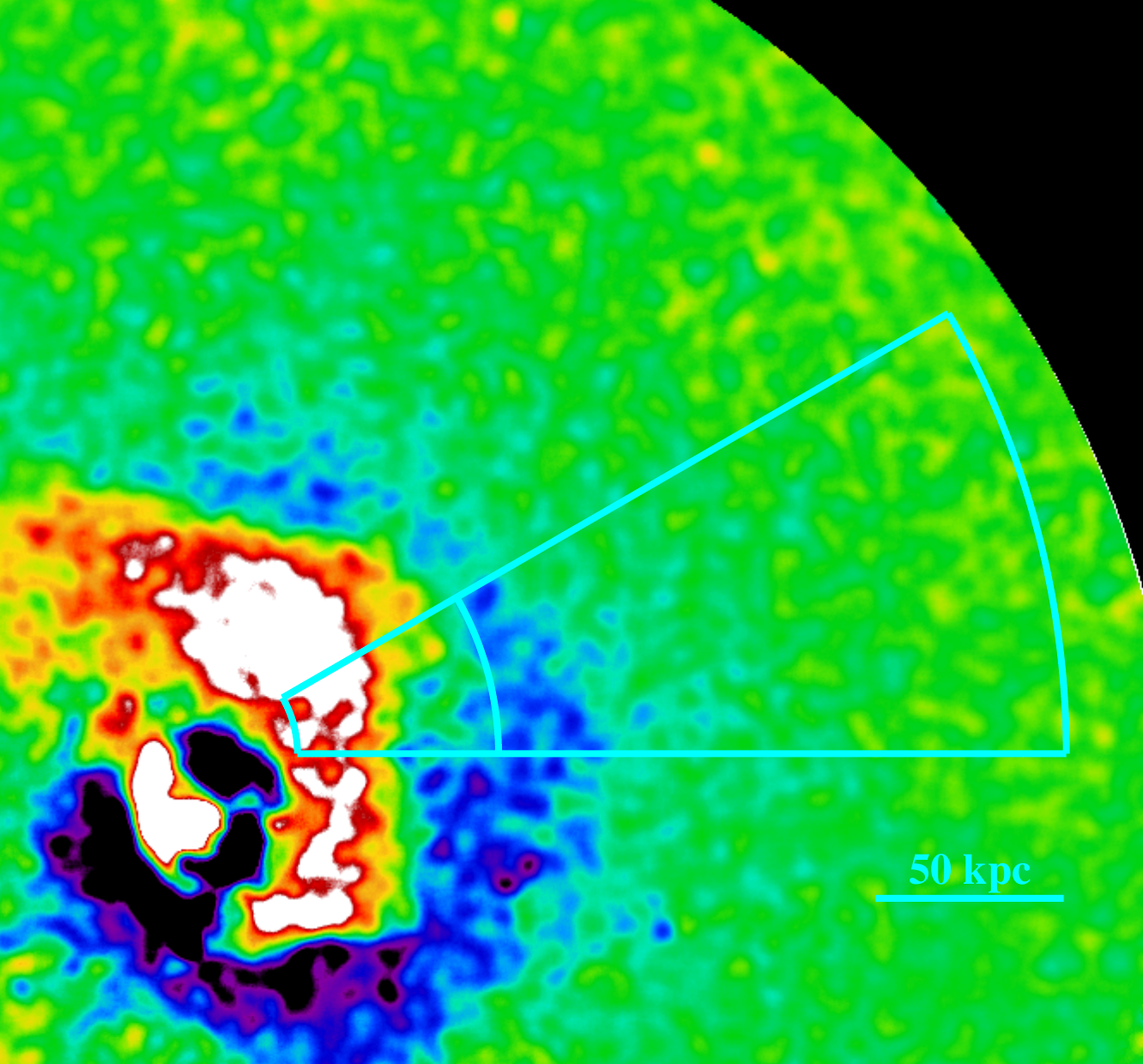}
        \includegraphics[height=6cm]{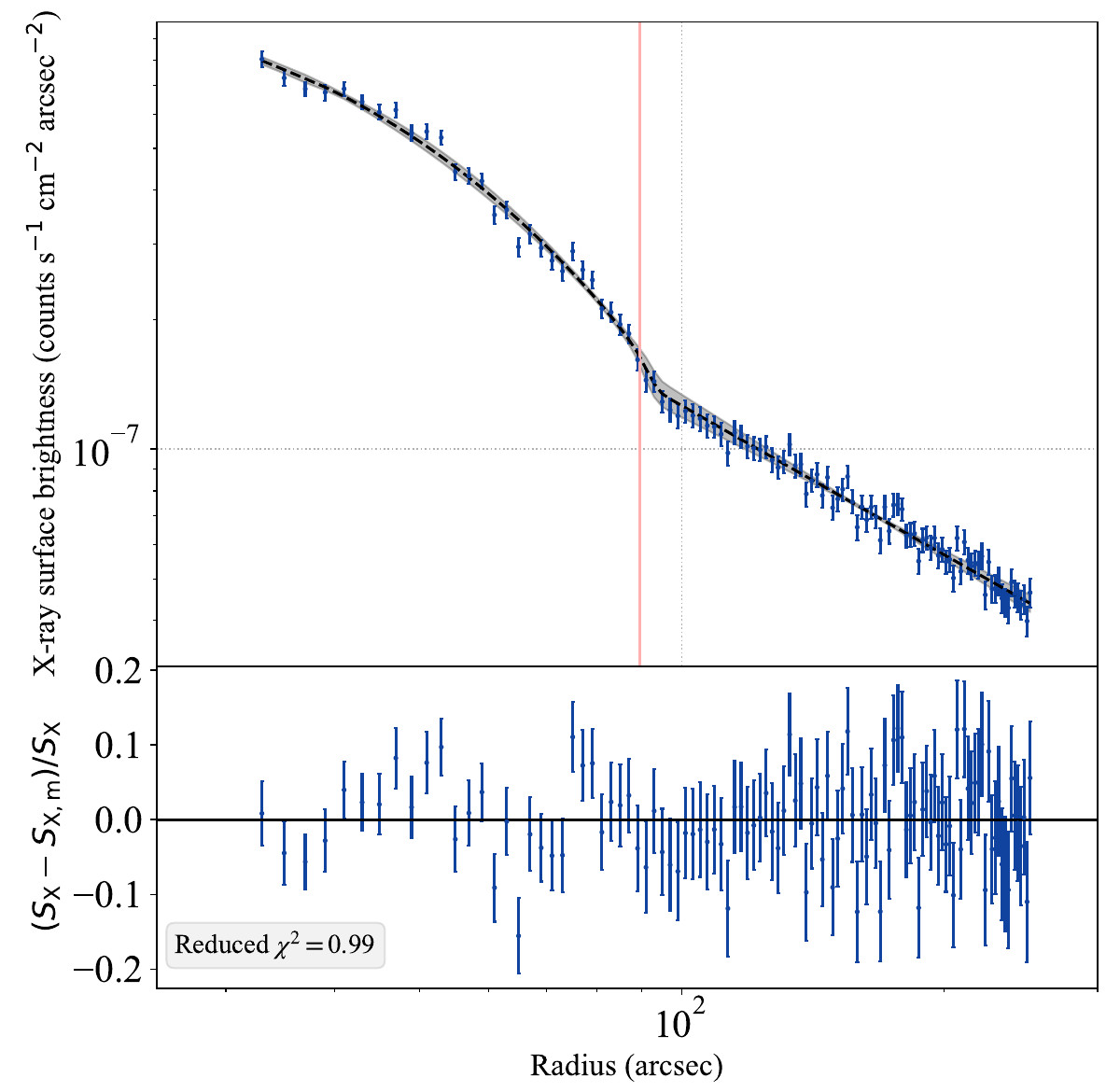}
    \end{center}
\caption{The residual X-ray surface brightness map of the northwestern region of A496 (left) and the X-ray surface brightness profile (right) extracted from the region enclosed by the cyan sector shown in the left panel. (Left): the cyan sector is aligned with the curvature of the spiral excess, as described in Appendix~\ref{sec:SysUncertainA496}, and the cyan arc shows the position of the detected edge determined from the X-ray surface brightness fitting. The horizontal bar denotes the spatial scale. (Right): same format as in Figure~\ref{fig:AnalysesSxA496}.}
\label{fig:SysUncertainA496_1}
\end{figure*}

\begin{figure*}
    \begin{center}
        \includegraphics[height=7cm]{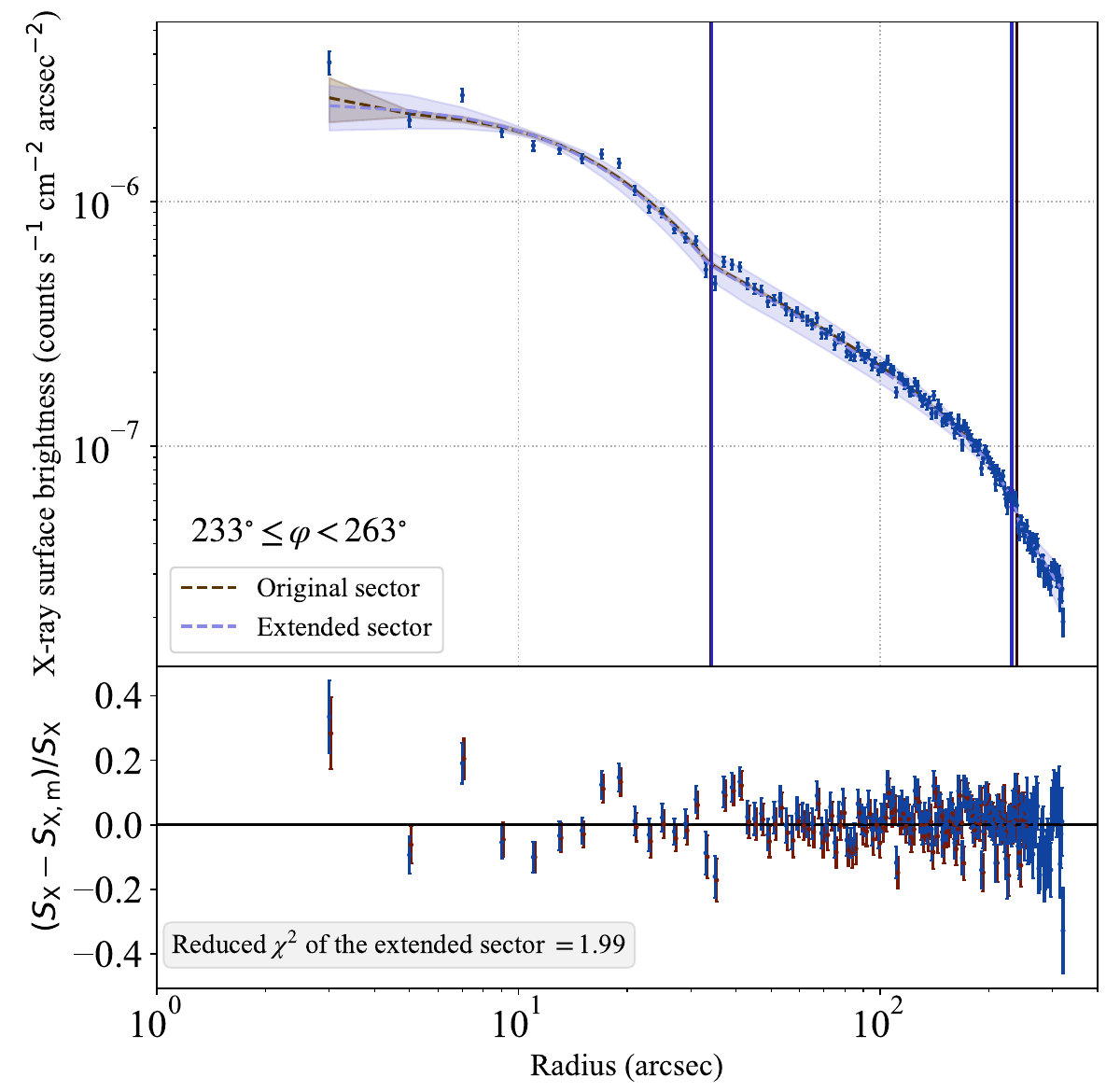}
        \includegraphics[height=7cm]{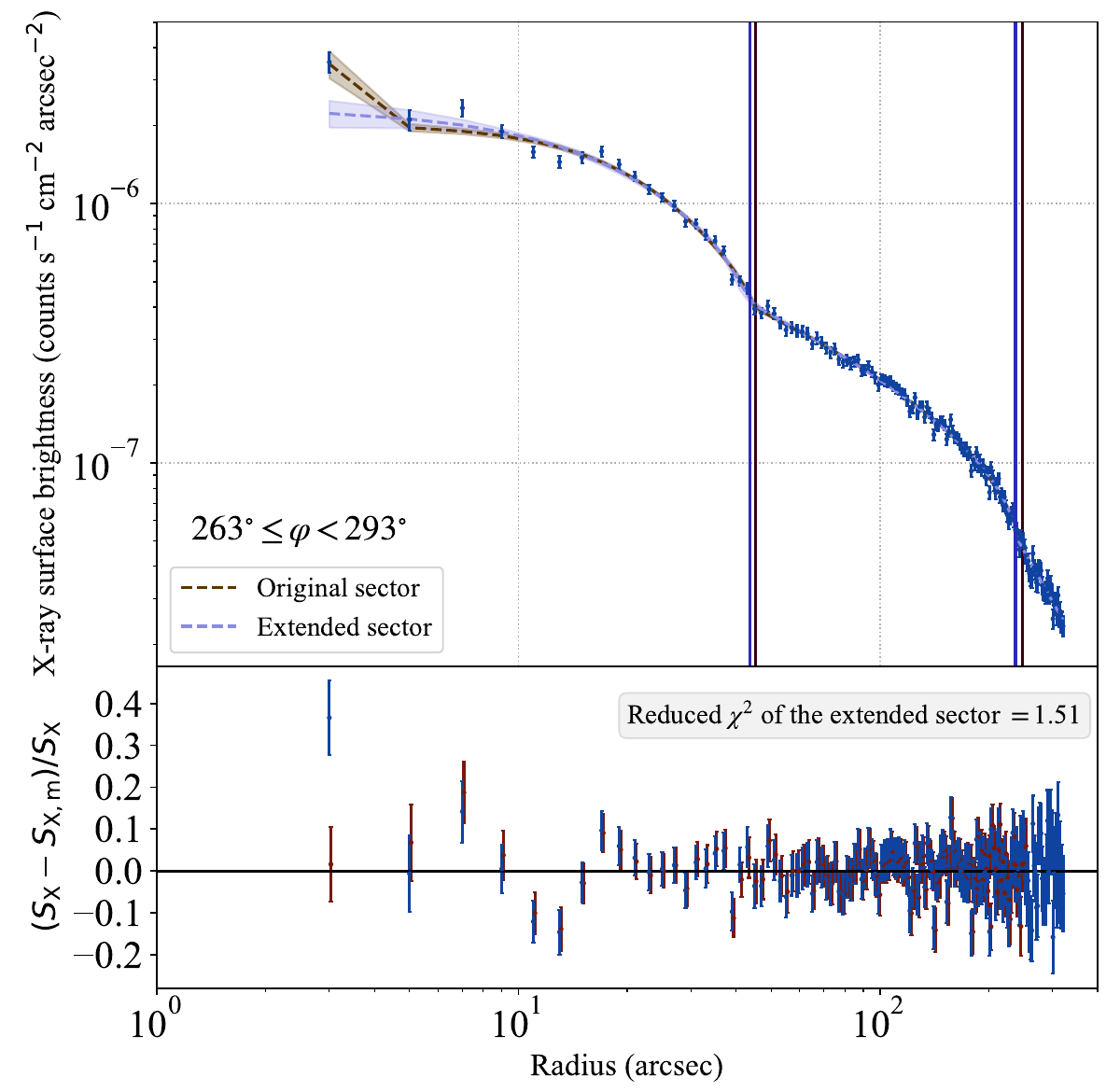}
    \end{center}
\caption{The X-ray surface brightness profiles extracted from sectors of $233^{\circ} \leq \phi < 263^{\circ}$ (left) and $263^{\circ} \leq \phi < 293^{\circ}$ (right) in A496, as described in Appendix~\ref{sec:SysUncertainA496}. The outer boundaries of these sectors are extended to $322''$ to allow a more accurate determination of the edge $r_{\rm{out}}$. Same format and symbols as used in Figure~\ref{fig:AnalysesSxA496}. The brown dashed line and vertical lines indicate the best-fit profile and the edge positions obtained from the original sector, which covers the same azimuthal range as the extended sector but adopts a smaller outer boundary ($252''$). The brown and blue data points in the lower panel show the corresponding residuals from the original and extended sectors, respectively.}
\label{fig:SysUncertainA496_2}
\end{figure*}

According to the {\em XMM-Newton} observation conducted by \cite{Ghizzardi2014}, a sharp cold front with a boxy morphology is located $100'' \left(65\,\mathrm{kpc}\right)$ northwest of the cluster center. We therefore define a sector centering at (4:33:38.4614, $-$13:15:18.242), scanning from $0^{\circ}$ to $30^{\circ}$, with a radial range of $32'' - 252''$. The curvature of this sector aligns with the southern part of the boxy morphology in the residual X-ray surface brightness map of A496, as shown in the left panel of Figure~\ref{fig:SysUncertainA496_1}. 
For this sector, we marginally detect a surface brightness edge at $89.5 \pm 12.9''$ from the center of the sector, with a density contrast of $\jn = 1.30 \pm 0.25$ and a temperature contrast of $j_{T} = 0.64^{+0.13}_{-0.10}$, which yield a thermal pressure contrast of $j_{p} = 0.83^{+0.23}_{-0.21}$. Given that the pressure contrast of the sector $4$, defined in Section~\ref{sec:AnalysesSx} and largely overlapping with the new sector, is $j_{p} = 1.31^{+0.42}_{-0.29}$, the difference in pressure contrast between the two sectors is $1.1\sigma$. This indicates that the two measurements are statistically consistent, supporting the reliability of the fiducial sector definition.

Another sloshing cold front with the observation from {\em XMM-Newton} is situated at $\approx 252'' \left(165\,\mathrm{kpc}\right)$ from the cluster center, within the azimuthal angle range of $240^{\circ}$ to $285^{\circ}$ \citep{Ghizzardi2014}. Since this cold front lies close to the outer boundaries of the sector $11$ and sector $12$,
we analyze X-ray surface brightness profiles extracted from sectors defined by extending the outer boundaries of the sector $11$ and sector $12$ to a radius of $322''$ from the X-ray peak.
The two X-ray surface brightness profiles are presented in Figure~\ref{fig:SysUncertainA496_2}. Using the three-component model described in Section~\ref{sec:AnalysesSx}, we detect an X-ray surface brightness edge ($\jn = 1.15 \pm 0.23$) at $231.5 \pm 29.4''$ for the sector scanning from $233^{\circ}$ to $263^{\circ}$. In the sector covering $263^{\circ}$ to $293^{\circ}$, the density contrast is slightly more prominent (\jn\ $= 1.22 \pm 0.07$) and the surface brightness edge positions at $237.1 \pm 6.5''$. 
To directly compare with \cite{Ghizzardi2014}, in which they conducted only projected spectral analysis, we also carry out the projected spectral analysis and infer that the thermal pressure contrasts \jp\ across the outer X-ray surface brightness edges $r_{\rm{out}}$ are $0.87\pm0.19$ in the sector covering $233^{\circ} \leq \phi < 263^{\circ}$ and $1.17^{+0.11}_{-0.10}$ in the sector covering $263^{\circ} \leq \phi < 293^{\circ}$. The mean thermal pressure contrast of these sectors differs from the pressure contrast measured in \cite{Ghizzardi2014} by approximately $1.5\sigma$, showing no statistically significant discrepancy.

\subsection*{A2029}
\label{sec:SysUncertainA2029}

\begin{figure*}[ht!]
    \begin{center}
        \includegraphics[height=6cm]{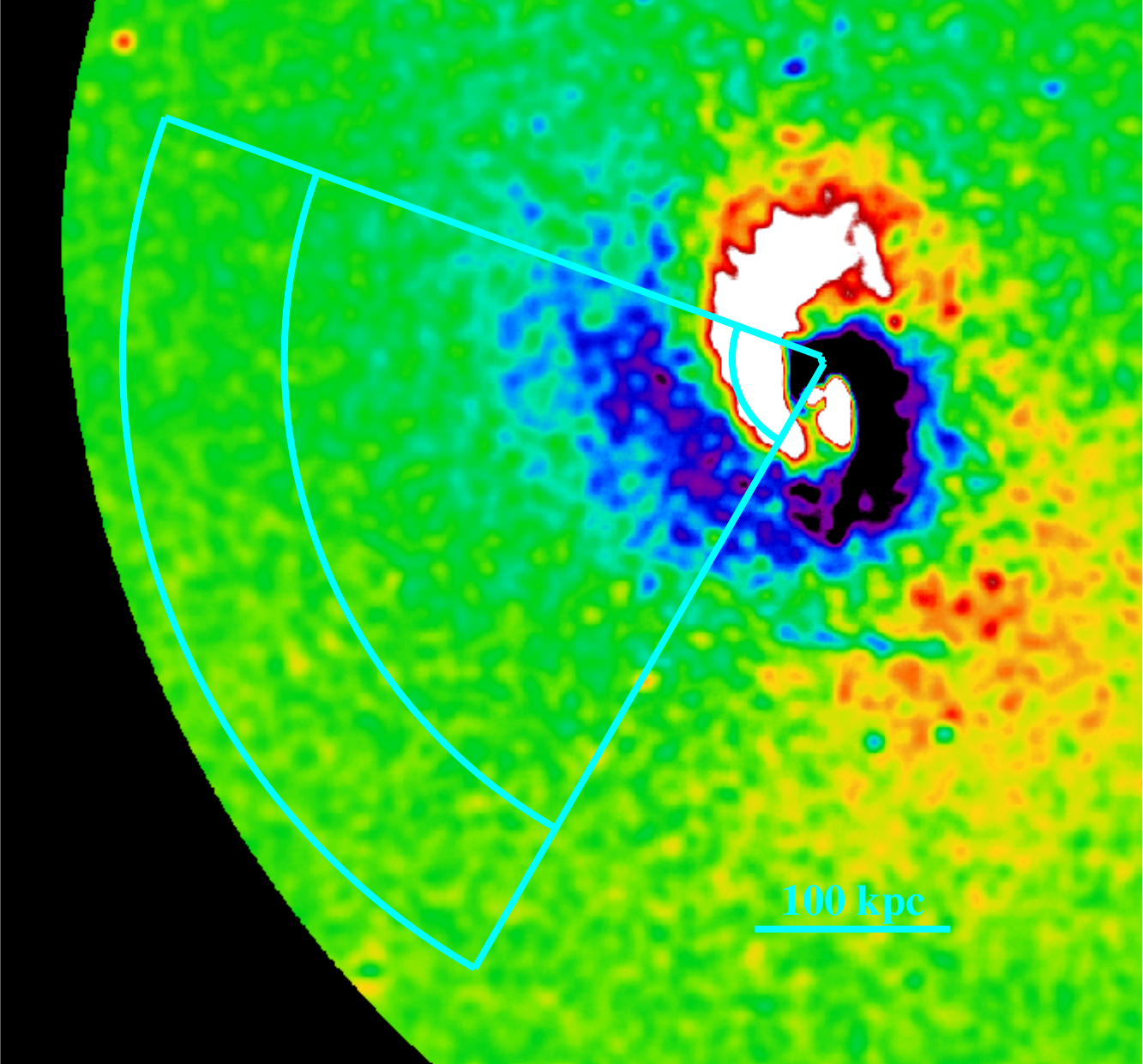}
        \includegraphics[height=6cm]{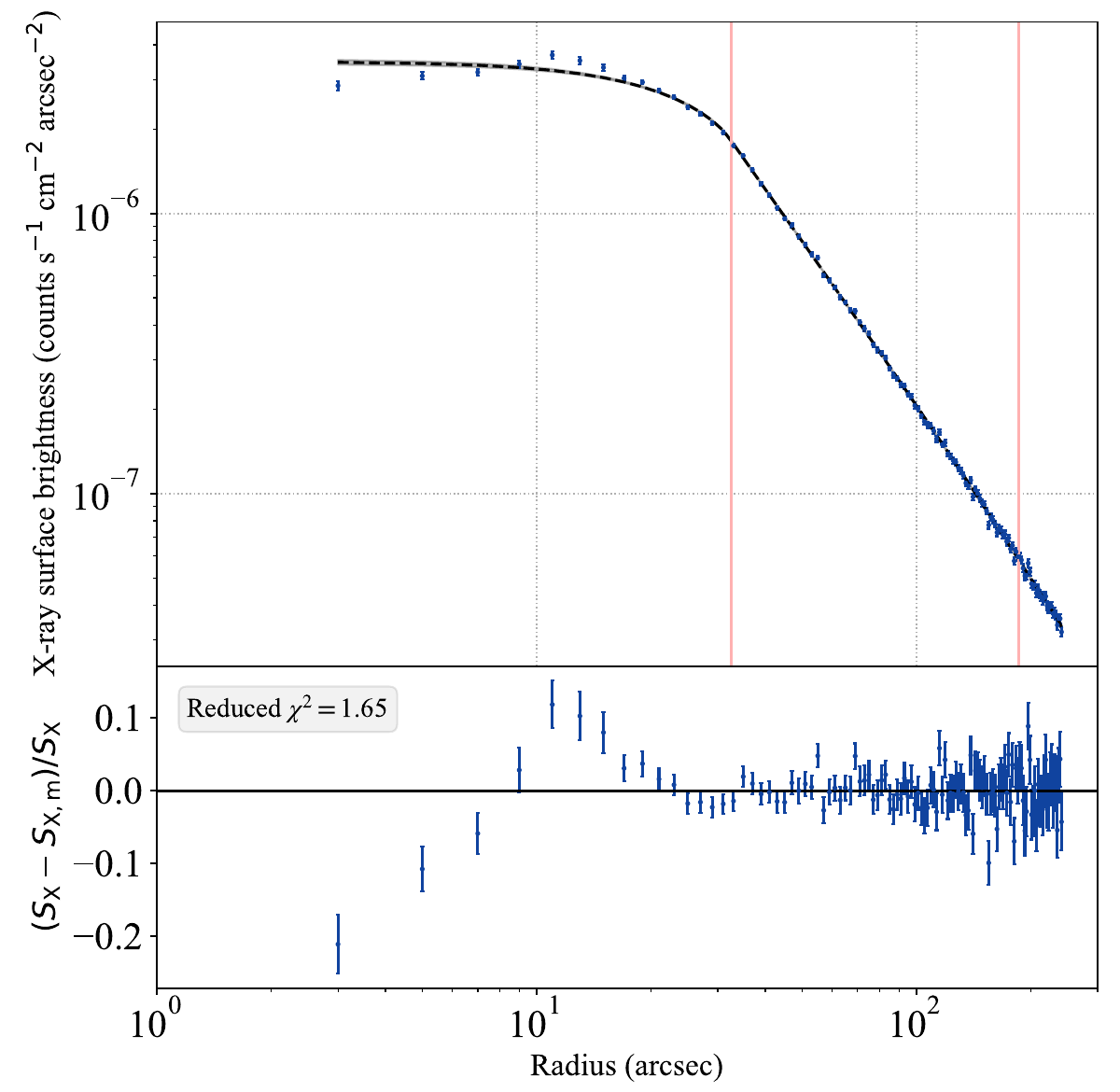}
    \end{center}
\caption{The residual X-ray surface brightness map of the eastern region of A2029 (left) and the X-ray surface brightness profile (right) extracted from the region enclosed by the cyan sector shown in the left panel. (Left): the cyan sector is aligned with the curvature of the spiral excess, as described in Appendix~\ref{sec:SysUncertainA2029}, and the cyan arc shows the position of the detected edge determined from the X-ray surface brightness fitting. The horizontal bar denotes the spatial scale. (Right): same format as in Figure~\ref{fig:AnalysesSxA496}.}
\label{fig:SysUncertainA2029}
\end{figure*}

In order to investigate the impact of the curvature-based sector definition in A2029, we analyze a sector region aligned with the most clearly visible spiral excess in the east of the residual X-ray surface brightness map. This sector centers at (15:10:55.8166, 5:44:53.451), scans from $160^{\circ}$ to $240^{\circ}$, and extends from $2''$ to $242''$. Through modeling the X-ray surface brightness profile (Figure~\ref{fig:SysUncertainA2029}) of this sector with the three-component density model expressed in Equation~(\ref{eq:AnalysesSx3ComModel}), we measure a subtle density contrast $j_{n} = 1.023 \pm 0.01654$ at $r_{\rm{in}} = 32.5 \pm 0.8''$. 
The density contrast of this curvature-based sector is marginally greater than $1$, in contrast to the density contrasts of the edges detected in the default analysis: as seen in Section~\ref{sec:AnalysesSx}, $r_{\rm{in}}$ of the sector $7$ and $8$, and $r_{\rm{out}}$ of the sector $9$, which roughly cover the same region, have density contrasts below $1$.
The density contrast measured in the curvature-based sector differs from the mean value of those sectors by $2.2\sigma$, indicating a moderate discrepancy.
Additionally, we perform the deprojected spectral analysis of the curvature-based sector to determine whether a cold front exists. 
Following the method described in Section~\ref{sec:AnalysesSpectra}, but increasing the mean photon counts in each radial bin to $4534$ per bin, we measure the temperature contrast to be $0.4379^{+0.0770}_{-0.0443}$, which is in disagreement with our default analysis.
Thus, we detect a cold front in this curvature-based sector according to our cold front criteria.
This finding indicates that, when the density contrast is subtle, carefully defining the sector could play a role in identifying cold fronts.
Since the sectors defined in Section~\ref{sec:AnalysesSx} have smaller opening angles, they are better suited to capture the azimuthal variation in cold front properties.

\subsection*{A1644}

\label{sec:SysUncertainA1644}
\begin{figure*}[ht!]
    \begin{center}
        \includegraphics[height=6cm]{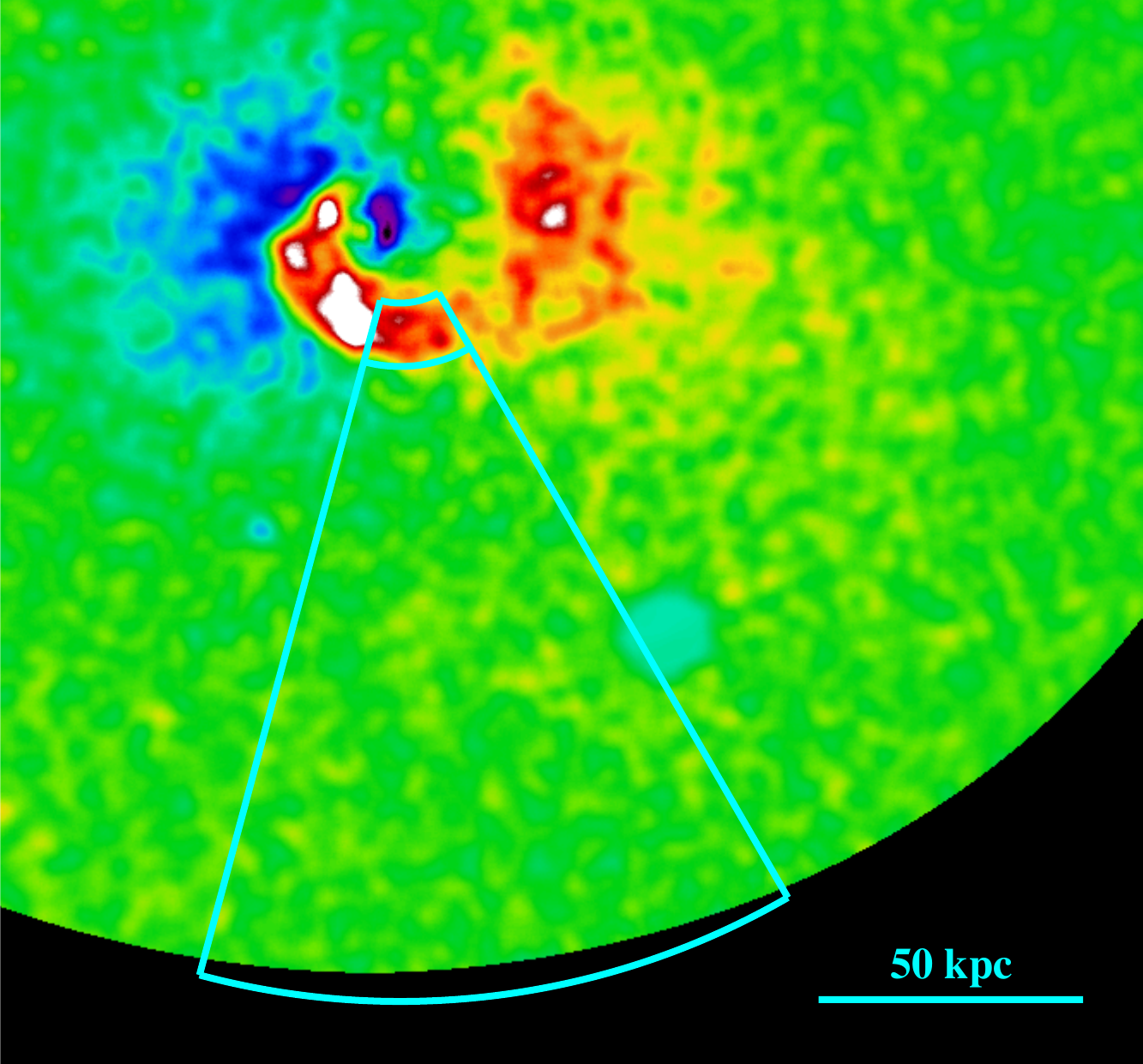}
        \includegraphics[height=6cm]{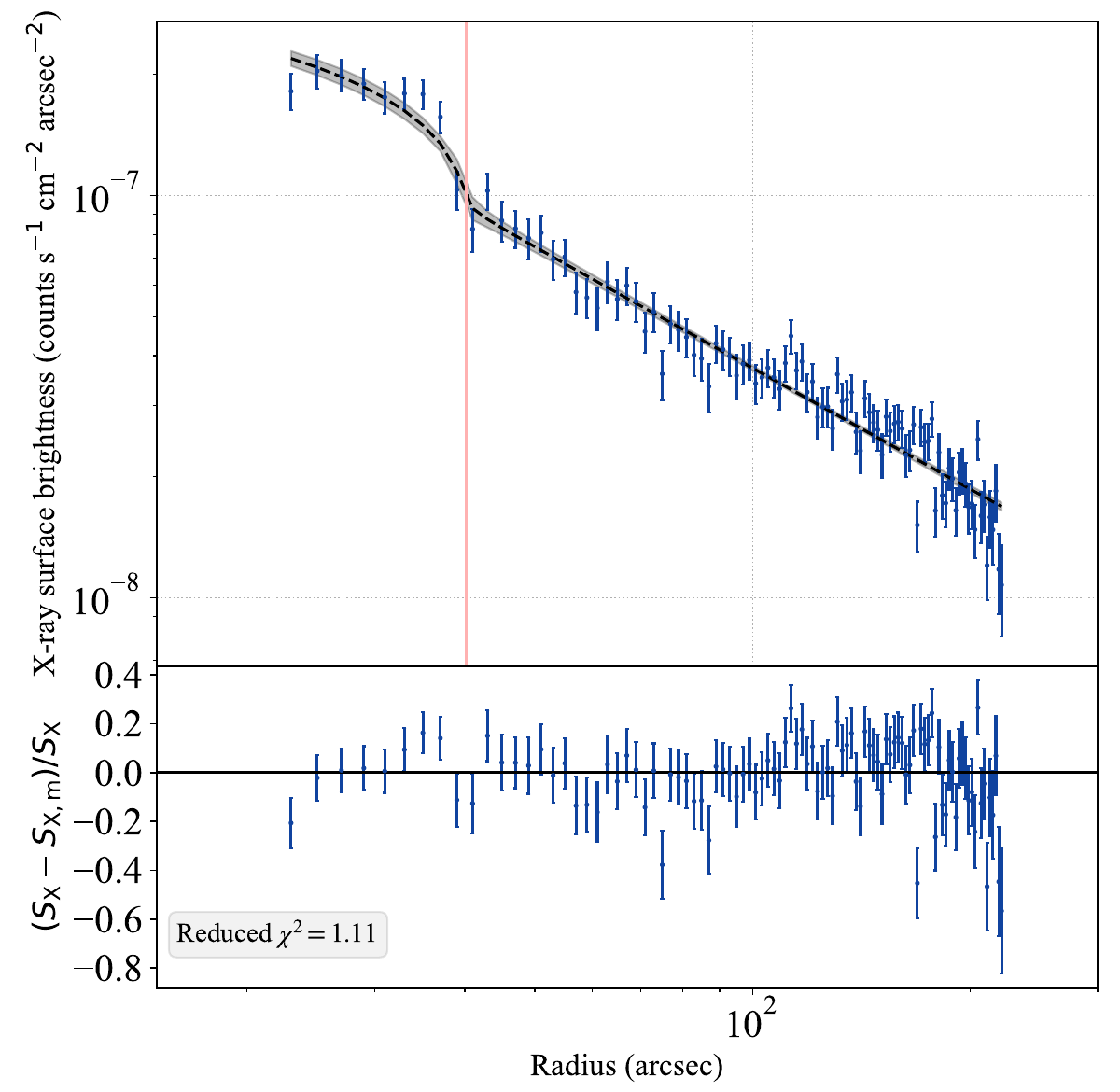}
    \end{center}
\caption{The residual X-ray surface brightness map of the southern region of A1644 (left) and the X-ray surface brightness profile (right) extracted from the region enclosed by the cyan sector shown in the left panel. (Left): the cyan sector is aligned with the curvature of the spiral excess, as described in Appendix~\ref{sec:SysUncertainA1644}, and the cyan arc shows the position of the detected edge determined from the X-ray surface brightness fitting. The horizontal bar denotes the spatial scale. (Right): same format as in Figure~\ref{fig:AnalysesSxA496}.}
\label{fig:SysUncertainA1644}
\end{figure*}

We analyze a sector aligned with the spiral excess in the residual X-ray surface brightness map for A1644. The sector is centered at (12:57:10.7043, $-$17:24:33.480) and spans an azimuthal angle from $255^{\circ}$ to $300^{\circ}$ with an open angle of $45^{\circ}$. 
We extract the X-ray surface brightness profile within this sector over the radial range $22'' - 222''$ (Figure~\ref{fig:SysUncertainA1644}), and fit it using the two-component density model described in Section~\ref{sec:AnalysesSx}.
The reason we switch from the three-component density model to the two-component density model for fitting this sector is that $r_{\rm{in}}$ of the sector $7$ in the same direction described in Section~\ref{sec:AnalysesSx} appears to align with the inner edge of the spiral excess. 
However, the curvature-based sector used here starts from the middle of the spiral excess and does not include the inner-edge region. 
Therefore, we expect that only one edge remains within this curvature-based sector.
The density contrasts \jn\ of the sector $7$ 
defined in Section~\ref{sec:AnalysesSx} and the curvature-based sector, $2.110 \pm 0.1554$ and $1.563 \pm 0.1459$ respectively, differ by $\approx 2.57\sigma$. 
Meanwhile, the density contrast in the curvature-based sector is lower than in the sector defined in Section~\ref{sec:AnalysesSx}. 
Such a reduction may result from the broader opening angle of the curvature-based sector, which has the potential to either smear out or amplify the density contrasts.

\section{Estimation of the velocity gradient using the stagnation-point model}
\label{app:StagPoinModel}

\begin{figure}
\centering
\resizebox{0.4\textwidth}{!}{\includegraphics[scale=1]{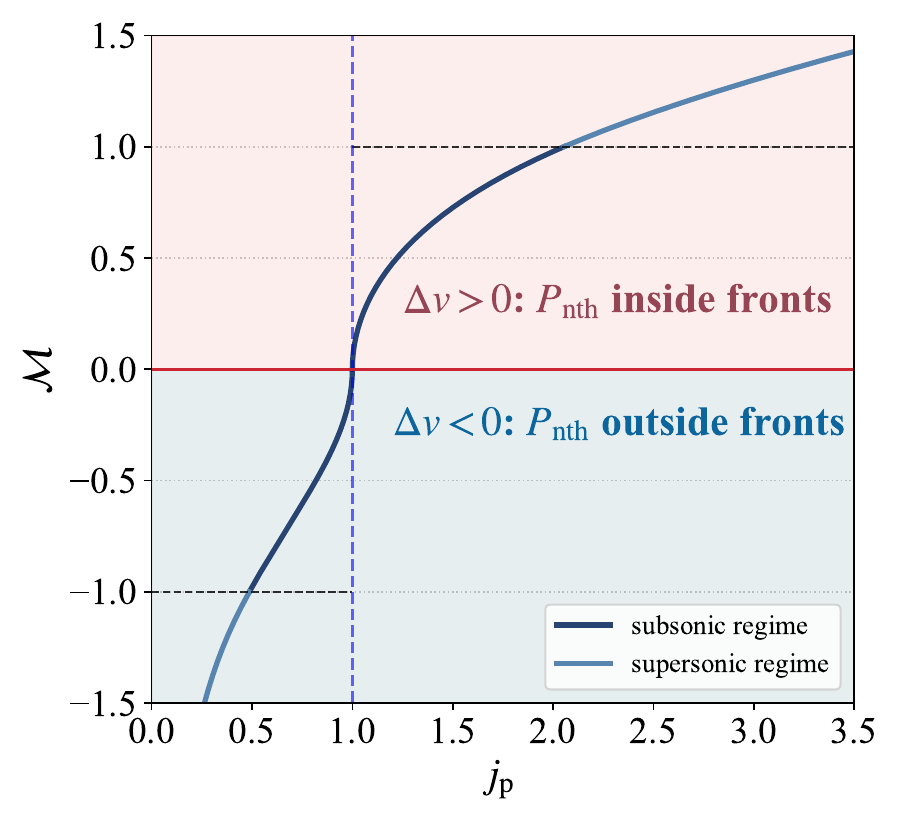}}
\caption{The relation between the thermal pressure contrast \jp\ and the Mach number $\mathcal{M}$ as defined by Equation~(\ref{eq:DiscussionV_p0_p1_M}). The horizontal dashed lines at $\mathcal{M} = 1.0$ and $\mathcal{M} = -1.0$ separate the plot into two regions: subsonic region, corresponding to $\left\vert\mathcal{M}\right\vert \leq 1$, and supersonic region, corresponding to $\left\vert\mathcal{M}\right\vert > 1$.}
\label{fig:DiscussionV_p0_p1_M}
\end{figure}

\begin{figure*}[ht!]
    \centering
        \resizebox{0.9\textwidth}{!}{\includegraphics[scale=1]{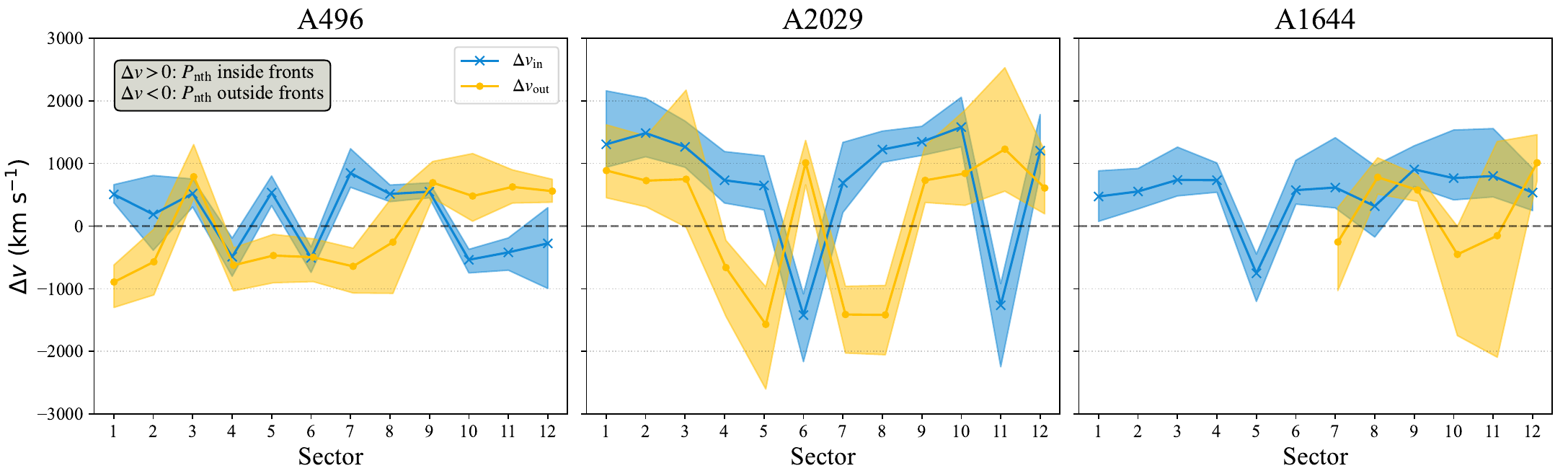}}
\caption{Same as Figure~\ref{fig:AnalysesSpectraT0_T1}, but for velocity gradients $\Delta v$.}
\label{fig:Azi_Del_v}
\end{figure*}

The stagnation-point model developed in \cite{Vikhlinin2001} relates the thermal pressure contrast at a stagnation point to the relative velocity in terms of the Mach number.
Therefore, this model best describes the fluid dynamics of stripping cold fronts, where the gas moves like a blunt body through an ambient gas and a stagnation point forms near the leading edge. Because the gas flow in sloshing cold fronts is predominantly tangential, a stagnation point is not expected to form, making this model less physically optimal for describing the sloshing cold fronts.

Nevertheless, we retain the stagnation-point model as a toy baseline to gauge the gas kinematics as a useful approximation. 
We express the amplitude of the bulk motion required to maintain the total pressure equilibrium in terms of an effective velocity gradient, denoted as $\Delta v$.
For sloshing cold fronts, the Mach number inferred from the stagnation-point model should be interpreted as an order-of-magnitude estimate of the effective Mach number, or equivalently as the amplitude of the gas motion required to maintain the total pressure equilibrium if the non-thermal pressure support entirely comes from the bulk motion \citep[see also Section 3.2 in][]{Ghizzardi2014}.

In the stagnation-point model \citep{Vikhlinin2001}, the relation associating the thermal pressure contrast with the ICM Mach number $\mathcal{M}$ is expressed as 
\begin{equation}
\label{eq:DiscussionV_p0_p1_M}
    j_{p} =
        \begin{cases}
            \left(1 + \frac{\gamma - 1}{2} \mathcal{M}^{2}\right)^{\frac{\gamma}{\gamma - 1}} & \mathrm{if}~\mathcal{M} \leq 1 \\
            \left(\frac{\gamma + 1}{2}\right)^{\frac{\gamma + 1}{\gamma - 1}} \mathcal{M}^{2} \left(\gamma - \frac{\gamma -1}{2 \mathcal{M}^{2}}\right)^{\frac{-1}{\gamma - 1}} & \mathrm{if}~\mathcal{M} > 1 \\
        \end{cases}
        \, ,
\end{equation}
where the adiabatic index of the monatomic gas $\gamma$ is set to $5/3$, and $\mathcal{M}$ is the effective Mach number of the gas flow across the edges.
By construction, we have $\Delta v \equiv v_{\rm{i}} - v_{\rm{o}}$, where $v_{\rm{i}}$ and $v_{\rm{o}}$ are the 
bulk velocities of the gas inside and outside the edge, respectively.
In this notation, the effective velocity gradient $\Delta v$ across a surface brightness edge is related to the effective Mach number $\mathcal{M}$ as
\begin{equation}
\label{eq:DiscussionV_DeltaV}
\Delta v = v_{\rm{i}} - v_{\rm{o}} = 
    \begin{cases}
    \mathcal{M} \times c_{\rm{s, \rm{i}}} & j_{p} < 1    \\
    -\mathcal{M} \times c_{\rm{s, \rm{o}}} & j_{p} \geq 1 \\
    \end{cases}
    \, ,
\end{equation}
in which $c_{\rm{s}}$ is the gas sound speed on the lower thermal pressure side. We compute the sound speed as $c_{\rm{s}} = \left(\gamma T/m_{\rm{p}} \mu\right)^{1/2}$, where $m_{\rm{p}}$ is the proton mass, and the mean molecular weight of the ICM $\mu$ is set to $0.6$. 
We show the relation between the thermal pressure contrast and the Mach number in Figure~\ref{fig:DiscussionV_p0_p1_M}. 

We stress that the sign of the effective velocity gradient $\Delta v$ does not necessarily refer to the direction of the gas flow across the cold front edges.
We cannot determine the direction of the gas flow from the observed thermal pressure contrasts, as $\Delta v$ is inferred as the velocity gradient associated with the effective Mach number.
Instead, the sign of $\Delta v$ indicates which side of the cold front requires additional non-thermal pressure support.
Specifically, the positive sign of $\Delta v$ indicates $v_{\rm{i}} > v_{\rm{o}}$ and, hence, requires a non-thermal pressure support radially inward of the front ($j_{p} < 1$).
In contrast, a negative $\Delta v$ implies an additional non-thermal pressure support radially outward of the front.

Azimuthally resolved studies of stripping cold fronts indicate that the highest thermal pressure contrasts exist at the leading edge of the cold front \citep{Ichinohe2017, Ueda2024}, which is due to the highest velocity gradient generated by the bulk motion of the infalling merging subcluster.
On the contrary, as shown in Figure~\ref{fig:Azi_Del_v}, the azimuthal variations of velocity gradients observed in our sample vary from cluster to cluster, strongly suggesting a different mechanism from the ram pressure as a source of the non-thermal pressure support.

It is noteworthy that simulations from \cite{ZuHone2013} found that the turbulent motion of the sloshing ICM is amplified, resulting in significantly higher turbulent velocities compared to that of the ambient ICM. 
This suggests a noticeable contribution from the turbulent motion of the ICM to the non-thermal pressure support in sloshing cold fronts.

We stress that the velocity gradient estimated in this work relies on an estimate from the thermal pressure contrast, under the assumption that all non-thermal pressure support arises from the bulk motion of ICM. Since the gas motion may not be the only source of the non-thermal pressure, our result can be considered as an upper limit on the effective velocity of the bulk motion.
To this end, the ongoing X-ray observatory \textit{XRISM} \citep{Tashiro2024} with high-resolution spectroscopy will enable us to directly measure the ICM motions to further explore the source of the non-thermal pressure support in sloshing cold fronts.
\bibliography{X-ray}{}
\bibliographystyle{aasjournal}



\end{document}